\documentclass{biometrika}

\usepackage{amsmath}
\usepackage{amsfonts}
\usepackage{mathtools}

\usepackage{newtxtext}
\usepackage[subscriptcorrection]{newtxmath}
\usepackage{bm}
\usepackage{bbm}

\usepackage{graphicx}
\usepackage{subcaption}
\usepackage{booktabs}
\usepackage[dvipsnames]{xcolor}
\usepackage{enumitem}
\usepackage{dirtytalk}

\usepackage{tikz}
\usetikzlibrary{arrows.meta,positioning,calc}
\usetikzlibrary{decorations.pathreplacing}
\usepackage{pgfplots}
\pgfplotsset{compat=1.18}
\usepgfplotslibrary{fillbetween}
\usepgfplotslibrary{groupplots}

\usepackage[unicode,bookmarks,hypertexnames=false]{hyperref}
\usepackage{bookmark}
\usepackage[capitalise]{cleveref}
\usepackage{flafter}
\usepackage{placeins}

\graphicspath{{./}{Figures/}}

\tikzstyle{dag}=[
    scale=0.9,
    transform shape,
    >=Latex,
    observed/.style={
        circle,
        draw,
        minimum size=0.75cm,
        font=\small
    },
    latent/.style={
        circle,
        draw,
        dashed,
        minimum size=0.75cm,
        font=\small
    },
    forbiddenlatent/.style={
        circle,
        draw=red!75!black,
        text=red!75!black,
        dashed,
        thick,
        minimum size=0.75cm,
        font=\small
    },
    arrow/.style={->, thick},
    conf/.style={->, thick, dashed},
    forbidden/.style={->, thick, dashed, red!75!black}
]

\DeclareMathOperator*{\argmax}{arg\,max}

\newcommand{\indep}{\perp \!\!\! \perp}

\makeatletter
\newcommand\paragraph{\@startsection{paragraph}{4}{\z@}%
  {2.25ex \@plus 1ex \@minus .2ex}{-1em}%
  {\normalfont\normalsize\itshape}}
\makeatother

\makeatletter
\newcommand\supplementarymaterial{%
  \clearpage
  \setcounter{page}{1}%
  \setcounter{section}{0}\setcounter{equation}{0}%
  \setcounter{figure}{0}\setcounter{table}{0}%
  \setcounter{theorem}{0}\setcounter{lemma}{0}\setcounter{corollary}{0}%
  \setcounter{proposition}{0}\setcounter{definition}{0}%
  \setcounter{assumption}{0}\setcounter{remark}{0}\setcounter{example}{0}%
  \renewcommand\thesection{S\arabic{section}}%
  \renewcommand\theequation{S\arabic{equation}}%
  \renewcommand\thefigure{S\arabic{figure}}%
  \renewcommand\thetable{S\arabic{table}}%
  \renewcommand\thetheorem{S\arabic{theorem}}%
  \renewcommand\thelemma{S\arabic{lemma}}%
  \renewcommand\thecorollary{S\arabic{corollary}}%
  \renewcommand\theproposition{S\arabic{proposition}}%
  \renewcommand\thedefinition{S\arabic{definition}}%
  \renewcommand\theassumption{S\arabic{assumption}}%
  \renewcommand\theremark{S\arabic{remark}}%
  \renewcommand\theexample{S\arabic{example}}%
  \global\let\@title\@savedtitle
  \@suppmaketitle
}
\makeatother

\theoremstyle{plain}
\theoremsymbol{}
\theoremheaderfont{\hskip\parindent\normalfont\scshape}
\theorembodyfont{\itshape}
\theoremseparator{.}
\theoremnumbering{arabic}

\newtheorem{innercustomthm}{Theorem}
\newenvironment{customthm}[1]
  {\renewcommand\theinnercustomthm{#1}\innercustomthm}
  {\endinnercustomthm}

\newtheorem{innercustomlem}{Lemma}
\newenvironment{customlem}[1]
  {\renewcommand\theinnercustomlem{#1}\innercustomlem}
  {\endinnercustomlem}

\newtheorem{innercustomcor}{Corollary}
\newenvironment{customcor}[1]
  {\renewcommand\theinnercustomcor{#1}\innercustomcor}
  {\endinnercustomcor}

\theoremheaderfont{\hskip\parindent\normalfont\itshape}
\theorembodyfont{\normalfont}

\newenvironment{myproof}{%
  \par\medskip\noindent
  \textit{Proof}.
}{%
\newline
\rightline{\QEDlogo}
}

\hypersetup{
    colorlinks=true,
    linkcolor={blue!50!black},
    bookmarksnumbered=true,
    bookmarksopen=true,
    bookmarksopenlevel=1,
    bookmarksdepth=2,
    pdfpagemode=UseOutlines,
    pdfstartview=FitH,
    citecolor={blue!50!black},
    urlcolor={blue!80!black}
}

\begin{document}

\jname{Biometrika}
\jyear{2026}
\jvol{113}
\jnum{1}
\cyear{2026}

\markboth{J. Bodik and N. Thams}{Extreme proxies for causal identification}

\title{Extreme proxies for causal identification: deconfounding with outliers}

\author{JURAJ BODIK}
\affil{Department of Statistics, Stanford University, California 94305, U.S.A.
\email{jbodik@stanford.edu}}

\author{\and NIKOLAJ THAMS}
\affil{Department of Mathematical Sciences, University of Copenhagen,\\ Universitetsparken 5, DK-2100 Copenhagen \O, Denmark
\email{thams@math.ku.dk}}

\makeatletter\let\@savedtitle\@title\makeatother
\maketitle
\pdfbookmark[0]{Main paper}{nav.main}

\begin{abstract}
Causal identification under unobserved confounding typically relies on additional structure, such as a valid instrument or proxy variables. We show that heavy tailed variables offer an alternative route to identification by revealing treatment variation for which confounding becomes asymptotically negligible. 
Our key assumption is \emph{extreme propagation}: when an unobserved confounder takes an extreme value, some observed variable in the system (called a \emph{safeguard}) will become extreme with large probability. This way, selecting extreme treatment observations while safeguards remain non-extreme makes confounding asymptotically negligible relative to the treatment. We call this \emph{certification}. 
We derive graphical conditions specifying which latent confounders must be monitored by safeguards and distributional conditions ensuring effective certification. 
The framework includes valid-instrument and double-proxy structures as special cases, as well as confounded instruments that are invalid in most approaches that do not exploit heavy tails. 
Geometrically, certification isolates the extremal ray whose slope equals the causal effect. This connects causal identification with multivariate extreme-value geometry and motivates a family of novel simple estimators. Simulations demonstrate performance gains under nonlinear confounding and heavy tails, while revealing limitations in light-tailed regimes (where cancellation can conceal extreme confounding from the safeguards). We validate our methodology in simulation studies and by recovering the ground-truth flow in a river discharge application.

\end{abstract}


\section{Introduction}

Causal inference seeks to distinguish the effects of interventions from statistical associations \citep{Fisher1935}. When randomized experiments are infeasible \citep{Imbens_Rubin_book}, one must rely on observational data, where unmeasured confounding can prevent identification. 

An instrumental variable can restore identification, but only if it is independent of the latent confounding and affects the outcome exclusively through the treatment \citep{Pearl_book}. Proxy-based methods weaken this requirement by using observed variables that carry information about the latent confounder, but typically require proxies informative enough to satisfy completeness or rank conditions \citep{MiaoGengTchetgenTchetgen2018,TchetgenTchetgenYingCuiShiMiao2020,KallusMaoUehara2021}.  We show that substantially less can suffice: rather than recovering the full distribution of the latent confounder, it is enough to detect when the confounder is extreme.

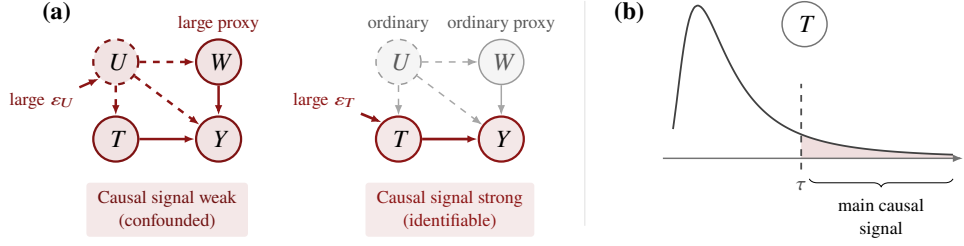
\begin{figure}[t]
\centering
\scalebox{0.85}{%
\begingroup
\definecolor{confoundorange}{RGB}{140,21,21}
\definecolor{causalblue}{RGB}{140,21,21}
\definecolor{proxyviolet}{RGB}{120,70,170}
\tikzset{
  obs/.style={circle,draw=black!55,fill=white,minimum size=7mm,
              inner sep=1pt,font=\normalsize,line width=.65pt},
  lat/.style={obs,dashed},
  quiet/.style={draw=black!40,fill=black!4,text=black!65},
  uhot/.style={draw=confoundorange!90!black,fill=confoundorange!12,line width=1pt},
  thot/.style={draw=causalblue,fill=causalblue!10,line width=1pt},
  arr/.style={-{Latex[length=1.6mm]},draw=black!35,line width=.65pt},
  uarr/.style={arr,draw=confoundorange,line width=1.1pt},
  tarr/.style={arr,draw=causalblue,line width=1.2pt},
  smallnote/.style={font=\footnotesize,align=center,inner sep=1.5pt},
  badge/.style={rounded corners=1.5pt,inner xsep=5pt,inner ysep=3.5pt,
                 font=\footnotesize,align=center},
  source/.style={font=\footnotesize,inner sep=2pt,fill=white},
  panel/.style={font=\bfseries,anchor=north west,inner sep=0pt},
}
\begin{tikzpicture}
\path[use as bounding box] (0,-0.1) rectangle (14.6,3.65);
\node[panel] at (0,3.6) {(a)};
\begin{scope}[shift={(1.95,2.05)}]
  \node[lat,uhot] (U) at (-0.8,.6) {$U$};
  \node[obs,uhot] (W) at (0.8,.6) {$W$};
  \node[obs,uhot] (T) at (-0.8,-.6) {$T$};
  \node[obs,uhot] (Y) at (0.8,-.6) {$Y$};
  \draw[uarr,dashed] (U)--(W);
  \draw[uarr,dashed] (U)--(T);
  \draw[uarr,dashed] (U)--(Y);
  \draw[uarr] (T)--(Y);
  \draw[uarr] (W)--(Y);
  \node[source,text=confoundorange!90!black] (eps) at (-1.95,0.0)
    {large $\varepsilon_U$};
  \draw[uarr] (eps.north east)--(U.south west);
  \node[smallnote,text=confoundorange!90!black,anchor=south] at (0.8,1.0)
    {large proxy};
\end{scope}
\node[badge,fill=confoundorange!10,text=confoundorange!80!black] at (1.95,0.35)
  {Causal signal weak\\(confounded)};
\begin{scope}[shift={(6.35,2.05)}]
  \node[lat,quiet] (U) at (-0.8,.6) {$U$};
  \node[obs,quiet] (W) at (0.8,.6) {$W$};
  \node[obs,thot] (T) at (-0.8,-.6) {$T$};
  \node[obs,thot] (Y) at (0.8,-.6) {$Y$};
  \draw[arr,dashed] (U)--(W);
  \draw[arr,dashed] (U)--(T);
  \draw[arr,dashed] (U)--(Y);
  \draw[arr] (W)--(Y);
  \draw[tarr] (T)--(Y);
  \node[source,text=causalblue] (eps) at (-1.95,0.0)
    {large $\varepsilon_T$};
  \draw[tarr] (eps.south east)--(T.north west);
  \node[smallnote,text=black!65,anchor=south] at (-0.8,1.0)
    {ordinary};
  \node[smallnote,text=black!65,anchor=south] at (0.8,1.0)
    {ordinary proxy};
\end{scope}
\node[badge,fill=causalblue!9,text=causalblue] at (6.35,0.35)
  {Causal signal strong\\(identifiable)};
\draw[black!12] (8.45,0.1)--(8.45,3.6);
\node[panel] at (8.9,3.6) {(b)};
\begin{scope}[shift={(9.35,1.15)},x=1.05cm,y=2.9cm,
              declare function={dens(\x)=2*\x^(-3)*exp(-\x^(-2));}]
  \pgfmathsetmacro{\tauX}{2.35}
  \pgfmathsetmacro{\xEnd}{4.6}
  \fill[causalblue!14]
    plot[domain=\tauX:\xEnd,samples=40,smooth] (\x,{dens(\x)})
    -- (\xEnd,0) -- (\tauX,0) -- cycle;
  \draw[-{Latex[length=1.6mm]},black!55,line width=.65pt]
    (0.3,0) -- (\xEnd+0.15,0);
  \draw[black!75,line width=.8pt]
    plot[domain=0.45:\xEnd,samples=80,smooth] (\x,{dens(\x)});
  \draw[dashed,thick,black!70] (\tauX,-0.06) -- (\tauX,0.40)
    node[below=1pt,at start,font=\small] {$\tau$};
  \node[obs] at (2.4,0.72) {$T$};
  \draw[decorate,decoration={brace,amplitude=3.5pt,mirror}]
    (\tauX+0.12,-0.11) -- (\xEnd,-0.11)
    node[midway,below=5pt,font=\footnotesize,align=center]
    {main causal\\signal};
\end{scope}
\end{tikzpicture}
\endgroup%
}
\caption{
\textbf{(a)} Two different origins of an extreme value in $T$
(extreme values are coloured red).
Left: the extreme originates in the latent \(U\), and
propagates to \(T\), $Y$ and \(W\). The effect $T\to Y$ is confounded
by the extreme $U$.
Right: the extreme originates in \(T\) and propagates to $Y$, but
\(U\) and the proxy \(W\) remain ordinary. The confounding of $T$
and $Y$ is negligible relative to the extreme signal $T \to Y$.
We identify the origin of the extreme from whether $W$ is extreme
(assuming extreme propagation).
\textbf{(b)} Extreme treatment observations carry the main causal
signal, since they are the observations in which the causal
contribution of $T$ can dominate the confounding.
}
\label{fig:extreme-propagation-dags-clean}
\end{figure}

We therefore study what instrument-like and proxy-like variables can identify through their extremes. Informally, an extreme is an observation lying far in a tail of its distribution (Fig.~\ref{fig:extreme-propagation-dags-clean}b); for a continuous variable, a value exceeding a high threshold. Extremeness could more generally describe rare or atypical configurations, such as an unusual object in an image, but we focus on right-tail extremes of continuous variables. This setting is natural when unusually large observations have a meaningful interpretation, such as extreme financial losses, environmental measurements, demand shocks, or physiological responses \citep{Coles,Resnick,EngelkeSparsity}.
Our central assumption is \emph{extreme propagation}: 
\begin{definition}[Extreme propagation (informal)]
    A cause variable $U$ and effect variable $S$ exhibit \emph{extreme propagation} if 
    \begin{equation}
        U \text{ takes extreme value} \implies 
        S \text{ takes extreme value with high probability}.
    \end{equation}
\end{definition}
That is, the causal mechanism from $U$ to $S$ does not ``hide'' extreme values of $U$. Such propagation arises naturally in many physical models and has been studied in recent work on causal models for extremes \citep{bodik_extremes, Gnecco, courgeau2021extreme, engelke2025extremesstructuralcausalmodels, leimenstoll2026identificationinferencecausaleffects} and information-theoretic outlier propagation \citep{Ebtekar, orchard2026falsifyingcausalgraphsoutlier}. 

We use observed variables, which we call \emph{safeguards}, to monitor whether an extreme treatment originated in treatment-specific noise $\varepsilon_T$ or in a hidden confounder $U$. 
When the extreme originates in the treatment, the causal signal dominates the latent confounding, and the treatment effect is approximately identified. 
In general causal structures, the extreme may originate in the treatment itself or in an observed driver of the treatment, and we may use one or several safeguards to screen out extremes in confounders. 

Unlike the proxies of classical proxy methods \citep{MiaoGengTchetgenTchetgen2018,TchetgenTchetgenYingCuiShiMiao2020}, a safeguard need not reconstruct the latent confounder or satisfy a completeness condition. It only needs to detect every latent ancestor that can produce a confounded source extreme. A valid instrument is the special case of a source that requires no safeguards. Conversely, an invalid instrument may still be useful as a safeguard, or as a source if other observed variables detect its confounded extremes.

Our main contribution is to formalise these notions and give sufficient conditions for identification of the causal effect. 
The conditions separate into a graphical part and a distributional part. 
The graphical part determines which latent causes of an extreme must be monitored, and the distributional part ensures that the system is heavy-tailed enough, and propagates extremes reliably enough, for the safeguards to monitor them.
When both hold, extreme observations reveal the effect of \(T\) on \(Y\) without reconstructing the latent confounders. 
Our intention is not to propose certification to replace instrumental-variable or proxy estimators when their assumptions hold; instead, we point out that a small number of certified data points carry enough information to identify the effect at comparable accuracy (and that it also does so in structures in which no valid instrument or bridge function exists). 

Section~\ref{sec:causal-setup} defines notation and reviews instruments, proxies and heavy tails. Section~\ref{Section_3} gives the formal certification and identification results, and Section~\ref{sec:causal-pot} their geometric interpretation in multivariate extreme value theory. Section~\ref{sec:estimation} derives estimators and convergence rates. Section~\ref{sec:simulations} reports simulations and Section~\ref{sec:application} an application to river discharge.

\section{Background: notation and latent confounding}
\label{sec:causal-setup}

We consider a scalar treatment $T$, an outcome $Y$, an unobserved variable $U$ collecting common causes of $T$ and $Y$, and observed auxiliary variables $W\in\mathbb{R}^{d_W}$.
We use the same symbols for random variables and their realisations. 
When sampling assumptions are needed, we consider independent copies of $(Y,T,W, U)$. All variables are real-valued, and the main text focuses on their upper tails; lower-tail extremes are analogous.

We use $\overset{d}{\to}$ to denote convergence in distribution. For random variables $X, L$ and a family of events $(A_v)_{v\in\mathbb{R}}$, we write $X\mid A_v \overset{d}{\to} L$ as $v\to\infty$ for weak convergence of the conditional laws $\mathbb{P}(X\in \cdot\mid A_v)$ to $\mathbb{P}(L \in \cdot\,)$, and $X\mid A_v \overset{d}{\to} \infty$ if $\mathbb P(X\le c\mid A_v)\to 0$ for every finite $c$. 

\subsection{Causal effects: potential outcomes and structural causal models}

A causal effect describes how the distribution of $Y$ changes when the
mechanism generating $T$ is externally modified. In intervention notation \citep{Pearl_book},
the mean response is $\mathbb E\{Y\mid \operatorname{do}(T=t)\}$, and in potential-outcomes notation \citep{rubin2005} it is $\mathbb{E}[Y(t)]$, where $Y(t)$ is the potential outcome under $T=t$. 

The simplest model has an additive linear causal response, $Y=\beta_0 T+R,$ where $R$ collects all outcome variation not transmitted through the edge $T\to Y$. $R$ may depend on $U$, on observed variables such as $W$, and on an outcome-specific disturbance. The coefficient $\beta_0$ is the average causal effect of a unit increase in $T$, $\mathbb E[Y(t+1)-Y(t)]=\beta_0$: changing $T$ by one unit while leaving the remaining outcome mechanism unchanged changes $Y$ by $\beta_0$. 

We use structural causal models over a directed acyclic graph $G$, that is, directed graphs and assignments of the form
\[
    V_j=f_j\bigl(V_{\operatorname{pa}(j)},\varepsilon_j\bigr), \qquad \varepsilon_j\text{ are mutually independent,} 
\]
to encode the causal ordering of the variables and the channels through which large values propagate. An arrow $V_k\to V_j$ indicates that
the mechanism for $V_j$ may depend on $V_k$, and $\varepsilon_j$ represents variation specific to that mechanism; standard references include
\citet{Pearl_book} and \citet{Peters2017}. Write $\mathrm{an}(X)$ for the ancestors of $X$, \emph{including} $X$ itself, and $\mathrm{an}_{G^-}(X)$ for the ancestors in the graph $G^-$ obtained from $G$ by removing the edge $T\to Y$; we identify a node $j$ with its variable $V_j$.

\subsection{From instruments to proxy variables}
An instrumental variable $Z$ is independent of latent confounding $U$ and affects $Y$ only through $T$ \citep{Imbens2014}. The causal effect of $T$ on $Y$ can then be estimated using $Z$ as a pseudo-randomisation of $T$. These conditions are often difficult to justify: strong drivers of treatment
often share unobserved causes with the outcome, and such a variable is predictive of $T$ but not a valid instrument.

Proxy-variable methods \citep{TchetgenTchetgenYingCuiShiMiao2020} use observed variables that are informative about $U$. A treatment-inducing proxy $Z$ is associated with $T$, possibly only through $U$, and satisfies $Z\indep Y\mid(T,U)$; this is the instrument assumption without independence from $U$. An outcome-inducing proxy $W$ is informative about $U$ and satisfies $W\indep(T,Z)\mid U$. These restrictions are closely related to negative-control exposure and outcome assumptions \citep{MiaoGengTchetgenTchetgen2018}. 

In proximal identification, the two proxies enter a conditional moment equation for an outcome bridge function, which identifies the causal effect under suitable completeness conditions. Extensions include minimax and kernel methods, partial identification, and identification using a single proxy under additional structure \citep{KallusMaoUehara2021,MastouriZhuGultchinKorbaSilvaKusnerGrettonMuandet2021,GhassamiShpitserTchetgenTchetgen2023,VollmerPfisterWeichwald2026}. 
We employ similar structural assumptions to proxy-variable methods, but use the proxies differently. 
A proxy need not reconstruct $U$ over its full range, and our results do not invoke a bridge function or completeness. 
What matters is \emph{tail informativeness}: a non-extreme proxy value must rule out, or make
asymptotically negligible, an extreme $U$.

\subsection{Extremes and causality}

Extreme value theory studies extreme observations such as exceedances $\{X>v\}$, typically through the limiting conditional law as $v\to\infty$; see \citet{Coles}. Multivariate extreme value theory studies which components can become extreme together and the sparse dependence structures in the tails \citep{EngelkeSparsity}. These
ideas are natural for studying whether an extreme input propagates to its descendants \citep{bai2026structuralcausalmodelsextremes}.

The intersection of causality and extremes has emerged only recently; this includes graphical models for extremal dependence
\citep{EngelkeGraphicalModels},  the extremal behaviour of structural
causal models \citep{engelke2025extremesstructuralcausalmodels}, and causal discovery from tail asymmetries and causal tail coefficients, also in heavy-tailed and time-series settings
\citep{Gnecco,Pasche,bodik_extremes,bodik_pasche,yin2025causaltailcoefficientcompound, routh2026heavytailedhomogeneousstructural}.
A separate literature studies treatment effects at extreme outcome quantiles
\citep{zhang2018extremal,Deuber} and extrapolation \citep{engelke2026extrapolationstatisticallearningextreme}.

Closest to our work, \citet{wang2025causaleffectidentificationinference} exploit light-tailed errors and tail separation to identify effects under endogeneity. They estimate the average causal effect as a difference between two extreme conditional quantiles, provided that the noise enters additively and the hidden confounder is light-tailed. We instead treat extremal observations as a source of quasi-exogenous variation, and rely on heavy rather than light tails. 

\subsection[Heavy tails and regular variation]{Heavy tails and regular variation}

Regular variation is the standard framework for describing the tails of random variables \citep{Resnick, Coles, RegularVariationBook, clanok_o_regularly_varying_densities}. 
Let \(F\) be a distribution function with unbounded upper support and write $\overline F=1-F$; \(F\) has a \emph{regularly varying upper tail} with tail index \(\alpha>0\), written $F\in RV(\alpha)$, if
\(
1- F(x)
    =
    x^{-\alpha}L(x), 
\)
where \(L\) is slowly varying; that is, $\lim_{t\to\infty} \frac{L(tx)}{L(t)} = 1$ for every $x>0$. 
A random variable $X$ is regularly varying with index $\alpha$ if $|X|$ has a regularly varying tail with index $\alpha$ and the two tails are balanced, $\mathbb P(X>x)/\mathbb P(|X|>x)\to p\in[0,1]$. The index $\alpha$ describes how fast the tail decays; the larger $\alpha$, the faster the decay. For example, if $\alpha<2$ then the variance is infinite. 

We also use a weaker notion of heavy tails. A distribution $F$ is \textit{subexponential} in its upper tail if $\frac{\overline F(x+t)}{\overline F(x)}\to 1$ as $x\to\infty$ for every fixed $t>0$, and if, for independent $X_1,X_2\sim F$, $\mathbb P(X_1+X_2>x) \sim 2\overline F(x)$. 
A sum is then asymptotically driven by one exceptionally large summand rather than by several moderately large ones \citep{FossKorshunovZachary2013}. 
Regularly varying distributions (such as Pareto and Student's $t$) are subexponential, and so are the lognormal distribution and Weibull distributions with shape parameter below one, which are not regularly varying. 

The Gaussian distribution is light-tailed and, for our purposes, essentially bounded: even in a sample of one million observations, it is unlikely that any observation lies more than five standard deviations above the mean. 

\section{Certifying events and extreme propagation}
\label{Section_3}

We give conditions under which a structural model generates \emph{certifying} events, that is, events on which the hidden confounding between $T$ and $Y$ is asymptotically negligible. Throughout this section, we assume a linear effect $\beta_0$; Section~\ref{sec:beyond-regular-variation-evt} discusses generalisations.

\begin{assumption}[Linear effect]
    \label{ass:constant-linear-causal-effect}
    The structural equation for $Y$ has the form $Y=\beta_0 T+R$ with $R=h_Y(V_{\mathrm{pa}(Y)\setminus\{T\}},\varepsilon_Y)$, for a continuous function $h_Y$; that is the treatment enters the assignment of $Y$ linearly, and the remainder $R$ does not depend on $T$.
\end{assumption}

The remainder $R$ may depend arbitrarily on latent confounders; for example, $R = h_Y(U, W, \varepsilon_Y)$ in Model~\mbox{\eqref{SCM_general}}. Since $R$ and $T$ are dependent, $\beta_0$ is not identified without further assumptions. Linearity is assumed only for the edge $T\to Y$, not the rest of the system. 
As motivated in the introduction, we focus on events that carry the strongest causal signal and are asymptotically unconfounded in the following sense.

\begin{definition}[Certifying sequence]
\label{def:certifying-sequence}
A family of events $(A_v)_{v\in\mathbb R}$ with $\mathbb P(A_v)>0$ is a \emph{certifying sequence} for a parameter $\beta$ if there exists a fixed measurable function $g$ such that
\begin{equation}
    \label{CONVERGENCE}
    g(Y,T)\mid A_v
    \overset{d}{\longrightarrow}
    \beta
    \qquad\text{as }v\to\infty.
\end{equation}
\end{definition}
In the linear setting $Y = \beta_0 T + R$, taking $g(y, t) = y/t$, a certifying sequence is obtained by requiring $T$ to be increasingly extreme while $R$ remains bounded, since $Y/T=\beta_0+R/T\approx \beta_0$. The next lemma formalizes this observation.

\begin{lemma}[Certifying sequence]
\label{lem:certifying-sequence}
\label{lemma}
Let Assumption~\ref{ass:constant-linear-causal-effect} hold, and let $(A_v)_{v\in\mathbb{R}}$ be a family of events with  $\mathbb P(A_v)>0$ for all $v$ satisfying:
\begin{enumerate}[leftmargin=2em]
    \item \textbf{$T$ is extreme.}  $T\,|\,A_v \overset{d}{\to} \infty$ as $v\to\infty$. That is, for every finite $c$,
    \begin{equation}
        \mathbb P(T\leq c\mid A_v)\longrightarrow0,
        \qquad v\to\infty.
        \label{eq:selected-treatment-diverges}
    \end{equation}
   
    \item \textbf{$R$ is not extreme.}  For every $\epsilon>0$, there is a finite $M$, not depending on $v$, such that
    \begin{equation}
        \limsup_{v\to\infty}
        \mathbb P\bigl(|R|>M\mid A_v\bigr)<\epsilon.
        \label{eq:selected-remainder-controlled}
    \end{equation}
\end{enumerate}
Then $(A_v)_{v\in\mathbb R}$ is a certifying sequence and $\beta_0$ is identified from \eqref{CONVERGENCE} with $g(y, t) = \frac{y}{t}$. 
\end{lemma}

The first condition ensures that the selected events contain increasingly large treatment values. 
The second formalizes the certification argument from the introduction: $R$ need not be bounded or have light tails, but the selection must prevent it from becoming extreme on the same scale as $T$. 
The lemma thus requires no moment assumptions on $R$, only conditional tightness along the selected events.

\subsection{Monitoring of confounders yields certification}\label{sec:certification-under-monitoring}

Consider an SCM over a DAG $G$ with vertices $V\supseteq\{T,Y\}$, observed vertices $O\subseteq V$, and assignments $V_j=f_j(V_{\mathrm{pa}(j)},\varepsilon_j)$ with mutually independent noises and continuous $f_j$, as in Section~\ref{sec:causal-setup}. The target is the causal effect $\beta_0$ along the edge $T \to Y$. 
Typical structures are shown in Fig.~\ref{fig:examples-of-structures-identified}.

We rely on a source $E$ whose extremes propagate to $T$, and an observed safeguard $S$ whose non-extreme values certify that the confounders are non-extreme. 
The certification is probabilistic: an extreme confounder can hide in a safeguard if a compensating noise cancels it (as $\varepsilon_W\approx-\psi_W U$ would in the linear model $W=\psi_W U+\varepsilon_W$). Our criterion requires only that such cancellation is asymptotically less probable than the source exceedance.

\begin{definition}[Source--safeguard pair]
\label{def:anchor-safeguard-prob}
A \emph{source--safeguard pair} $(E,S)$ consists of an observed scalar \emph{source} $E\in O$ and an observed \emph{safeguard set} $S\subseteq O$, together with a compact set $\mathcal B$ in the outcome space of $S$ with $\mathbb P(S\in\mathcal B)>0$. The pair generates the \textit{candidate certifying events}
\[
    A_v^{E,S}:=\{E>v,\ S\in\mathcal B\},\qquad v\in\mathbb{R}.
\]
If $S=\emptyset$, we set $A_v^{E,S} := \{E > v\}$. 
\end{definition}

The source and the safeguard play complementary roles: the source creates the extreme causal signal, and the safeguard rules out confounded explanations of that signal. At this stage, $A_v^{E,S}$ is only a candidate certifying sequence: its validity depends on whether the safeguard actually controls the relevant latent variables.

\begin{definition}[Monitoring]
\label{def:prob-monitoring}
Assume $\mathbb P(A_v^{E,S})>0$ for all $v$. A node $V_j$ is \emph{monitored almost surely} along $(A_v^{E,S})$ if there is a compact $K$, not depending on $v$, with $\mathbb P(V_j\notin K\mid A_v^{E,S})=0$ for all $v$. $V_j$ is \emph{monitored in probability} along $(A_v^{E,S})$ if the conditional distributions given $A_v^{E,S}$ are tight: for every $\epsilon>0$ there is a compact $K$, not depending on $v$, with
\[
    \limsup_{v\to\infty}\ \mathbb P\bigl(V_j\notin K\ \big|\ A_v^{E,S}\bigr)\ <\ \epsilon.
\]
\end{definition}
\begin{figure}[t]
\centering
\tikzset{condrow/.style={dag, scale=0.7, baseline=(current bounding box.south)}}
\captionsetup[sub]{labelformat=empty,justification=centering,margin=4pt,skip=2pt}
\captionsetup{skip=4pt}
\vspace{-4pt}

\scalebox{0.85}{%
\begin{subfigure}[t]{0.33\textwidth}%
    \centering
    \begin{tikzpicture}[condrow]
        \node[observed] (E) at (0,0) {$E$};
        \node[observed] (T) at (2,0) {$T$};
        \node[observed] (Y) at (4,0) {$Y$};
        \node[observed] (S) at (4,1.15) {$S$};
        \node[latent] (U) at (2,1.15) {$U$};
        \draw[arrow, line width=1.6pt, ForestGreen] (E) -- (T);
        \draw[arrow] (T) -- node[above] {$\beta_0$} (Y);
        \draw[conf] (U) -- (T);
        \draw[conf] (U) -- (Y);
        \draw[conf] (U) -- (S);
    \end{tikzpicture}
    \caption{\labelcref{ass:E-affects-T}: $E$ is an ancestor of $T$}
    \label{fig:condition-C1}
\end{subfigure}\hfill
\begin{subfigure}[t]{0.33\textwidth}%
    \centering
    \begin{tikzpicture}[condrow]
        \node[observed] (E) at (0,0) {$E$};
        \node[observed] (T) at (2,0) {$T$};
        \node[observed] (Y) at (4,0) {$Y$};
        \node[observed] (S) at (4,1.15) {$S$};
        \node[latent] (U) at (2,1.15) {$U$};
        \draw[arrow] (E) -- (T);
        \draw[arrow] (T) -- node[above] {$\beta_0$} (Y);
        \draw[conf] (U) -- (T);
        \draw[conf] (U) -- (Y);
        \draw[conf] (U) -- (S);
        \draw[forbidden] (E) to[bend left=75] (S);
        \draw[forbidden] (E) to[bend right=30] (Y);
    \end{tikzpicture}
    \caption{\labelcref{ass:E-not-affects-SY}: no edge $E \to S$, no directed path $E$ to $Y$ avoiding $T$}
    \label{fig:condition-C2}
\end{subfigure}\hfill
\begin{subfigure}[t]{0.33\textwidth}%
    \centering
    \begin{tikzpicture}[condrow]
        \node[observed] (E) at (0,0) {$E$};
        \node[observed] (T) at (2,0) {$T$};
        \node[observed] (Y) at (4,0) {$Y$};
        \node[observed] (S) at (5,1.15) {$S$};
        \node[latent] (U) at (3,1.15) {$U$};
        \node[forbiddenlatent] (U2) at (1,1.15) {$U'$};
        \draw[arrow] (E) -- (T);
        \draw[arrow] (T) -- node[above] {$\beta_0$} (Y);
        \draw[conf] (U) -- (E);
        \draw[conf] (U) -- (Y);
        \draw[conf, line width=1.6pt, ForestGreen] (U) -- (S);
        \draw[forbidden] (U2) -- (E);
        \draw[forbidden] (U2) -- (Y);
    \end{tikzpicture}
    \caption{\labelcref{ass:no-unmonitored-confounding}: every confounder $U$ of $E$ and $Y$ is an ancestor of $S$}
    \label{fig:condition-C3}
\end{subfigure}%
}

\caption{Illustrating the graphical conditions of \cref{thm:unified-certification-prob}. Dashed nodes denote latent variables; red dashed nodes and arrows are structures that the condition excludes. Bold green edges indicate assumed extreme propagation. 
}
\label{fig:graphical-conditions}
\vspace{-10pt}
\end{figure}
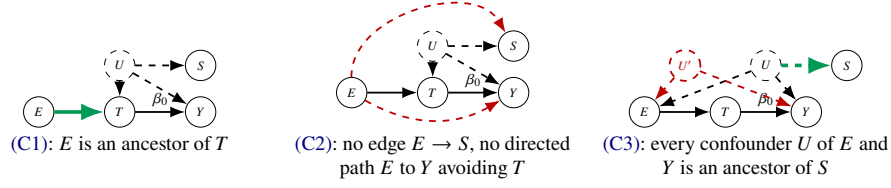

Intuitively, monitoring in probability means that when $S$ is observed to be non-extreme, $V_j$ is unlikely to be extreme, even as $E$ becomes arbitrarily large. Almost-sure monitoring implies monitoring in probability. Three examples, each with a single latent $U$ and a scalar safeguard $S$, as in Fig.~\ref{fig:extreme-propagation-dags-clean}(a) with $E=T$ and $S=W$:
\begin{enumerate}[label=\alph*.]
    \item \emph{Positive linear:} Let $S=\psi_S U+\varepsilon_S$ and $E=\psi_E U+\varepsilon_E$ with $U, \psi_S, \psi_E, \varepsilon_S, \varepsilon_E>0$, and $\mathcal B=[0,\tau]$. Then $S>\psi_SU$, so $\mathbb P(U \notin [0, \tau/\psi_S] \, |\, S \leq \tau) = 0$; monitoring holds almost surely, not just in probability.
    \item \emph{Gaussian proxy noise:} For the same model except that $\varepsilon_S\sim N(0,\sigma^2)$ and $\mathcal B=[-\tau,\tau]$, an extreme $U$ can hide inside $\{|S|\leq\tau\}$ through the cancellation $\varepsilon_S\approx-\psi_SU$. This has positive probability, but only of order $e^{-cU^2}$ for some $c>0$. If $\varepsilon_E$ is sufficiently heavy-tailed, monitoring holds in probability but not almost surely. 
    Section~\ref{sec:extreme-propagation-monitoring} makes ``sufficiently heavy-tailed'' precise, but as the next example shows, a Gaussian $\varepsilon_E$ is not.
    \item \emph{Fully Gaussian:} If $E, U, S$ are all Gaussian (and $\psi_E,\psi_S\neq0$), cancellation becomes the \emph{most plausible} explanation of $\{E>v,S\in\mathcal B\}$ and monitoring in probability fails; see Example~\ref{ex:necessity-tail-race}.
\end{enumerate}

We now give general graphical and distributional conditions for certification. The graphical conditions determine which ancestors of the source also affect the non-causal part of the outcome; monitoring then ensures that these ancestors remain non-extreme on the selected events.

\begin{theorem}[Certification by graphical coverage under monitoring]
\label{thm:unified-certification-prob}

    Let Assumption~\ref{ass:constant-linear-causal-effect} hold, and let $(E,S)$ be a source--safeguard pair. Suppose that:
    \begin{enumerate}[label=\textup{(C\arabic*)},leftmargin=2.6em]
        \item \label{ass:E-affects-T}
        \emph{$E\in\mathrm{an}(T)$ and extremes of $E$ propagate to $T$ along $(A_v^{E,S})$:} $\mathbb P(A_v^{E,S})>0$ for all $v$ and, for every finite $c$,
        \[
            \mathbb P\!\left(
                T\leq c
                \,\middle|\,
                A_v^{E,S}
            \right)
            \overset{v\to\infty}{\longrightarrow}0.
        \]

        \item \label{ass:E-not-affects-SY}
        \emph{The source does not cause the safeguard and affects $Y$ only through $T$:}
        $E\notin\mathrm{an}(S)$ and
        $E\notin\mathrm{an}_{G^-}(Y)$.

        \item \label{ass:no-unmonitored-confounding}
        \emph{Every confounder of $E$ and $Y$ is an ancestor of $S$:}
        $\mathrm{an}(E)\cap\mathrm{an}_{G^-}(Y)
        \subseteq\mathrm{an}(S)$.
    \end{enumerate}
    If additionally 
    \begin{equation}
        \tag{M}
        \label{ass:confounders-monitored-in-probability}
        \text{every node in } \mathcal{C}_S
        := \operatorname{an}_G(S)\cap \operatorname{an}_{G^-}(Y) 
        \text{ is monitored in probability along } (A_v^{E,S})_v,
    \end{equation}
    then $(A_v^{E,S})_v$ is a certifying sequence for $\beta_0$ with $g(y,t)=y/t$, and $\beta_0$ is identified.
\end{theorem}
\Cref{fig:graphical-conditions} illustrates the three conditions.
When $E=T$, the distributional part of \labelcref{ass:E-affects-T} is immediate. 

The theorem reduces certification to a monitoring problem: its graphical conditions specify which latent variables must be monitored, and the next subsection gives structural and tail conditions under which the safeguard monitors them.

\subsection{Sufficient conditions for monitoring}\label{sec:extreme-propagation-monitoring}
We now outline conditions to achieve the monitoring assumption \eqref{ass:confounders-monitored-in-probability}.

We fix a source--safeguard pair $(E,S)$ and let $U\in\mathbb{R}^k$ denote the vector of the nodes in $\mathcal C_S\setminus S$, the latent variables that must be monitored. For simplicity, suppose that the reduced structural equations are
\begin{equation}
    U = f_U(\varepsilon_U),\qquad
    S = f_S(U,\varepsilon_S),\qquad
    E = f_E(U,S,\varepsilon_E),\qquad
    Y = \beta_0T+h_Y(U,S,\varepsilon_Y),
    \label{SCM_general}
\end{equation}
where $\varepsilon_U,\varepsilon_S,\varepsilon_E,\varepsilon_Y$ are mutually independent noise blocks and $h_Y$ is continuous.

\begin{definition}[Extreme propagation from $U$ to $S$]
\label{def:extreme-propagation}
\label{ass:structural-proxy-screening}
We say that \textit{weak extreme propagation from $U\to S$} holds if 
\begin{equation}
    \lim_{x\to\infty}\mathbb P\!\left(S\in\mathcal B\,\middle|\,\|U\|>x\right)=0.
\end{equation}
We say that \textit{strong extreme propagation from $U\to S$} holds if at least one of the following is satisfied:
\begin{enumerate}[label=(\roman*),leftmargin=2.5em]
    \item \emph{Negation form.} There exists a compact set $B_U$ such that
    \begin{equation}
        S\in\mathcal B
        \quad\Longrightarrow\quad
        U\in B_U
        \qquad\text{almost surely}.
        \label{eq:structural-proxy-screening}
    \end{equation}

    \item \emph{Structural form.} The function $f_S$ from \eqref{SCM_general} is continuous and
    \begin{equation}
        \lim_{\substack{\lVert u\rVert\to\infty\\ u\in\operatorname{supp}(U)}}  %
        \inf_{s\in\operatorname{supp}(\varepsilon_S)}
        \bigl\lVert f_S(u,s)\bigr\rVert
        =
        \infty.
        \label{eq:uniform-structural-propagation}
    \end{equation}

    \item \emph{Bounding form.}
    There exist a nondecreasing function
    $\gamma:[0,\infty)\to[0,\infty)$ with
    $\gamma(r)\to\infty$ as $r\to\infty$, and a random variable
    $\Lambda\geq0$, such that
    \begin{equation}
        \|S\|
        \geq
        \gamma(\|U\|)-\Lambda
        \qquad\text{almost surely},
        \label{eq:bounding-extreme-propagation}
    \end{equation}
    where $\Lambda$ is non-extreme in the sense that
    $\Lambda\mid A_v^{E,S}=O_{\mathbb P}(1)$.
\end{enumerate}
\end{definition}

Weak propagation only makes cancellation unlikely: a large $U$ may still be offset by a large negative $\varepsilon_S$, so that $S$ remains bounded. We therefore additionally assume that every extreme event is generated by one exogenous noise variable, which we call the single-big-jump property.

\begin{definition}[Single-big-jump SCM]
    \label{def:single-big-jump}
An SCM with reduced forms $V_j=g_j(\varepsilon_1,\ldots,\varepsilon_m)$ is a \emph{single-big-jump SCM} (SBJ-SCM) if
(i) $\varepsilon_1,\ldots,\varepsilon_m$ are i.i.d.\ and regularly varying with index $\alpha$, with $\lim_{x\to\infty}\mathbb P(\varepsilon_1>x)/\mathbb P(|\varepsilon_1|>x)>0$; and
(ii) every $g_j$ is continuous, \emph{tail-sublinear}, that is, $g_j=\phi_j+O(1)$ at infinity for some positively homogeneous and subadditive $\phi_j$, and nondegenerate in its own noise, $g_j(e_{-j},e_j)\to\infty$ as $e_j\to\infty$ for every $e_{-j}$.
\end{definition}

An SBJ-SCM formalizes the principle that an extreme endogenous value is asymptotically generated by one extreme exogenous noise, rather than by several coordinated moderate deviations. The i.i.d.\ assumption may be weakened to independence, provided that the source noise is regularly varying and not lighter-tailed than any other noise. 
Gaussian linear SCMs fall outside this class, since a Gaussian exceedance is more likely to be produced by two coordinated moderate deviations rather than by one big jump (\cref{ex:necessity-tail-race}); some heavy-tailed noise is therefore needed for weak extreme propagation to suffice for monitoring.

There is a route to monitoring that requires no propagation at all: $U$ is automatically monitored if its tail, after conditioning on the safeguard, is lighter than that of the source noise.
\begin{definition}[Tail dominance]
\label{tail_race}
We say that $E$ has tail dominance over $U$ if at least one of the following is satisfied:
\begin{enumerate}[label=(\roman*),leftmargin=2.5em]
\item \label{cond:tail-race-conditional}
\emph{$\varepsilon_E$ has heavier tails than $U\mid S$:}
The source has the form $ E=g_E(U,S)+\varepsilon_E,$ where \(\varepsilon_E\) is independent of \((U,S)\), \(g_E\) is globally Lipschitz, $\varepsilon_E$ has a subexponential upper tail, and, for every $a>0$,
\begin{equation*}
    \mathbb P\!\left(
        \|U\|>av
        \,\middle|\,
        S\in\mathcal B
        \right)
        =
        o\!\left(\mathbb P(\varepsilon_E>v)\right),
        \qquad v\to\infty.
\end{equation*}

\item \label{cond:tail-race-primitive}
\emph{\(\varepsilon_E\) has a heavier tail than the joint tail of $(\varepsilon_U,\varepsilon_S)$:}
The model is $U = \varepsilon_U$, $S = BU+\varepsilon_S$ with $B\in\mathbb{R}^{d_S\times k}$ of full column rank and $E=g_E(U,S)+\varepsilon_E$ with $g_E$ globally Lipschitz and $\varepsilon_E$ independent of $(U,S)$ with a subexponential upper tail, such that, for every $a,b>0$,
\[
    \mathbb P\!\left(
        \|\varepsilon_U\|>av,\,
        \|\varepsilon_S\|>bv
    \right)
    =
    o\!\left(\mathbb P(\varepsilon_E>v)\right),
    \qquad v\to\infty.
\]
\end{enumerate}
\end{definition}

We can now characterise sufficient conditions for \eqref{ass:confounders-monitored-in-probability}. 
\begin{theorem}[Extreme propagation or tail dominance $\implies$ monitoring]
\label{thm:extreme-propagation-tail-race-monitoring}
Consider model \eqref{SCM_general}. Let $(E,S)$ be a source--safeguard pair with $\mathbb P(A_v^{E,S})>0$ for all $v$. If one of the following holds:
\begin{enumerate}[label=(\roman*),leftmargin=2.5em]
    \item Strong extreme propagation from $U\to S$ (Definition~\ref{def:extreme-propagation}),
    \item Weak extreme propagation from $U\to S$ (Definition~\ref{def:extreme-propagation}) and the SCM is an SBJ-SCM (Definition~\ref{def:single-big-jump}),
    \item $E$ has tail dominance over $U$ (Definition~\ref{tail_race}),
\end{enumerate}
then $U$ is monitored in probability along $(A_v^{E,S})_v$, so that \eqref{ass:confounders-monitored-in-probability} holds.
\end{theorem}

The following two examples illustrate the tail-dominance route and show why weak extreme propagation alone is insufficient. Throughout, consider
\[
    W=\psi_WU+\varepsilon_W,
    \qquad
    T=\psi_TU+\varepsilon_T,
    \qquad
    Y=\beta_0T+\psi_YU+\chi W+\varepsilon_Y,
\]
where \(U,\varepsilon_W,\varepsilon_T,\varepsilon_Y\) are mutually independent and \(\psi_W\neq0\). 
Let \(E=T\) and \(S=\{W\}\). The graphical conditions of \cref{thm:unified-certification-prob} are satisfied, so certification of $A_v=\{T>v,\ |W|\le\tau\}$ reduces to whether \(W\) monitors \(U\).

\begin{example}[Two-sided heavy-tailed proxy]\label{ex:pareto-W-boundary-prob}
    If \(U\) and \(\varepsilon_W\) are Gaussian and $\varepsilon_T$ is regularly varying, then the tail-dominance condition \labelcref{cond:tail-race-primitive} holds. 
    Suppose instead that \(|U|\), \(|\varepsilon_W|\), and \(\varepsilon_T\) have regularly varying tails with indices $\alpha_U$, $\alpha_W$, and $\alpha_T$. Then $\alpha_U+\alpha_W>\alpha_T$ suffices for condition \labelcref{cond:tail-race-primitive}.
    At equality, the slowly varying factors decide. Heavy two-sided proxy noise can thus destroy monitoring even when the graph criterion holds: heaviness is required of the source noise, and lightness of the cancelling noise, not of the confounder itself. 
\end{example}

\begin{example}[Weak propagation alone is insufficient; necessity of tail dominance]
\label{ex:necessity-tail-race}
    Suppose that
    \(U,\varepsilon_W,\varepsilon_T,\varepsilon_Y\overset{\mathrm{i.i.d.}}{\sim}
    N(0,1)\) and \(\psi_T,\psi_Y\neq0\). Weak extreme propagation from
    \(U\) to \(W\) holds, but strong propagation, the single-big-jump
    property, and the tail dominance all fail. A large-deviation computation shows that, conditionally on $A_v$, the confounder grows linearly, $U\approx c^{*}v$ with $c^{*}=\psi_T/(1+\psi_T^2+\psi_W^2)$, so that $Y/T\mid A_v\to\beta_0+\psi_Y c^{*}\neq\beta_0$. Identification fails even though the graph criterion holds and cancellation is marginally unlikely. Because the exceedance $\varepsilon_T>v$ is as unlikely as a cancellation of the same size, the cheapest explanation of $A_v$ is a cooperative deviation (a moderately large $U$ cancelled by a moderate $\varepsilon_W$) rather than a large $\varepsilon_T$ alone. For $U$ of size $\delta v$ with $\delta$ small, hiding $U$ inside the safeguard is more likely than the exceedance $\varepsilon_T>v$.
\end{example}

\subsection{Examples of graphical structures covered}\label{sec:examples-of-graphical-structures-covered}
\begin{figure}[t]
\centering
\tikzset{dagrow/.style={dag, node distance=1.6cm, scale=0.8,
    baseline=(current bounding box.south)}}
\captionsetup[sub]{justification=centering,margin=4pt}

\scalebox{0.85}{%
\begin{subfigure}[t]{0.205\textwidth}%
    \centering
    \begin{tikzpicture}[dagrow]
        \node[latent] (U) at (0,0) {$U$};
        \node[observed, left of=U] (Z) {$Z$};
        \node[observed, right of=U] (W) {$W$};
        \node[observed, below left of=U] (T) {$T$};
        \node[observed, below right of=U] (Y) {$Y$};

        \draw[conf] (U) -- (W);
        \draw[conf] (U) -- (Z);
        \draw[conf] (U) -- (T);
        \draw[conf] (U) -- (Y);
        \draw[arrow] (T) -- node[above] {$\beta_0$} (Y);
        \draw[arrow] (Z) -- (T);
        \draw[arrow] (W) -- (Y);
    \end{tikzpicture}
    \caption{Double proxy structure (identified)}
    \label{fig:double-proxy-structure}
\end{subfigure}\hfill
\begin{subfigure}[t]{0.19\textwidth}%
    \centering
    \begin{tikzpicture}[dagrow]
        \node[latent] (U) at (0,0) {$U$};
        \node[observed, below left of=U] (T) {$T$};
        \node[observed, below right of=U] (Y) {$Y$};
        \node[observed, left of=U] (Z) {$Z$};

        \draw[conf] (U) -- (T);
        \draw[conf] (U) -- (Y);
        \draw[arrow] (T) -- node[above] {$\beta_0$} (Y);
        \draw[arrow] (Z) -- (T);
    \end{tikzpicture}
    \caption{IV structure (identified)}
    \label{fig:iv-structure}
\end{subfigure}\hfill
\begin{subfigure}[t]{0.19\textwidth}%
    \centering
    \begin{tikzpicture}[dagrow]
        \node[latent] (U) at (0,0) {$U$};
        \node[observed, below left of=U] (T) {$T$};
        \node[observed, below right of=U] (Y) {$Y$};
        \node[observed, left of=U] (Z) {$Z$};

        \draw[conf] (U) -- (Z);
        \draw[conf] (U) -- (T);
        \draw[conf] (U) -- (Y);
        \draw[arrow] (T) -- node[above] {$\beta_0$} (Y);
        \draw[arrow] (Z) -- (T);
    \end{tikzpicture}
    \caption{Invalid IV structure (identified)}
    \label{fig:invalid-iv-structure}
\end{subfigure}\hfill
\begin{subfigure}[t]{0.19\textwidth}%
    \centering
    \begin{tikzpicture}[dagrow]
        \node[latent] (U1) at (0,0) {$U_1$};
        \node[observed, below left of=U1] (T) {$T$};
        \node[latent, above=0.4cm of U1] (U2) {$U_2$};
        \node[observed, below right of=U1] (Y) {$Y$};
        \node[observed, left of=U1] (Z) {$Z$};

        \draw[conf] (U1) -- (T);
        \draw[conf] (U1) -- (Y);
        \draw[conf] (U2) -- (Z);
        \draw[conf] (U2) -- (Y);
        \draw[arrow] (T) -- node[above] {$\beta_0$} (Y);
        \draw[arrow] (Z) -- (T);
    \end{tikzpicture}
    \caption{Two distinct confounders (not identified)}
    \label{fig:double-confounded-structure}
\end{subfigure}\hfill
\begin{subfigure}[t]{0.205\textwidth}%
    \centering
    \begin{tikzpicture}[dagrow]
        \node[latent] (U1) at (0,0) {$U_1$};
        \node[observed, below left of=U1] (T) {$T$};
        \node[latent, above=0.4cm of U1] (U2) {$U_2$};
        \node[observed, below right of=U1] (Y) {$Y$};
        \node[observed, left of=U1] (Z) {$Z$};
        \node[observed, right of=U1] (W) {$W$};

        \draw[conf] (U1) -- (T);
        \draw[conf] (U1) -- (Y);
        \draw[conf] (U2) -- (Z);
        \draw[conf] (U2) -- (Y);
        \draw[arrow] (W) -- (Y);
        \draw[conf] (U2) -- (W);
        \draw[arrow] (T) -- node[above] {$\beta_0$} (Y);
        \draw[arrow] (Z) -- (T);
    \end{tikzpicture}
    \caption{Two distinct confounders with proxy (identified)}
    \label{fig:double-confounded-proxy-structure}
\end{subfigure}%
}

\caption{Examples of graphical structures. We label which ones are identified
by \cref{thm:unified-certification-prob}, that is, whether there exist choices
of $E$ and $S$ that satisfy the graphical criteria in
\labelcref{ass:E-affects-T,ass:E-not-affects-SY,ass:no-unmonitored-confounding} in
\cref{thm:unified-certification-prob}. Beyond graphical structure,
\cref{thm:unified-certification-prob} also requires the distributional
assumption that confounders are monitored in probability,
\eqref{ass:confounders-monitored-in-probability}.
}
\label{fig:examples-of-structures-identified}
\end{figure}
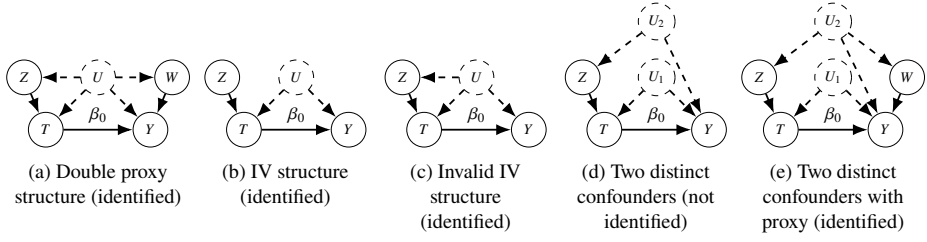

We also explore how to satisfy the graphical conditions \labelcref{ass:E-affects-T,ass:E-not-affects-SY,ass:no-unmonitored-confounding} of \cref{thm:unified-certification-prob} in the causal structures of \cref{fig:examples-of-structures-identified}; for \cref{thm:unified-certification-prob} to apply, the distributional condition \eqref{ass:confounders-monitored-in-probability} must hold in addition.

\textbf{Double proxy structure}: In \cref{fig:double-proxy-structure}, we consider the double proxy structure of \cite{MiaoGengTchetgenTchetgen2018,TchetgenTchetgenYingCuiShiMiao2020}, with a treatment-side proxy $Z$ and an outcome-side proxy $W$. Setting $E = Z$ and $S = \{W\}$ meets the graphical conditions \labelcref{ass:E-affects-T,ass:E-not-affects-SY,ass:no-unmonitored-confounding} of \cref{thm:unified-certification-prob}, and so does $E=T$ with $S=\{W\}$ or $S=\{W,Z\}$.

\textbf{IV structure}: In \cref{fig:iv-structure}, the instrument $Z$ acts on $T$, and we take $E = Z$, $S = \emptyset$.
The graphical conditions of \cref{thm:unified-certification-prob} reduce to the usual IV assumptions:
\labelcref{ass:E-affects-T} is relevance,
\labelcref{ass:E-not-affects-SY} is exclusion, and
\labelcref{ass:no-unmonitored-confounding} reads $\mathrm{an}(Z) \cap \mathrm{an}_{G^-}(Y) = \emptyset$, that is $Z$ and $Y$ are unconfounded.
The distributional condition \eqref{ass:confounders-monitored-in-probability} is vacuous when $S=\emptyset$. What remains beyond the IV assumptions is the requirement in \labelcref{ass:E-affects-T} that $Z$ generates extremes that propagate to $T$; the resulting estimator also differs from the usual method-of-moments estimators \citep{hansen1982large}. 

\textbf{Invalid IV structure}: In \cref{fig:invalid-iv-structure},
the candidate instrument $Z$ is confounded with the outcome through $U$. 
With $E=Z$, the only admissible safeguard is $S=\emptyset$, and \labelcref{ass:no-unmonitored-confounding} fails since $U\in\mathrm{an}(Z)\cap\mathrm{an}_{G^-}(Y)$. Instead, we use $Z$ as a safeguard against $U$, taking $E = T$ and $S = \{Z\}$. \labelcref{ass:E-affects-T} is trivial; \labelcref{ass:E-not-affects-SY} holds because $T\notin\mathrm{an}(Z)$ and $T\notin \mathrm{an}_{G^-}(Y)$; and \labelcref{ass:no-unmonitored-confounding} holds because $\mathrm{an}(T)\cap \mathrm{an}_{G^-}(Y) = \{U\}\subseteq\mathrm{an}(Z)$. The graphical criterion is thus met in a structure that admits no valid instrument, with certifying events $\{T>v,\ Z\in\mathcal B\}$ as in \cref{sec:extreme-propagation-monitoring}, with $Z$ in the role of $W$.
The distributional condition \eqref{ass:confounders-monitored-in-probability} is not vacuous here: it demands that a large $U$ hiding behind a moderate $Z$ is asymptotically improbable, and the effect is not identified in a Gaussian linear model (\cref{ex:necessity-tail-race}). 
Extreme certification thus does not relax the IV assumptions but replaces the structural assumption $U \indep Z$ with a distributional assumption on the tails.

\textbf{Two distinct confounders}: In \cref{fig:double-confounded-structure}, $U_1$ confounds $T$ and $Y$, and $U_2$ confounds $Z$ and $Y$. Neither source works. With $E = Z$, the only admissible safeguard is $S = \emptyset$, and $U_2\in\mathrm{an}(Z)\cap\mathrm{an}_{G^-}(Y)$ is unmonitored. 
With $E = T$, the only admissible safeguard is $S = \{Z\}$, which monitors $U_2$ but not $U_1$: $\mathrm{an}(T)\cap \mathrm{an}_{G^-}(Y) = \{U_1, U_2\}\not\subseteq\mathrm{an}(Z)$. Either way, an exceedance of the source admits a confounded explanation that no observed variable rules out. \Cref{fig:invalid-iv-structure,fig:double-confounded-structure} have the same observed variables and edges and differ only in whether one or two latents generate the confounding. This highlights that our theory requires assumptions on the latent structure: every unobserved confounder of $E$ and $Y$ must be an ancestor of a safeguard, \labelcref{ass:no-unmonitored-confounding}.

\textbf{Two distinct confounders with proxy}: Adding an observed proxy $W$ of $U_2$ (\cref{fig:double-confounded-proxy-structure}) restores identification with $E = Z$ and $S = \{W\}$, since $Z\in\mathrm{an}(T)$; $Z\notin\mathrm{an}(W)$ and $Z\notin \mathrm{an}_{G^-}(Y)$; and $\mathrm{an}(Z)\cap \mathrm{an}_{G^-}(Y) = \{U_2\}\subseteq\mathrm{an}(W)$. The confounder $U_1$ needs no safeguard, as it is not an ancestor of the source: it is independent of $(Z, W)$, so conditioning on $\{Z > v,\ W\in\mathcal B\}$ does not change its distribution.

The above choices of $(E, S)$ use knowledge of the latent structure, which can partly be replaced by the classical proxy assumptions of \cref{sec:causal-setup}: a treatment-inducing proxy $Z\indep Y \mid (T, U)$ and an outcome-inducing proxy $W\indep(T, Z)\mid U$ give \labelcref{ass:E-not-affects-SY} and, if $Z$ is relevant for $T$, the graphical part of \labelcref{ass:E-affects-T} for the pair $(Z, \{W\})$. They do not give \labelcref{ass:no-unmonitored-confounding}, since they permit a component such as $U_2$ in \cref{fig:double-confounded-structure} that leaves no trace in $W$. Conditions \labelcref{ass:no-unmonitored-confounding} and \eqref{ass:confounders-monitored-in-probability} are thus our analogue of completeness: rather than requiring $W$ to vary richly enough with $U$ for a bridge function to identify the effect, we require only that no confounding ancestor of the source is extreme while $W$ stays non-extreme.

\section{Extreme-value geometry under certification}
\label{sec:causal-pot}

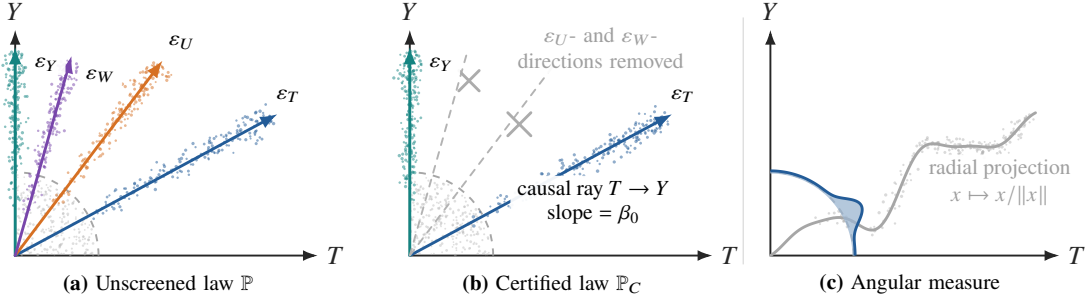
\begin{figure}[t]
\centering

\begingroup

\definecolor{causalblue}{RGB}{35,92,152}
\definecolor{confoundorange}{RGB}{214,112,35}
\definecolor{outcometeal}{RGB}{20,132,129}
\definecolor{proxyviolet}{RGB}{120,70,170}

\tikzset{
  axis/.style={
    ->, thick, black!85,
    -{Latex[length=2.2mm,width=1.6mm]}
  },
  causalray/.style={
    very thick, causalblue,
    -{Latex[length=2.6mm,width=1.9mm]},
    line cap=round
  },
  confoundray/.style={
    very thick, confoundorange,
    -{Latex[length=2.6mm,width=1.9mm]},
    line cap=round
  },
  outcomeray/.style={
    very thick, outcometeal,
    -{Latex[length=2.6mm,width=1.9mm]},
    line cap=round
  },
  proxyray/.style={
    very thick, proxyviolet,
    -{Latex[length=2.6mm,width=1.9mm]},
    line cap=round
  },
  removedray/.style={
    thick, gray!60, dashed, line cap=round
  },
  thresholdarc/.style={
    dashed, gray!70, thick
  },
  labelbox/.style={
    fill=white,
    fill opacity=.94,
    text opacity=1,
    rounded corners=1.2pt,
    inner xsep=3pt,
    inner ysep=2pt,
    font=\small,
    align=center
  },
  rayname/.style={
    labelbox,
    font=\small\itshape
  },
  panelcaption/.style={
    font=\small,
    align=center,
    text width=4.5cm
  }
}

\newcommand{\RayCloud}[6]{%
  \pgfmathsetseed{#6}%
  \pgfmathsetmacro{\raylength}
    {sqrt((#2)*(#2)+(#3)*(#3))}%
  \foreach \i in {1,...,#4}{%
    \pgfmathsetmacro{\rr}{0.20+0.78*pow(rnd,0.58)}%
    \pgfmathsetmacro{\zz}{rnd+rnd+rnd+rnd-2}%
    \pgfmathsetmacro{\jj}{(#5)*(0.50+0.50*\rr)*\zz}%
    \pgfmathsetmacro{\xx}
      {\rr*(#2)-\jj*(#3)/\raylength}%
    \pgfmathsetmacro{\yy}
      {\rr*(#3)+\jj*(#2)/\raylength}%
    \pgfmathsetmacro{\dotsize}{0.42+0.32*rnd}%
    \fill[#1,opacity=.50]
      (\xx,\yy) circle[radius=\dotsize pt];%
  }%
}

\newcommand{\BodyCloud}[2]{%
  \pgfmathsetseed{#2}%
  \foreach \i in {1,...,#1}{%
    \pgfmathsetmacro{\rr}{0.97*\Rad*sqrt(rnd)}%
    \pgfmathsetmacro{\tt}{90*rnd}%
    \pgfmathsetmacro{\xx}{\rr*cos(\tt)}%
    \pgfmathsetmacro{\yy}{\rr*sin(\tt)}%
    \pgfmathsetmacro{\dotsize}{0.34+0.28*rnd}%
    \fill[gray!55,opacity=.45]
      (\xx,\yy) circle[radius=\dotsize pt];%
  }%
}

\newcommand{\CurveCloud}[6]{%
  \pgfmathsetseed{#6}%
  \foreach \i in {1,...,#4}{%
    \pgfmathsetmacro{\tt}{0.18+0.79*pow(rnd,0.58)}%
    \pgfmathsetmacro{\baseX}{#2}%
    \pgfmathsetmacro{\baseY}{#3}%
    \pgfmathsetmacro{\jx}{#5*(rnd+rnd+rnd+rnd-2)}%
    \pgfmathsetmacro{\jy}{#5*(rnd+rnd+rnd+rnd-2)}%
    \pgfmathsetmacro{\dotsize}{0.40+0.30*rnd}%
    \fill[#1,opacity=.48]
      ({\baseX+\jx},{\baseY+\jy})
      circle[radius=\dotsize pt];%
  }%
}

\newcommand{\FilledDensityArc}[4]{%
  \fill[causalblue!45,opacity=.75]
    plot[samples=180,domain=0:90,variable=\aa]
    (
      {(#1+0.02+#2*exp(-((\aa-#3)^2)/(2*(#4)^2)))*cos(\aa)},
      {(#1+0.02+#2*exp(-((\aa-#3)^2)/(2*(#4)^2)))*sin(\aa)}
    )
    --
    plot[samples=180,domain=90:0,variable=\aa]
    (
      {(#1-0.02)*cos(\aa)},
      {(#1-0.02)*sin(\aa)}
    )
    -- cycle;

  \draw[causalblue,very thick]
    plot[samples=180,domain=0:90,variable=\aa]
    (
      {(#1+0.02+#2*exp(-((\aa-#3)^2)/(2*(#4)^2)))*cos(\aa)},
      {(#1+0.02+#2*exp(-((\aa-#3)^2)/(2*(#4)^2)))*sin(\aa)}
    );
}
\resizebox{\linewidth}{!}{%
\begin{tikzpicture}[x=1cm,y=1cm]

  \path[use as bounding box]
    (-0.55,-0.62) rectangle (15.85,3.79);

  \def\Rad{1.225}

  \def\CausalY{%
    2.10*\tt*(
      1
      +0.75*(1-\tt)*sin(720*\tt)
      +0.20*(1-\tt)*sin(1440*\tt+25)
    )%
  }

  \begin{scope}

    \draw[axis] (0,0) -- (4.45,0) node[right] {$T$};
    \draw[axis] (0,0) -- (0,3.32) node[above] {$Y$};

    \draw[thresholdarc]
      (\Rad,0)
      arc[start angle=0,end angle=90,radius=\Rad];

    \BodyCloud{260}{8101}

    \RayCloud{causalblue}
      {3.86}{2.08}{170}{0.18}{1401}
    \RayCloud{confoundorange}
      {2.17}{2.87}{170}{0.17}{2309}
    \RayCloud{outcometeal}
      {0.00}{3.07}{170}{0.15}{3571}
    \RayCloud{proxyviolet}
      {0.83}{2.94}{160}{0.14}{4891}

    \draw[causalray]    (0,0) -- (3.86,2.08);
    \draw[confoundray]  (0,0) -- (2.17,2.87);
    \draw[outcomeray]   (0,0) -- (0,3.07);
    \draw[proxyray]     (0,0) -- (0.83,2.94);

    \node[rayname,anchor=west]
      at (3.73,2.35) {$\varepsilon_T$};

    \node[rayname,anchor=south west]
      at (2.15,2.96) {$\varepsilon_U$};

    \node[rayname,anchor=west]
      at (0.18,2.81) {$\varepsilon_Y$};

    \node[rayname,anchor=west]
      at (0.93,2.64) {$\varepsilon_W$};

    \node[panelcaption] at (2.10,-0.42)
      {\textbf{(a)} Unscreened law $\mathbb P$};

  \end{scope}

  \begin{scope}[xshift=5.80cm]

    \draw[axis] (0,0) -- (4.45,0) node[right] {$T$};
    \draw[axis] (0,0) -- (0,3.32) node[above] {$Y$};

    \draw[thresholdarc]
      (\Rad,0)
      arc[start angle=0,end angle=90,radius=\Rad];

    \BodyCloud{260}{9103}

    \RayCloud{causalblue}
      {3.86}{2.08}{190}{0.18}{4517}
    \RayCloud{outcometeal}
      {0.00}{3.07}{190}{0.15}{6673}

    \draw[causalray]  (0,0) -- (3.86,2.08);
    \draw[outcomeray] (0,0) -- (0,3.07);

    \draw[removedray]
      (0,0) -- (2.17,2.87);

    \draw[gray!65,very thick,line cap=round]
      (1.44,1.76) -- (1.78,2.10)
      (1.78,1.76) -- (1.44,2.10);

    \draw[removedray]
      (0,0) -- (0.83,2.94);

    \draw[gray!65,very thick,line cap=round]
      (0.72,2.42) -- (1.02,2.72)
      (1.02,2.42) -- (0.72,2.72);

    \node[rayname,anchor=west]
      at (3.73,2.35) {$\varepsilon_T$};

    \node[rayname,anchor=west]
      at (0.18,2.81) {$\varepsilon_Y$};

    \node[labelbox,anchor=north] at (2.70,1.18)
      {
        causal ray $T\to Y$\\[-1pt]
        $\operatorname{slope}=\beta_0$
      };

    \node[
      labelbox,
      text=gray!75,
      anchor=west,
      text width=2.75cm
    ] at (1.32,3.05)
      {
        $\varepsilon_U$- and $\varepsilon_W$-directions removed
      };

    \node[panelcaption] at (2.10,-0.42)
      {\textbf{(b)} Certified law $\mathbb P_C$};

  \end{scope}

  \draw[gray!35,thin]
    (10.70,-0.13) -- (10.70,3.50);

  \begin{scope}[xshift=11.10cm]

    \draw[axis] (0,0) -- (4.25,0) node[right] {$T$};
    \draw[axis] (0,0) -- (0,3.32) node[above] {$Y$};

    \draw[thresholdarc]
      (\Rad,0)
      arc[start angle=0,end angle=90,radius=\Rad];

    \CurveCloud{gray!60}
      {3.90*\tt}
      {\CausalY}
      {210}{0.115}{4201}

    \draw[
      very thick,
      gray!70,
      line cap=round,
      samples=300,
      domain=0:1,
      variable=\tt
    ]
      plot ({3.90*\tt},{\CausalY});

    \FilledDensityArc{\Rad}{0.28}{29}{7}

    \node[
      labelbox,
      text=gray!75,
      anchor=west
    ] at (2.22,1.08)
      {
        radial projection\\
        $x\mapsto x/\lVert x\rVert$
      };

    \node[panelcaption] at (2.05,-0.42)
      {\textbf{(c)} Angular measure};

  \end{scope}

\end{tikzpicture}%
}
\endgroup

\caption{%
Extremal ray geometry before and after proxy certification.
\textbf{(a)} Extremes may arise along the
directions generated by
$\varepsilon_T,\varepsilon_U,\varepsilon_W$, and $\varepsilon_Y$.
\textbf{(b)} Conditioning on $C=\{W\in\mathcal B\}$ removes the
$\varepsilon_U$- and $\varepsilon_W$-directions, leaving the causal
treatment-shock ray, with slope $\beta_0$, and the vertical
outcome-shock ray.  The $\varepsilon_Y$ ray is additionally removed by conditioning on $T>v$. 
\textbf{(c)} In case of non-linear relations, the radial projection
$x\mapsto x/\lVert x\rVert$ induces an angular measure.
}
\label{fig:certified-extremal-rays}
\end{figure}

Certification has a geometric interpretation connected to multivariate extreme-value theory \citep{Resnick}. Under certain assumptions, each sufficiently large structural shock $\varepsilon_j$ propagates into the treatment--outcome plane along an ``extremal ray'' (Figure~\ref{fig:certified-extremal-rays}).  Certification removes the non-causal rays and leaves the causal one.
The causal effect is then the slope of an atom of an angular measure, and estimators of angular measures apply to the certified sample (\cref{sec:beyond-regular-variation-evt}).

\subsection{Extremal rays and angular measure}

Let $\boldsymbol X=(T,Y)^\top$ and define  $\boldsymbol\Theta=\boldsymbol X/\lVert\boldsymbol X\rVert$ where $\lVert\cdot\rVert$ is any norm on $\mathbb R^2$ with $\lVert(0,1)^\top\rVert=1$ (and we define $\boldsymbol\Theta=0$ when $\boldsymbol X=0$). Write $\mathbb S=\{\boldsymbol\theta\in\mathbb{R}^2:\lVert\boldsymbol\theta\rVert=1\}$.

\begin{definition}[Atomic extremal ray]
\label{def:evt-extremal-ray}
Under a probability law $\mathbb Q$, suppose that
$\mathcal L_{\mathbb Q}(\boldsymbol\Theta\mid \lVert\boldsymbol X\rVert>v)$ converges weakly to a
probability measure $H_{\mathbb Q}$ on $\mathbb S$. We call $H_{\mathbb Q}$ an \textit{angular measure} (quarter-circle on Fig.~\ref{fig:certified-extremal-rays}). If $H_{\mathbb Q}\{\boldsymbol\omega\}>0$ for some $\boldsymbol\omega = (\omega_T,  \omega_Y)\in \mathbb S$, we call $\mathcal R(\boldsymbol\omega)=\{r\boldsymbol\omega:r>0\}$ an \emph{atomic extremal ray} with slope $\omega_Y/\omega_T$.
\end{definition}

Consider the proxy certificate $C=\{S\in\mathcal B\}$ and write $\mathbb P_C(\,\cdot\,)=\mathbb P(\,\cdot\mid C)$, with $\mathcal L_C$ denoting laws under $\mathbb P_C$. In a linear SCM $\boldsymbol X$ has the representation $\boldsymbol X=\sum_{j\in  V}\boldsymbol a_j\varepsilon_j$, where $\boldsymbol a_j=(a_{Tj},a_{Yj})^\top\in\mathbb R^2$ collects the total effects of $\varepsilon_j$ on $T$ and on $Y$; for the treatment shock, $\boldsymbol a_T=(1,\beta_0)^\top$. If a shock $\varepsilon_j$ is sufficiently heavy-tailed, it creates an atomic extremal ray in the direction $\boldsymbol a_j/\lVert\boldsymbol a_j\rVert$, with slope $a_{Yj}/a_{Tj}$. However, only the shock from $\varepsilon_T$ corresponds to the true causal ray; we show that certification isolates it.

\begin{theorem}[Certification isolates the causal ray]
\label{thm:certified-extremal-rays}
Consider a linear SCM with i.i.d. regularly varying noise variables. Then $\boldsymbol\Theta$ concentrates on a finite number of extremal rays (Fig.~\ref{fig:certified-extremal-rays}a):
\begin{equation}
    \mathcal L\!\left(
        \boldsymbol\Theta
        \,\middle|\,
        \lVert\boldsymbol X\rVert>v
    \right)
    \overset{d}{\longrightarrow}
    \sum_{j\in\mathcal{J}}
    \sum_{\sigma\in\{-1,1\}}
        \pi_{j,\sigma}\,\delta_{\boldsymbol\omega_{j,\sigma}},
    \qquad v\to\infty,
    \label{eq:uncertified-ray-limit}
\end{equation}
where the weights \(\pi_{j,\sigma}\geq 0\), the directions \(\boldsymbol\omega_{j,\sigma}\in\mathbb{S}\), one for each sign $\sigma$ of a two-sided shock, and the finite index set $\mathcal{J}$ are given explicitly in the Supplementary Material (\cref{thm_certified-extremal-rays_appendix_version}).

Let \((E,S)\) satisfy the conditions of Theorem~\ref{thm:unified-certification-prob}. Then, writing  \(
\boldsymbol\omega_{\beta_0}
=(1,\beta_0)^\top/\lVert(1,\beta_0)^\top\rVert
\), we have, for some weights $\widetilde\pi_T,\widetilde\pi_Y\geq0$ with $\widetilde\pi_T+\widetilde\pi_Y=1$: 
\begin{align}
\mathcal L_C\!\left(
    \boldsymbol\Theta
    \,\middle|\,
    \lVert\boldsymbol X\rVert>v
\right)
&\ \overset{d}{\longrightarrow}\
    \widetilde\pi_T\delta_{\boldsymbol\omega_{\beta_0}}
    +\widetilde\pi_Y\delta_{(0,1)},
\qquad v\to\infty,
\label{eq:conditional-ray-limit}
\\
\mathcal L\!\left(
    \boldsymbol\Theta
    \,\middle|\,
    A_v^{E,S}
\right)
&\ \overset{d}{\longrightarrow}\
    \delta_{\boldsymbol\omega_{\beta_0}},
\qquad v\to\infty,
\label{eq:certified-ray-limit}
\end{align}
where \eqref{eq:conditional-ray-limit} additionally requires \(E=T\) and $\boldsymbol X\in\mathbb R_+^2$ almost surely (if this nonnegativity assumption is dropped, additional limiting rays may appear in the other quadrants).
\end{theorem}

This gives a geometric interpretation of certification. If weak extreme propagation holds from each confounder $U\in\mathcal{C}_S$ to $S$ (in a linear case it simply means that the coefficients are non-zero), then certification removes extremes generated by the latent confounder and by the proxy. The vertical rays correspond to $\varepsilon_Y$ and cannot generate $T>v$. The only disturbance surviving the certification is the treatment shock,
whose propagation vector is $(1,\beta_0)^\top$.

\subsection{Estimation and extensions}
\label{sec:beyond-regular-variation-evt}

For estimation, let $(T_i,Y_i,E_i,S_i)$, $i=1,\ldots,n$, be independent copies of $(T,Y,E,S)$ (\cref{sec:causal-setup}), as above write $\boldsymbol X_i=(T_i,Y_i)^\top$ and $\boldsymbol\Theta_i=\boldsymbol X_i/\lVert\boldsymbol X_i\rVert$, and let $v_n$ be a threshold sequence. The indicator $I_{i,n}=\mathbbm 1\{E_i>v_n,\ S_i\in\mathcal B\}$ records whether observation $i$ lies in the certifying event $A_{v_n}^{E,S}$; $N_n=\sum_{i=1}^n I_{i,n}$ is the number of certified observations and $m_n=\mathbb E(N_n)=n\,\mathbb P(A_{v_n}^{E,S})$ its mean. Several parametric, semiparametric and nonparametric estimators of the angular measure $H$ are available
\citep{EinmahlSegers2009,AvellaMedinaDavisSamorodnitsky2024}. 
Since the limit in \cref{thm:certified-extremal-rays} is finite-atomic, spherical or spectral clustering of the certified directions \citep{JanssenWan2020,DengTangBai2025} is possible too. 
Restricting to certified data and averaging the directions gives the mean certified direction $\widehat{\boldsymbol\omega}_\beta=(\widehat\omega_{\beta,T},\widehat\omega_{\beta,Y})^\top=N_n^{-1}\sum_{i=1}^n I_{i,n}\boldsymbol\Theta_i$, whose slope is the estimate
\begin{equation}
\widehat\beta_n^{\mathrm{ray}}
=
\frac{\widehat\omega_{\beta, Y}}
     {\widehat\omega_{\beta, T}}=
\frac{\sum_{i=1}^n I_{i,n}Y_i/\lVert\boldsymbol X_i\rVert}
     {\sum_{i=1}^n I_{i,n}T_i/\lVert\boldsymbol X_i\rVert}.
\label{beta_ray_estimator}
\end{equation}
If $v_n\to\infty$ and $m_n\to\infty$, the law of large numbers and Theorem~\ref{thm:certified-extremal-rays} give
$\widehat\beta_n^{\mathrm{ray}}\overset{\mathbb P}{\to}\beta_0$. Unlike ordinary least squares, each observation enters only through its direction, so a single unusually large radial magnitude cannot dominate the estimator. 
The empirical distribution of the certified directions also gives a diagnostic: additional modes may indicate incomplete certification, multiple extremal mechanisms, or effect heterogeneity.

Exact linearity is not essential. If $Y=f(T)+R$ with $R/T$ negligible on the certifying sequence and if the limit $f(t)/t\to\beta_\infty$ exists, then the certified ray has slope $\beta_\infty$: certification identifies the first-order response to an extreme treatment rather than the response surface (Fig.~\ref{fig:certified-extremal-rays}c).

\section{Estimation and convergence rates}
\label{sec:estimation}

Until now, we have considered identifiability of the causal effect with certification. We now consider concrete certified estimators and their convergence rates. 
In this section, we assume the linear effect of Assumption~\ref{ass:constant-linear-causal-effect}, $Y=\beta_0T+R$, where certification keeps the remainder $R$ non-extreme while $T$ becomes large. 
We keep the sampling notation of \cref{sec:beyond-regular-variation-evt}: $\boldsymbol X_i=(T_i,Y_i)^\top$ is the $i$th of $n$ independent copies, $\lVert\cdot\rVert$ a norm on $\mathbb R^2$, $I_{i,n}$ indicates membership in $A_{v_n}$, $N_n=\sum_{i=1}^n I_{i,n}$ the number of certified observations and $m_n=n\,\mathbb P(A_{v_n})$ its mean. The only change is that $(A_v)_v$ may now be any certifying sequence of observable events (\cref{def:certifying-sequence}) along which $T$ is extreme; our leading example remains $A_v^{T,S}=\{T>v,\ S\in\mathcal B\}$.

\subsection{Direct certified estimators}
\label{subsec:naive-estimators}

Writing $\overline T_n=N_n^{-1}\sum_i I_{i,n}T_i$ and
$\overline Y_n=N_n^{-1}\sum_i I_{i,n}Y_i$, define the certified OLS slopes
through the origin and with an intercept by
\begin{equation*}
\begin{aligned}
    \widehat\beta_n^{\mathrm{OLS}}=
    \frac{\sum_{i=1}^n I_{i,n}T_iY_i}
         {\sum_{i=1}^n I_{i,n}T_i^2},
\qquad 
    \widehat\beta_n^{\mathrm{OLS+intercept}}
    =
    \frac{\sum_{i=1}^n I_{i,n}(T_i-\overline T_n)(Y_i-\overline Y_n)}
         {\sum_{i=1}^n I_{i,n}(T_i-\overline T_n)^2}.
\end{aligned}
\end{equation*}
The least-squares slope may also be replaced by the median of the ratios $Y_i/T_i$ over the certified sample, which requires no moments of $R$; \cref{prop:certified-rates-extended} in the Supplementary Material gives its rate.
Finally, recall the ray estimator $\widehat\beta_n^{\mathrm{ray}}$ of \eqref{beta_ray_estimator}, which averages the certified slopes $Y_i/T_i$ with the bounded weights $T_i/\lVert\boldsymbol X_i\rVert$.
All estimators may be defined arbitrarily when a denominator vanishes (under the conditions below, these events have probability tending to zero).

\begin{theorem}[Convergence rates under finite moments]
\label{thm:naive-certified-estimator-rate}
Let $v_n\to\infty$, $m_n\to\infty$, and
$A_v\subseteq\{T\geq v\}$ for all sufficiently large $v$.

\begin{enumerate}[label=(\roman*),leftmargin=2.5em]
    \item \emph{OLS.} Put
    $\mu_v(T)=\mathbb E(R\mid T,A_v)$. If, uniformly for large $v$,
    $|\mu_v(T)|\leq M$ and
    $\operatorname{Var}(R\mid T,A_v)\leq\sigma^2$ almost surely, then
    \begin{equation}
        \widehat\beta_n^{\mathrm{OLS}}-\beta_0
        =O_{\mathbb P}(v_n^{-1})
         +O_{\mathbb P}\!\left((v_n\sqrt{m_n})^{-1}\right).
        \label{eq:certified-ols-rate}
    \end{equation}
    If $\mu_v(T)=0$ almost surely for all sufficiently large $v$, the first
    term is absent. 

     \item \emph{Ray estimator.} Let $\lVert\cdot\rVert$ be the $\ell_1$ norm, $\lVert\boldsymbol X\rVert=|T|+|Y|$. Define $b_v^{\mathrm{ray}} := v\,\mathbb E(R/\lVert\boldsymbol X\rVert\mid A_v)/\mathbb E(T/\lVert\boldsymbol X\rVert\mid A_v)$.
    Suppose, uniformly for large $v$,
    $\mathbb E(R^2\mid T,A_v)\leq M_2$ almost surely. Then
    $b_v^{\mathrm{ray}}=O(1)$ and
    \begin{equation}
        \widehat\beta_n^{\mathrm{ray}}-\beta_0
        =
        \frac{b_{v_n}^{\mathrm{ray}}}{v_n}
        +
        O_{\mathbb P}\!\left((v_n\sqrt{m_n})^{-1}\right)
        =
        O(v_n^{-1})
        +
        O_{\mathbb P}\!\left((v_n\sqrt{m_n})^{-1}\right).
        \label{eq:certified-ray-rate}
    \end{equation}
    If $\mathbb E(R\mid T,A_v)=0$ almost surely, then
    $b_v^{\mathrm{ray}}=O(v^{-1})$, so the first term improves to
    $O(v_n^{-2})$. If $\mathbb E(R/\lVert\boldsymbol X\rVert\mid A_v)=0$, it is absent.
    
\end{enumerate}
\end{theorem}

Notice the bias-variance trade-off: raising the threshold improves the approximation $Y/T\simeq\beta_0$, but
simultaneously reduces the number of certified observations.

\subsection{Heavier tails make convergence rates faster}
\label{subsec:tail-index-estimation-rates}

In the following, we show that the convergence rates of the certified estimators \textit{improve} as the treatment tail becomes heavier. This can be surprising, since classical OLS only works well with finite second moments. Smaller $\alpha$ means a heavier certified treatment tail. For a fixed number $k_n$ of exceedances, this produces a larger threshold $v_n\asymp(n/k_n)^{1/\alpha}$ and therefore faster suppression of $R/T$. 

\begin{corollary}[Tail-index-dependent rates]
\label{thm:tail-index-certified-rates}
\label{thm:ray-tail-index-rate}
Assume that the certifying events $A_v$ are nested, satisfying $A_v\subseteq\{T\geq v\}$ for all $v$, and being regularly varying:
\[
    p(v):=\mathbb P(A_v)\sim c_Av^{-\alpha},
    \qquad c_A>0,\quad \alpha>0.
\]
Let $k_n\to\infty$, $k_n/n\to0$, and \(
    v_n:=\left(\frac{c_An}{k_n}\right)^{1/\alpha}.
\)

\begin{enumerate}[label=(\roman*),leftmargin=2.5em]
    \item \emph{OLS estimator}.
    Under the conditions of Theorem~\ref{thm:naive-certified-estimator-rate} it satisfies $  \widehat\beta_n^{\mathrm{OLS}}-\beta_0
        =
        O_{\mathbb P}\!\left(
            \left(\frac{k_n}{n}\right)^{1/\alpha}
        \right).$

    \item \emph{Ray estimator.}
    Suppose that $A_v=A_v^{T,S}=\{T>v,\ S\in\mathcal B\}$ and the conditions of
    Theorem~\ref{thm:naive-certified-estimator-rate} for the ray estimator hold, and write the limit $(T/v_n,R)\mid A_{v_n}\overset{d}{\to}(Z_\alpha,R_\infty)$. Then the radial limit is standard Pareto,
    \(
        \mathbb P(Z_\alpha>z)=z^{-\alpha},z\geq1,
    \) and, writing $b_n^{\mathrm{ray}}:=b_{v_n}^{\mathrm{ray}}$, $b_n^{\mathrm{ray}}\to
    b^{\mathrm{ray}}
    :=\mathbb E(R_\infty/Z_\alpha)$, and    \begin{equation}
        \widehat\beta_n^{\mathrm{ray}}-\beta_0
        =
        b_n^{\mathrm{ray}}
        \left(\frac{k_n}{c_An}\right)^{1/\alpha}
        +
        O_{\mathbb P}\!\left(
            n^{-1/\alpha}k_n^{1/\alpha-1/2}
        \right).
        \label{eq:ray-tail-index-expansion}
    \end{equation}
    Thus its generic rate is also $O_{\mathbb P}((k_n/n)^{1/\alpha})$, but if $\sqrt{k_n}\,b_n^{\mathrm{ray}}\to0$, its rate improves to    $O_{\mathbb P}(n^{-1/\alpha}k_n^{1/\alpha-1/2})$.
\end{enumerate}
\end{corollary}

\begin{remark}[Infinite variance and a fixed number of certified observations]
\label{rem:infinite-variance-fixed-k}
    The moment conditions of \cref{thm:naive-certified-estimator-rate} concern $R$ given the certifying event. They fail for example when the outcome noise $\varepsilon_Y$ is regularly varying with $\alpha\leq2$. 
    When the number of certified observations is fixed, $k_n=k$, and $v_n$ is the $(k+1)$th largest source value among the safeguarded observations, we prove that $\widehat\beta_n-\beta_0=O_{\mathbb P}(n^{-1/\alpha})$ in \cref{prop:fixed-k-rate} in the Supplementary Material with mild assumptions that don't include moment conditions. 
\end{remark}

\begin{remark}[Estimation only from one data-point]\label{Remark_max_estimator}
Perhaps surprisingly, sufficiently heavy-tailed treatments can yield a consistent estimator based on a single observation. Let
\begin{equation}
    \widehat\beta_n^{\max}
    =\frac{Y_{i_n^\star}}{T_{i_n^\star}}, \qquad \text{where}\quad i_n^\star\in\argmax\{T_i:1\leq i\leq n,\,S_i\in\mathcal B\}.
    \label{eq:single-maximum-estimator}
\end{equation}
Under regular variation (see \cref{prop:single-certified-maximum} in the Supplementary Material), it holds  that $$\widehat\beta_n^{\max}-\beta_0=O_{\mathbb P}(n^{-1/\alpha}).$$

    It is even $o_{\mathbb P}(n^{-1/2})$ when $\alpha<2$, so a single extreme observation can give a consistent estimator of $\beta_0$. In practice this has obvious shortcomings: a single mis-recorded flood peak in \cref{sec:application} would determine the estimate entirely, and one observation admits no confidence interval, whereas the rank interval of \cref{app:confidence-intervals} needs only a few certified observations. 
\end{remark}

In the Supplementary Material, we outline practical considerations. 
    First, we propose confidence intervals (\cref{app:confidence-intervals}). With a handful of certified observations, and a bias of the same order as the standard deviation in \cref{thm:naive-certified-estimator-rate}, a Wald interval is unreliable. Instead, since certification makes the certified treatment values asymptotically independent of the remainders, we invert Kendall's rank test of association between the $T_i$ and the residuals $Y_i-\beta T_i$. The resulting interval does not require finite moments and for a fixed number of certified observations its coverage converges to the exact level of the rank test (\cref{prop:rank-interval}). 
Second, we discuss using the discarded observations from the bulk of the distribution. When the causal structure allows a classical estimator, we show in \cref{app:combining} that the certified sample and the body of the data are conditionally independent and that we can combine the two estimators by precision weighting. 

\section{Simulations}
\label{sec:simulations}

We examine the certified estimator in two simulation studies. 
First we compare to classical instrumental-variable and proxy estimators, both in designs where the classical assumptions hold and in designs where they fail, while the assumptions of \cref{thm:unified-certification-prob} hold throughout. 
Second, we remove the distributional assumption \eqref{ass:confounders-monitored-in-probability} and show the certificate failing, as \cref{ex:necessity-tail-race} predicts. 

\subsection{Setup}
\label{sec:sim-setup}

We consider $Y = \beta_0 T + R$ with $\beta_0 = 1$, and the remaining structural coefficients drawn independently from the uniform distribution on $[0.2, 1]$ in every repetition. 
Noise variables are independent and Pareto distributed with tail index $\alpha$, $\mathbb{P}(\varepsilon_j > x) = (1 + x)^{-\alpha}$ for $x \geq 0$, so that all variables are non-negative. This is the positive linear case in \cref{sec:certification-under-monitoring}, so \eqref{ass:confounders-monitored-in-probability} holds by construction. 
The structural equations of all designs are written out in \cref{app:simulations}.
We use $\alpha \in \{1.5, 2.5\}$, sample sizes $n$ between $10^4$ and $10^6$, and $1{,}000$ repetitions per setting.

For a source--safeguard pair $(E, S)$, the certified observations are the $k = 3$ observations with the largest $E$ among those with $S$ in the $\ell_1$-ball $\mathcal{B}$ whose radius is the $10\%$ quantile of $\|S\|_1$ (among all observations when $S = \emptyset$). 
On these we use the certified OLS $\widehat\beta_n^{\mathrm{OLS}}$ of \cref{subsec:naive-estimators}, least squares of $Y$ on $T$ through the origin, as in the application of \cref{sec:application}. 
Fixing the number of certified observations rather than a quantile of $E$ makes the certification threshold grow like $n^{1/\alpha}$, so that both terms of \eqref{eq:certified-ols-rate} are of order $n^{-1/\alpha}$, 
see \cref{rem:infinite-variance-fixed-k}.
We compare to classical two-stage least squares with instrument $Z$ in the instrumental-variable graphs, and proximal two-stage least squares \citep{TchetgenTchetgenYingCuiShiMiao2020} with treatment-inducing proxy $Z$ and outcome-inducing proxy $W$ in the double-proxy graph; the latter fits a linear bridge by regressing $W$ on $(T, Z)$ and then regresses $Y$ on $T$ and the fitted bridge. Since the classical estimators have infinite variance when $\alpha \leq 2$, we display the median over repetitions of $|\hat\beta - \beta_0|$. 
We report all numbers, error ratios and fitted slopes in \Cref{app:simulations}.

\subsection{Instruments and proxies, valid and invalid}
\label{sec:sim-iv-proxy}

\begin{figure}[t]
\centering
\begin{tikzpicture}
\definecolor{appblue}{HTML}{2A78D6}
\definecolor{apporange}{HTML}{EB6834}
\definecolor{appgreen}{HTML}{1BAF7A}
\definecolor{appviolet}{HTML}{4A3AA7}
\definecolor{appgrey}{HTML}{8A8A8A}
\definecolor{applight}{HTML}{CCCCCC}
\begin{groupplot}[
  group style={group size=4 by 1, horizontal sep=0.055\textwidth,
    xlabels at=edge bottom, ylabels at=edge left},
  scale only axis, width=0.17\textwidth, height=0.15\textwidth,
  xmode=log, ymode=log, xtick={10000,100000,1000000}, xminorticks=false, yminorticks=false,
  xmajorgrids, ymajorgrids, grid style={applight!50}, axis line style={applight}, tick style={applight},
  tick label style={font=\footnotesize}, label style={font=\small},
  title style={font=\small, yshift=-3pt}, tick label style={font=\scriptsize}, title style={font=\footnotesize},
  xlabel={sample size $n$}, ylabel={median absolute error},
  every axis plot/.append style={line width=0.9pt, mark size=1.3pt, mark options={solid}},
]
\nextgroupplot[title={(a) Valid instrument}, ytickten={-4,-3,-2,-1}]
\addplot[apporange, solid, mark=*, forget plot] coordinates {(10000,0.00761403) (50000,0.00302669) (100000,0.00176696) (500000,0.000654678) (1000000,0.000404663)};
\addplot[apporange, dashed, mark=square*, forget plot] coordinates {(10000,0.0130104) (50000,0.00547921) (100000,0.00376097) (500000,0.00185768) (1000000,0.00120529)};
\addplot[appblue, solid, mark=*, forget plot] coordinates {(10000,0.0072297) (50000,0.0023353) (100000,0.00136976) (500000,0.000493865) (1000000,0.000275879)};
\addplot[appblue, dashed, mark=square*, forget plot] coordinates {(10000,0.0358569) (50000,0.0192093) (100000,0.0154489) (500000,0.00769035) (1000000,0.00593801)};
\nextgroupplot[title={(b) Double proxy, linear}, ytickten={-4,-3,-2,-1}]
\addplot[apporange, solid, mark=*, forget plot] coordinates {(10000,0.00813073) (50000,0.00367325) (100000,0.00227899) (500000,0.000962528) (1000000,0.00052762)};
\addplot[apporange, dashed, mark=square*, forget plot] coordinates {(10000,0.0271826) (50000,0.012619) (100000,0.0084108) (500000,0.00415933) (1000000,0.00298778)};
\addplot[appblue, solid, mark=*, forget plot] coordinates {(10000,0.0195345) (50000,0.0070097) (100000,0.00419252) (500000,0.00140113) (1000000,0.000890124)};
\addplot[appblue, dashed, mark=square*, forget plot] coordinates {(10000,0.0685748) (50000,0.0352684) (100000,0.0272401) (500000,0.013384) (1000000,0.011236)};
\nextgroupplot[title={(c) Invalid instrument}, ytickten={-4,-3,-2,-1,0}]
\addplot[apporange, solid, mark=*, forget plot] coordinates {(10000,0.260541) (50000,0.270324) (100000,0.273371) (500000,0.26683) (1000000,0.266541)};
\addplot[apporange, dashed, mark=square*, forget plot] coordinates {(10000,0.277678) (50000,0.281814) (100000,0.280319) (500000,0.281483) (1000000,0.281965)};
\addplot[appblue, solid, mark=*, forget plot] coordinates {(10000,0.00980319) (50000,0.00335479) (100000,0.00199801) (500000,0.000713799) (1000000,0.000463734)};
\addplot[appblue, dashed, mark=square*, forget plot] coordinates {(10000,0.0356234) (50000,0.0173648) (100000,0.0127467) (500000,0.00714795) (1000000,0.0052993)};
\addplot[appgreen, solid, mark=*, forget plot] coordinates {(10000,0.235964) (50000,0.256929) (100000,0.267352) (500000,0.216867) (1000000,0.235024)};
\addplot[appgreen, dashed, mark=square*, forget plot] coordinates {(10000,0.146698) (50000,0.0847425) (100000,0.116768) (500000,0.0415839) (1000000,0.0257932)};
\nextgroupplot[title={(d) Double proxy, nonlinear}, ytickten={-7,-6,-5,-4,-3,-2,-1,0,1}, legend columns=3, legend style={at={(-1.4853,0)}, anchor=north, yshift=-1cm,
  draw=none, fill=none, font=\footnotesize, legend cell align=left,
  /tikz/every even column/.append style={column sep=0.45cm}}]
\addplot[apporange, solid, mark=*, forget plot] coordinates {(10000,1.0075) (50000,1.00918) (100000,1.00919) (500000,1.00919) (1000000,1.00919)};
\addplot[apporange, dashed, mark=square*, forget plot] coordinates {(10000,1.01145) (50000,1.00961) (100000,1.01001) (500000,1.00916) (1000000,1.00912)};
\addplot[appblue, solid, mark=*, forget plot] coordinates {(10000,0.000124719) (50000,1.38491e-05) (100000,6.65635e-06) (500000,7.0403e-07) (1000000,2.83031e-07)};
\addplot[appblue, dashed, mark=square*, forget plot] coordinates {(10000,0.00328476) (50000,0.000942131) (100000,0.000487044) (500000,0.000122492) (1000000,7.93225e-05)};
\addplot[appviolet, solid, mark=*, forget plot] coordinates {(10000,2.94902e-05) (50000,4.06872e-06) (100000,1.65532e-06) (500000,2.68471e-07) (1000000,1.16936e-07)};
\addplot[appviolet, dashed, mark=square*, forget plot] coordinates {(10000,0.00140832) (50000,0.000391059) (100000,0.000248383) (500000,7.21852e-05) (1000000,4.60496e-05)};
\addlegendimage{apporange, solid, line width=1pt}\addlegendentry{2SLS / proximal bridge}
\addlegendimage{appblue, solid, line width=1pt}\addlegendentry{certified OLS}
\addlegendimage{appgreen, solid, line width=1pt}\addlegendentry{certified OLS, disqualified pair}
\addlegendimage{appviolet, solid, line width=1pt}\addlegendentry{proximal bridge, correctly specified}
\addlegendimage{black, solid, line width=1pt}\addlegendentry{$\alpha = 1.5$}
\addlegendimage{black, dashed, line width=1pt}\addlegendentry{$\alpha = 2.5$}
\end{groupplot}
\end{tikzpicture}
\caption{Median absolute error $|\hat\beta - \beta_0|$ over $1000$ repetitions against sample size, on logarithmic axes. In (a) and (b) the assumption of the classical method holds; in (c) and (d) it fails. Orange: the classical estimator, two-stage least squares in (a) and (c) and proximal two-stage least squares in (b) and (d). Blue: the certified OLS with the pair selected by \cref{thm:unified-certification-prob}, $(E, S) = (Z, \emptyset)$ in (a), $(Z, \{W\})$ in (b) and (d), and $(T, \{Z\})$ in (c). Green: the certified OLS with the pair $(Z, \emptyset)$, which condition \labelcref{ass:no-unmonitored-confounding} disqualifies in (c). Violet: proximal two-stage least squares with the correctly specified quadratic bridge in (d), computed in exact arithmetic. Solid circle lines: $\alpha = 1.5$; dashed lines with squares: $\alpha = 2.5$.}
\label{fig:sim-iv-proxy}
\end{figure}
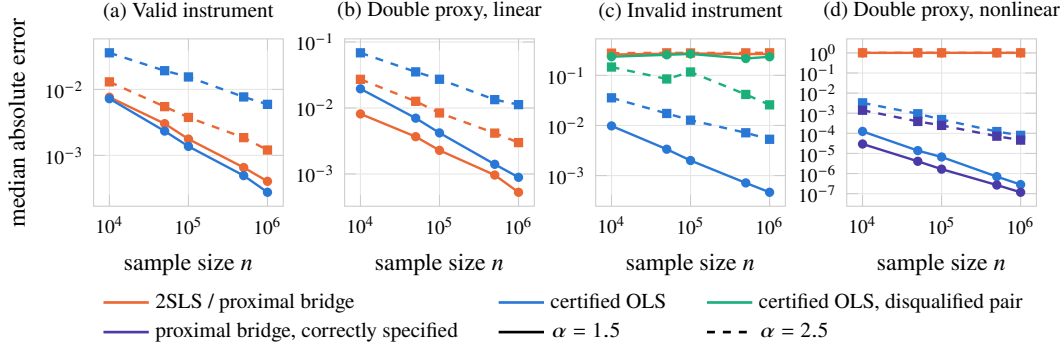

\Cref{fig:sim-iv-proxy} reports four designs. In (a) and (b) the assumption of the classical method holds; in (c) and (d) it fails, while the graphical conditions of \cref{thm:unified-certification-prob} continue to hold for a suitable pair $(E, S)$.

\emph{(a) Valid instrument.} On the graph of \cref{fig:iv-structure} the certificate uses $E = Z$ and $S = \emptyset$, for which conditions \labelcref{ass:E-affects-T,ass:E-not-affects-SY,ass:no-unmonitored-confounding} are the instrumental-variable assumptions (\cref{sec:examples-of-graphical-structures-covered}). \emph{(b) Double proxy, linear.} On the graph of \cref{fig:double-proxy-structure} the certificate uses $E = Z$ and $S = \{W\}$. The model is linear, so a linear bridge function exists and proximal two-stage least squares is consistent. \emph{(c) Invalid instrument.} Adding the edge $U \to Z$ (\cref{fig:invalid-iv-structure}) violates condition \labelcref{ass:no-unmonitored-confounding} for the pair $(Z, \emptyset)$; the criterion instead selects $E = T$ with $Z$ as safeguard, $S = \{Z\}$. We run both pairs. \emph{(d) Double proxy, nonlinear.} The same graph and the same pair $(Z, \{W\})$, but every structural assignment other than the edge $T \to Y$ is quadratic in its parents, $V_j = \varepsilon_j + \sum_{k} a_{jk} V_k^2$; in particular $R = \varepsilon_Y + a_{YU} U^2 + a_{YW} W^2$. Assumption~\ref{ass:constant-linear-causal-effect} still holds, the graphical conditions are unchanged, and $W \geq a_{WU} U^2$ still bounds $U$ on the safeguard event, but no linear bridge solves the bridge equation.

In (a) and (b), where the classical assumption holds, all estimators are consistent at comparable rates. The classical estimators are more accurate at $\alpha = 2.5$, where the noise has finite variance and method-of-moments estimators work well, whereas at $\alpha = 1.5$ the variance is infinite, the certified rate $n^{-1/\alpha}$ is faster than $n^{-1/2}$, and the certified OLS is on par with two-stage least squares although it uses three observations where the classical estimators use all $n$. 
When the classical assumptions hold, the classical estimator is a natural choice. Our intention is not to propose certification as a replacement, but point out that three certified observations carry a comparable amount of information. In practice, the classical and the certified estimators use disjoint parts of the sample and can be combined (\cref{app:combining}).
The classical estimates scatter around $\beta_0$, whereas the certified OLS overshoots, since $R \geq 0$ here, and its error is the finite-sample bias of \cref{thm:naive-certified-estimator-rate}. In (c) and (d) the classical estimators are inconsistent and miss $\beta_0$ by a wide margin in most repetitions, whereas the certified OLS converges as before. 
For illustration, we add a certified estimator in (c) with an invalid source--safeguard pair $(Z, \emptyset)$: With no safeguard, an extreme $Z$ may be an extreme $U$ in disguise, and conditioning on an extreme source alone does not remove the confounding. 
Our method does not require assumptions on other edges than $T\to Y$, so it works in (d) without modification. When the functional form of the whole system is known, the proximal estimator can be repaired: In (d) we add a correctly specified quadratic bridge, which is consistent. 
    Other proxy estimators, such as a nonparametric bridge \citep{MastouriZhuGultchinKorbaSilvaKusnerGrettonMuandet2021,KallusMaoUehara2021} in (d) or single-proxy methods \citep{XuGretton2023,VollmerPfisterWeichwald2026} in (c), could be added, but their behaviour under infinite variance is not well understood, and we do not intend to rank the certified OLS against every proxy method, only to show that certification does not depend on the functional form of the rest of the system.

The slopes (see \cref{tab:sim-slopes} for exact figures) of the certified curves match $-1/\alpha$ of \cref{thm:tail-index-certified-rates} in (a)--(c) and $-2/\alpha$ in (d), where $T \geq a_{TZ} Z^2$ makes the certified treatment values grow like $n^{2/\alpha}$. Heavier tails give faster convergence, and the certified OLS is most accurate in the nonlinear, heavy-tailed design in which the bridge estimator fails. 

\begin{figure}[t]
\centering
\begin{tikzpicture}
\definecolor{appblue}{HTML}{2A78D6}
\definecolor{apporange}{HTML}{EB6834}
\definecolor{appgreen}{HTML}{1BAF7A}
\definecolor{appviolet}{HTML}{4A3AA7}
\definecolor{appgrey}{HTML}{8A8A8A}
\definecolor{applight}{HTML}{CCCCCC}

\begin{groupplot}[
  group style={
    group size=2 by 1,
    horizontal sep=0.08\textwidth
  },
  scale only axis,
  width=0.23\textwidth,
  height=0.15\textwidth,
  xmode=log,
  ymode=log,
  xminorticks=false,
  yminorticks=false,
  xmajorgrids,
  ymajorgrids,
  grid style={applight!50},
  axis line style={applight},
  tick style={applight},
  tick label style={font=\scriptsize},
  label style={font=\small},
  title style={font=\footnotesize, yshift=-3pt},
  every axis plot/.append style={
    line width=0.9pt,
    mark size=1.3pt,
    mark options={solid}
  },
]

\nextgroupplot[
  title={(a) Gaussian noise},
  xtick={10000,100000,1000000},
  xlabel={sample size $n$},
  ylabel={median absolute error},
  ytickten={-3,-2,-1,0}
]
\addplot[apporange, mark=*, forget plot] coordinates {
  (10000,0.0352505) (50000,0.0148167) (100000,0.0106505)
  (500000,0.00470472) (1000000,0.00332005)
};
\addplot[appblue, mark=*, forget plot] coordinates {
  (10000,0.254025) (50000,0.245794) (100000,0.241372)
  (500000,0.23476) (1000000,0.237207)
};

\nextgroupplot[
  title={(b) Student-$t$ noise},
  xtick={1.5,2.5,5,10,30},
  xticklabels={1.5,2.5,5,10,30},
  xlabel={degrees of freedom $\nu$},
  ytickten={-4,-3,-2,-1,0},
  legend columns=1,
  legend style={
    at={(1.08,0.5)},
    anchor=west,
    draw=none,
    fill=none,
    font=\footnotesize,
    legend cell align=left,
    row sep=1pt
  }
]
\addplot[apporange, densely dotted, mark=triangle*, forget plot] coordinates {
  (1.5,0.0106925) (2,0.0245837) (2.5,0.0295686)
  (3,0.032044) (5,0.0329382) (10,0.0316272) (30,0.0315631)
};
\addplot[apporange, dashed, mark=square*, forget plot] coordinates {
  (1.5,0.00252582) (2,0.00744168) (2.5,0.00927607)
  (3,0.0100295) (5,0.0100915) (10,0.0102957) (30,0.010614)
};
\addplot[apporange, solid, mark=*, forget plot] coordinates {
  (1.5,0.000647955) (2,0.00244359) (2.5,0.00340081)
  (3,0.00318129) (5,0.00324206) (10,0.0031491) (30,0.00315977)
};
\addplot[appblue, densely dotted, mark=triangle*, forget plot] coordinates {
  (1.5,0.0160749) (2,0.0374252) (2.5,0.0554683)
  (3,0.0837017) (5,0.135136) (10,0.183511) (30,0.218028)
};
\addplot[appblue, dashed, mark=square*, forget plot] coordinates {
  (1.5,0.00300739) (2,0.0119453) (2.5,0.0230048)
  (3,0.0356404) (5,0.0904148) (10,0.158514) (30,0.199648)
};
\addplot[appblue, solid, mark=*, forget plot] coordinates {
  (1.5,0.000760868) (2,0.00367334) (2.5,0.00884195)
  (3,0.016394) (5,0.0519173) (10,0.111978) (30,0.196313)
};

\addlegendimage{apporange, solid, line width=1pt}
\addlegendentry{proximal bridge}
\addlegendimage{appblue, solid, line width=1pt}
\addlegendentry{certified OLS}
\addlegendimage{black, densely dotted, line width=1pt}
\addlegendentry{$n = 10^{4}$}
\addlegendimage{black, dashed, line width=1pt}
\addlegendentry{$n = 10^{5}$}
\addlegendimage{black, solid, line width=1pt}
\addlegendentry{$n = 10^{6}$}

\end{groupplot}
\end{tikzpicture}
\caption{Design (b) of \cref{fig:sim-iv-proxy} with light-tailed noise; median absolute error over $1000$ repetitions on logarithmic axes. (a) Gaussian noise, against sample size: the certificate plateaus while proximal two-stage least squares converges. (b) Student-$t$ noise, against the degrees of freedom $\nu$, for $n = 10^4$ (dotted), $10^5$ (dashed) and $10^6$ (solid): the certificate's error increases in $\nu$ towards the Gaussian value, while the bridge estimator is indifferent to $\nu$.}
\label{fig:sim-distributional}
\end{figure}
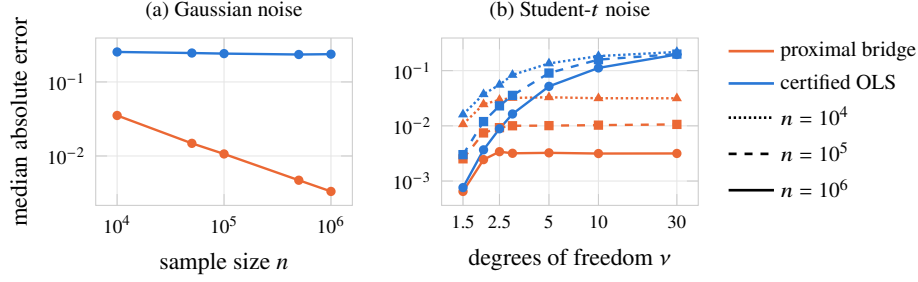

\subsection{Failure of the distributional assumption}
\label{sec:sim-distributional}

The graphical conditions are not sufficient on their own. \Cref{thm:unified-certification-prob} also requires \eqref{ass:confounders-monitored-in-probability}, that a large $U$ hiding behind a moderate safeguard is asymptotically improbable, and \cref{ex:necessity-tail-race} shows that this fails in a Gaussian linear model. We repeat design (b) of \cref{sec:sim-iv-proxy} where we replace the Pareto noise variables first by standard Gaussian noise and then by Student-$t$ noise with $\nu \in \{1.5, 2, 2.5, 3, 5, 10, 30\}$ degrees of freedom. The noise is now two-sided, so $W = a_{WU} U + \varepsilon_W$ no longer bounds $U$ almost surely, and whether $W$ monitors $U$ in probability is decided by the tail dominance of \cref{thm:extreme-propagation-tail-race-monitoring}. 

For Gaussian noise, \cref{ex:necessity-tail-race} shows that the certified estimator is not expected to work. 
For Student-$t$ noise with any finite $\nu$, tail dominance holds. For $Z$ to be extreme, say $Z \approx x$, while $W$ stays moderate, either the source's own noise is extreme, $\varepsilon_Z \approx x$, which has probability of order $x^{-\nu}$, or $U$ is extreme and its contribution to $W$ is cancelled, $\varepsilon_W \approx -a_{WU} U$, which requires two independent Student-$t$ variables to be extreme at once and has probability of order $x^{-2\nu}$. The cancellation is rarer by a factor $x^{-\nu}$, so it becomes negligible once the certified values of $Z$ are large. 
The $k$ largest of $n$ Student-$t$ draws are of order $n^{1/\nu}$, which grows quickly for heavy tails but hardly at all for light ones, $n^{1/\nu} \approx 1.6$ at $\nu = 30$ and $n = 10^6$. The theory therefore applies at every finite $\nu$, but the sample size at which cancellation stops being a plausible explanation of the certifying event grows without bound as $\nu \to \infty$, and in the Gaussian limit it is never reached.

\Cref{fig:sim-distributional}(a) shows the Gaussian case: the certified OLS does not converge, its error and its bias stay essentially constant in $n$, while proximal two-stage least squares, which requires no assumption on the tails, converges on the same data. 
On the certifying event the most likely explanation of an extreme $Z$ is a moderately large $U$ whose contribution to $W$ is cancelled by $\varepsilon_W$, so $U$ grows with the threshold and $Y/T$ converges to a value other than $\beta_0$, as in \mbox{\cref{ex:necessity-tail-race}}.

\Cref{fig:sim-distributional}(b) shows the Student-$t$ case. At every $\nu$ the certified OLS is consistent and its error decreases in $n$, but more slowly as the tail lightens, and at any fixed $n$ it increases in $\nu$ towards the Gaussian value. The composition of the error changes along the way: for small $\nu$ the certified OLS is approximately median-unbiased and its error is dispersion, whereas for large $\nu$ a bias appears, due to possible cancellations. The bridge estimator is unaffected by $\nu$. 
With moderately heavy tails, a formally valid certificate may be of little use at the sample sizes available, and a classical estimator whose structural assumptions hold is then preferable. Conversely, in designs (c) and (d) the classical estimator is inconsistent and the certificate is not.

\section{River discharge application}
\label{sec:application}

We apply the certified estimator to a dataset of river discharges (i.e. water flow) from the U.S. Geological Survey's National Water Information System \citep{USGS_NWIS}, covering 1930 to 2024 at 184 locations in the Susquehanna, Wabash and Iowa/Cedar basins.

\subsection{Design}
\label{sec:app-design}

\begin{figure}[t]
\centering
\scalebox{0.85}{%
\begin{subfigure}[b]{0.5\textwidth}
    \centering
    \begin{tikzpicture}[x=1cm, y=1cm]
\definecolor{appblue}{HTML}{2A78D6}
\definecolor{apporange}{HTML}{EB6834}
\definecolor{appgreen}{HTML}{1BAF7A}
\definecolor{appviolet}{HTML}{4A3AA7}
\definecolor{appgrey}{HTML}{8A8A8A}
\definecolor{applight}{HTML}{CCCCCC}
\begin{scope}
\clip (0,0) rectangle (7.400,4.030);
\draw[applight, line width=0.45pt] (6.722,4.460) -- (6.463,1.095);
\draw[applight, line width=0.45pt] (5.756,4.647) -- (6.463,1.095);
\draw[applight, line width=0.45pt] (6.350,3.163) -- (6.463,1.095);
\draw[applight, line width=0.45pt] (0.737,2.304) -- (1.526,2.430);
\draw[applight, line width=0.45pt] (0.995,3.171) -- (1.526,2.430);
\draw[black, line width=1.2pt] (1.526,2.430) -- (1.891,1.893);
\draw[applight, line width=0.45pt] (1.736,2.656) -- (1.891,1.893);
\draw[applight, line width=0.45pt] (-0.539,0.032) -- (0.452,0.869);
\draw[applight, line width=0.45pt] (0.452,0.869) -- (1.479,2.091);
\draw[applight, line width=0.45pt] (1.040,1.648) -- (1.479,2.091);
\draw[applight, line width=0.45pt] (1.479,2.091) -- (1.891,1.893);
\draw[applight, line width=0.45pt] (1.891,1.893) -- (5.402,2.392);
\draw[applight, line width=0.45pt] (1.933,0.850) -- (5.402,2.392);
\draw[applight, line width=0.45pt] (3.425,3.416) -- (5.402,2.392);
\draw[applight, line width=0.45pt] (4.140,2.607) -- (5.402,2.392);
\draw[applight, line width=0.45pt] (5.402,2.392) -- (6.463,1.095);
\draw[applight, line width=0.45pt] (5.231,1.493) -- (6.463,1.095);
\draw[applight, line width=0.45pt] (5.874,1.080) -- (6.463,1.095);
\draw[black, line width=0.6pt, -{Stealth[length=1.7mm, width=1.4mm]}] (1.658,2.236) -- (1.759,2.087);
\draw[appgrey, line width=0.6pt, -{Stealth[length=2.2mm, width=1.9mm]}] (3.528,2.126) -- (3.765,2.160);
\draw[appgrey, line width=0.6pt, -{Stealth[length=2.2mm, width=1.9mm]}] (5.857,1.836) -- (6.009,1.651);
\fill[appgrey] (6.350,3.163) circle (0.5mm);
\fill[appgrey] (0.737,2.304) circle (0.5mm);
\fill[appgrey] (0.995,3.171) circle (0.5mm);
\fill[appgrey] (1.736,2.656) circle (0.5mm);
\fill[appgrey] (0.452,0.869) circle (0.5mm);
\fill[appgrey] (1.040,1.648) circle (0.5mm);
\fill[appgrey] (1.479,2.091) circle (0.5mm);
\fill[appgrey] (3.425,3.416) circle (0.5mm);
\fill[appgrey] (4.140,2.607) circle (0.5mm);
\fill[appgrey] (5.402,2.392) circle (0.5mm);
\fill[appgrey] (6.463,1.095) circle (0.5mm);
\end{scope}
\begin{scope}[shift={(1.350,3.710)}, every node/.style={draw, circle, minimum size=3.2mm, inner sep=0.5pt, font=\scriptsize, fill=white}]
\fill[white, fill opacity=0.85, rounded corners=2pt] (-1.020,-0.970) rectangle (1.020,0.250);
\node[dashed] (U) at (0.000,0.000) {$U$};
\node[] (P1) at (-0.720,0.000) {$P_1$};
\node[] (P2) at (0.720,0.000) {$P_2$};
\node[] (T) at (-0.360,-0.720) {$T$};
\node[] (Y) at (0.360,-0.720) {$Y$};
\draw[-{Stealth[length=1.4mm]}, line width=0.5pt] (U) -- (P1);
\draw[-{Stealth[length=1.4mm]}, line width=0.5pt] (U) -- (P2);
\draw[-{Stealth[length=1.4mm]}, line width=0.5pt] (U) -- (T);
\draw[-{Stealth[length=1.4mm]}, line width=0.5pt] (U) -- (Y);
\draw[-{Stealth[length=1.4mm]}, line width=0.9pt] (T) -- node[draw=none, fill=none, below, font=\tiny, inner sep=1pt] {$\beta_0$} (Y);
\end{scope}
\fill[black] (1.526,2.430) ++(-1.1mm,-1.1mm) rectangle ++(2.2mm,2.2mm);
\node[anchor=east, font=\scriptsize, inner sep=1pt] at ($(1.526,2.430)+(-1.7mm,0.0mm)$) {T: Huntingdon};
\fill[black] (1.891,1.893) ++(-1.1mm,-1.1mm) rectangle ++(2.2mm,2.2mm);
\node[anchor=west, font=\scriptsize, inner sep=1pt] at ($(1.891,1.893)+(1.7mm,0.0mm)$) {Y: Mapleton Depot};
\fill[appblue] (1.933,0.850) ++(0,-1.5mm) -- ++(1.5mm,1.5mm) -- ++(-1.5mm,1.5mm) -- ++(-1.5mm,-1.5mm) -- cycle;
\node[anchor=north east, font=\scriptsize, inner sep=1pt] at ($(1.933,0.850)+(-1.3mm,-1.3mm)$) {P1: Three Springs};
\fill[appblue] (5.231,1.493) ++(0,-1.5mm) -- ++(1.5mm,1.5mm) -- ++(-1.5mm,1.5mm) -- ++(-1.5mm,-1.5mm) -- cycle;
\node[anchor=north west, font=\scriptsize, inner sep=1pt] at ($(5.231,1.493)+(1.3mm,1.3mm)$) {P2: Shermans Dale};
\fill[appblue] (5.874,1.080) ++(0,-1.5mm) -- ++(1.5mm,1.5mm) -- ++(-1.5mm,1.5mm) -- ++(-1.5mm,-1.5mm) -- cycle;
\node[anchor=north west, font=\scriptsize, inner sep=1pt] at ($(5.874,1.080)+(1.3mm,-1.3mm)$) {P3: Hogestown};
\coordinate (leg) at (3.700,-0.400);
\node[anchor=center, inner sep=0pt, font=\footnotesize] at (leg) {%
  \begin{tabular}{@{}c@{\,}l@{\hspace{3.2mm}}c@{\,}l@{\hspace{3.2mm}}c@{\,}l@{}}
  \tikz\fill[black] (0,0) rectangle (2.2mm,2.2mm); & $T$ (treatment) &
  \tikz\fill[black] (0,0) rectangle (2.2mm,2.2mm); & $Y$ (outcome) &
  \tikz\fill[appblue] (0,0) -- (1.5mm,1.5mm) -- (0,3.0mm) -- (-1.5mm,1.5mm) -- cycle; & $P$: proxy safeguard
  \end{tabular}};
\end{tikzpicture}
    \caption{A single $(T,Y)$-pair on the river network}
    \label{fig:app-map}
\end{subfigure}
\hfill
\begin{subfigure}[b]{0.42\textwidth}
    \centering
    \input{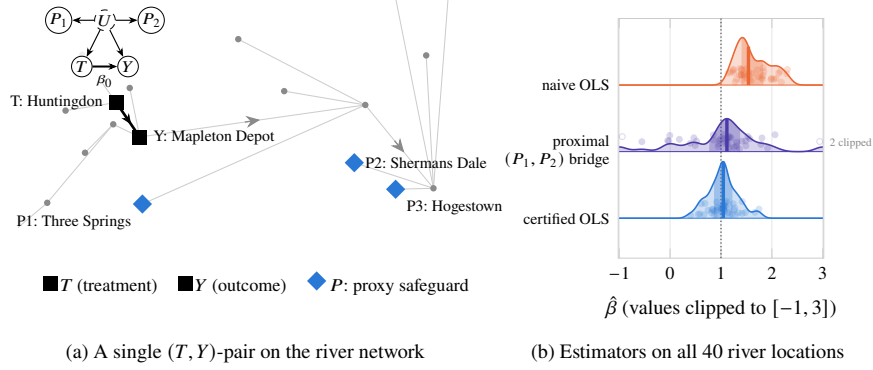}
    \caption{Estimators on all $40$ river locations}
    \label{fig:app-pairs}
\end{subfigure}%
}
\caption{(a) Juniata and lower Susquehanna river network, highlighting $T$ = Huntingdon and $Y$ = Mapleton Depot. The safeguards $P$ (blue) are the nearest locations that are hydrologically disconnected from both: they might share a confounding storm $U$, but neither feed nor descend from $T$ and $Y$. 
(b) Naive OLS, the proximal bridge and the certified OLS on the $40$ location pairs. For each estimator we plot a kernel density estimate of the estimates over pairs with the pairs as dots and a vertical line at the median. Estimates outside $[-1, 3]$ are set to the nearest edge before the dots and the density are drawn, and their number is printed at the edge. The dotted line is $\beta_0 = 1$.
}
\label{fig:app-design-and-results}
\end{figure}

Let $T$ be the daily mean discharge at an upstream location and $Y$ the discharge at a downstream location on the same river, one travel time later. 
The discharge at the upstream location should, up to some attenuation, be preserved at the downstream location, so the ground-truth causal effect of $T$ on $Y$ is $\beta_0 = 1$.
In this experiment, we test our estimator's ability to reach that conclusion. 

Additional water may appear at $Y$ both from rainfall between the locations of $T$ and $Y$ and from other tributaries than that of $T$. A basin-wide storm $U$ simultaneously raises both the flow at $T$ as well as other inflow into $Y$, so a plain regression of $Y$ on $T$ will attribute all the inflow to $T$ and bias the estimate upwards.
We construct a safeguard $P$ as the three locations nearest to $Y$ that are hydrologically disconnected (neither upstream nor downstream) from both $T$ and $Y$. The locations in $P$ might see a confounding storm, but feed neither $T$ nor $Y$, so $T \notin \mathrm{an}(P)$, condition \labelcref{ass:E-not-affects-SY} holds with $E = T$, and $P$ acts only through what it monitors about $U$. This is the double-proxy structure of \cref{sec:sim-iv-proxy}. The source is $T$ and the certifying event is $A_v = \{T > v,\ P \in \mathcal B\}$ of \cref{sec:estimation}, with $\mathcal B$ bounding each proxy location by its own $q_P$-quantile over all days: a large flood wave at $T$ on a day on which the neighbouring catchments are not in flood, hence a storm local to $T$'s catchment. \Cref{fig:app-map} shows one pair and the general causal structure we consider.
We ignore that flood peaks flatten as they travel downstream (which would make $\beta_0 < 1$), that there will be rainfall that arrives at $Y$ but not at any upstream location (which would make $\beta_0 > 1$), as well as a number of other minor hydrological factors. 

The data are daily discharges for $184$ locations in the Susquehanna, Wabash and Iowa/Cedar basins, restricted to the warm season; we filter the locations with relevance criteria, including an estimated heavy tail index of the in-season daily discharge at the upstream gauge of at most $2$ (see \cref{app:application} for all selection criteria). This leaves $40$ $(T,Y)$-pairs ($26$ Wabash, $13$ Susquehanna, $1$ Iowa/Cedar) with a median of $10{,}400$ aligned in-season days each (range $6{,}600$--$18{,}200$) and tail indices between $1.1$ and $2.0$.
We take $q_P = 0.75$, so that a day is safeguarded when each proxy location lies below its $0.75$-quantile, and take $v$ to be the $0.999$-quantile of $T$ over the safeguarded days; and a median of $6$ days per pair are certified (range $4$--$12$).
On the certified days we fit least squares $Y$ on $T$, the certified OLS; when a single day is certified, it is the ratio $Y/T$ at that day. 
We compare to a naive OLS regression of $Y$ on $T$ over all days and to the proximal bridge estimator of \cref{sec:sim-iv-proxy} with $(P_1, P_2)$ as proxies for $U$, fitted on the same days.
Further details in \cref{app:application}.

\subsection{Results}
\label{sec:app-results}
\Cref{fig:app-pairs} shows the estimated flows for every estimator and every $(T, Y)$ location.
Naive OLS is biased upwards from the ground-truth target $\beta_0 = 1$, with a median estimate over locations of $1.54$: the day-to-day covariation of the two locations is dominated by shared base flow and seasonality rather than by routed flood water. The certified OLS has a median of $1.05$, removing a median $94\%$ of the naive bias, and it is closer to the target than naive OLS on $35$ of the $40$ locations. The proximal bridge with $(P_1, P_2)$ from \cref{sec:sim-iv-proxy} has median $1.12$, and ranges from $-2.1$ to $10.1$, which is the heavy-tailed instability of \cref{fig:sim-iv-proxy}. 

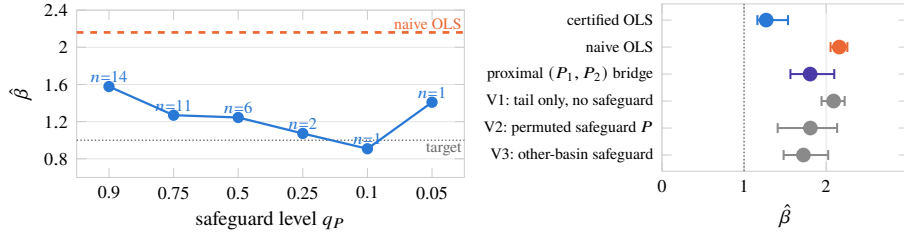
\begin{figure}[t]
\centering
\scalebox{0.85}{%
\begin{subfigure}[b]{0.49\textwidth}
    \centering
    \begin{tikzpicture}
\definecolor{appblue}{HTML}{2A78D6}
\definecolor{apporange}{HTML}{EB6834}
\definecolor{appgreen}{HTML}{1BAF7A}
\definecolor{appviolet}{HTML}{4A3AA7}
\definecolor{appgrey}{HTML}{8A8A8A}
\definecolor{applight}{HTML}{CCCCCC}
\pgfmathsetlengthmacro{\appaxw}{\linewidth-1.2cm}
\begin{axis}[
  scale only axis, width=\appaxw, height=0.36\linewidth,
  symbolic x coords={0.9,0.75,0.5,0.25,0.1,0.05}, xtick={0.9,0.75,0.5,0.25,0.1,0.05}, enlarge x limits=0.1,
  xlabel={safeguard level $q_P$}, ylabel={$\hat\beta$},
  ymin=0.6, ymax=2.4, ytick={0.8,1.2,...,2.4},
  ymajorgrids, grid style={applight!50},
  axis line style={applight}, tick style={applight},
  tick label style={font=\footnotesize}, label style={font=\small},
  ylabel style={yshift=-2pt}, xlabel style={yshift=2pt},
  clip=false,
]
\draw[densely dotted, black!60, line width=0.6pt] ({rel axis cs:0,0}|-{axis cs:0.9,1.0}) -- ({rel axis cs:1,0}|-{axis cs:0.9,1.0});
\draw[dashed, apporange, line width=1.2pt] ({rel axis cs:0,0}|-{axis cs:0.9,2.1600}) -- ({rel axis cs:1,0}|-{axis cs:0.9,2.1600});
\node[anchor=south east, font=\scriptsize , apporange, inner sep=1pt] at ({rel axis cs:1,0}|-{axis cs:0.9,2.1600}) {naive OLS};
\node[anchor=north east, font=\scriptsize , black!60, inner sep=1pt] at ({rel axis cs:1,0}|-{axis cs:0.9,1.0}) {target};
\addplot[appblue, line width=1pt, mark=*, mark size=2pt,
  nodes near coords={$n{=}\pgfplotspointmeta$}, point meta=explicit symbolic,
  visualization depends on={value \thisrow{anchor} \as \appanchor},
  every node near coord/.style={font=\scriptsize , color=appblue, anchor=\appanchor, inner sep=1.5pt}]
  table[x=q, y=beta, meta=n] {Figures/application/app_headline_sweep.dat};
\end{axis}
\end{tikzpicture}
\end{subfigure}
\hfill
\begin{subfigure}[b]{0.49\textwidth}
    \centering
    \raisebox{5mm}{\begin{tikzpicture}
\definecolor{appblue}{HTML}{2A78D6}
\definecolor{apporange}{HTML}{EB6834}
\definecolor{appgreen}{HTML}{1BAF7A}
\definecolor{appviolet}{HTML}{4A3AA7}
\definecolor{appgrey}{HTML}{8A8A8A}
\definecolor{applight}{HTML}{CCCCCC}
\pgfmathsetlengthmacro{\appaxw}{\linewidth-3.4cm}
\begin{axis}[
  scale only axis, width=\appaxw, height=0.36\linewidth,
  xmin=0, xmax=3, xtick={0,1,2,3},
  xlabel={$\hat\beta$}, xlabel style={yshift=2pt},
  ymin=0.4, ymax=6.6, y dir=reverse,
  ytick={1,...,6}, yticklabels={{certified OLS},{naive OLS},{proximal $(P_1, P_2)$ bridge},{V1: tail only, no safeguard},{V2: permuted safeguard $P$},{V3: other-basin safeguard}},
  xmajorgrids, grid style={applight!50},
  axis line style={applight}, tick style={applight},
  tick label style={font=\footnotesize}, label style={font=\small},
  y tick label style={font=\scriptsize},
  clip=false,
]
\draw[densely dotted, black!60, line width=0.6pt] ({axis cs:1,0}|-{rel axis cs:0,0}) -- ({axis cs:1,0}|-{rel axis cs:0,1});
\addplot[appblue, only marks, mark=*, mark size=2.8pt, line width=0.9pt,
  error bars/.cd, x dir=both, x explicit, error bar style={line width=0.9pt},
  error mark options={rotate=90, mark size=2.2pt, line width=0.9pt}]
  coordinates {(1.2695,1) -= (0.1070,0) += (0.2660,0)};
\addplot[apporange, only marks, mark=*, mark size=2.8pt, line width=0.9pt,
  error bars/.cd, x dir=both, x explicit, error bar style={line width=0.9pt},
  error mark options={rotate=90, mark size=2.2pt, line width=0.9pt}]
  coordinates {(2.1600,2) -= (0.1061,0) += (0.0987,0)};
\addplot[appviolet, only marks, mark=*, mark size=2.8pt, line width=0.9pt,
  error bars/.cd, x dir=both, x explicit, error bar style={line width=0.9pt},
  error mark options={rotate=90, mark size=2.2pt, line width=0.9pt}]
  coordinates {(1.8048,3) -= (0.2398,0) += (0.2938,0)};
\addplot[appgrey, only marks, mark=*, mark size=2.8pt, line width=0.9pt,
  error bars/.cd, x dir=both, x explicit, error bar style={line width=0.9pt},
  error mark options={rotate=90, mark size=2.2pt, line width=0.9pt}]
  coordinates {(2.0902,4) -= (0.1464,0) += (0.1351,0)};
\addplot[appgrey, only marks, mark=*, mark size=2.8pt, line width=0.9pt,
  error bars/.cd, x dir=both, x explicit, error bar style={line width=0.9pt},
  error mark options={rotate=90, mark size=2.2pt, line width=0.9pt}]
  coordinates {(1.8080,5) -= (0.3983,0) += (0.3258,0)};
\addplot[appgrey, only marks, mark=*, mark size=2.8pt, line width=0.9pt,
  error bars/.cd, x dir=both, x explicit, error bar style={line width=0.9pt},
  error mark options={rotate=90, mark size=2.2pt, line width=0.9pt}]
  coordinates {(1.7233,6) -= (0.2414,0) += (0.3015,0)};
\end{axis}
\end{tikzpicture}
}
\end{subfigure}%
}
\vspace{-0.3cm}
\caption{
Estimators at a single location, Huntingdon to Mapleton Depot. Left: The certified OLS estimate, with the number of certified days at each point; the dotted line the ground-truth target. Right: The estimators and the ablations V1--V3 at $q_P = 0.75$, with $90\%$ year-level bootstrap intervals for every estimator and ablation.
}
\label{fig:app-headline}
\end{figure}

    \Cref{fig:app-headline} examines a single location, the Juniata river from Huntingdon to Mapleton Depot (the pair in \cref{fig:app-map}).
Naive OLS gives $2.16$ and the bridge $1.80$, while the certified OLS gives $1.27$ on $11$ certified days. Three ablations show that the safeguard drives the improvement: the tail restriction on $T$ alone (V1) gives $2.09$, a safeguard permuted in time (V2) $1.81$ and a safeguard from another basin (V3) $1.72$. As the safeguard level $q_P$ tightens, the certified OLS moves towards the target until only one or two days are certified (\cref{fig:app-headline}a). The bootstrap intervals, the ablations and the sweep are detailed in \cref{app:application}.

\subsection{Negative controls: disconnected pairs}
\label{sec:app-placebo}

\begin{figure}[htp]
    \centering
    \input{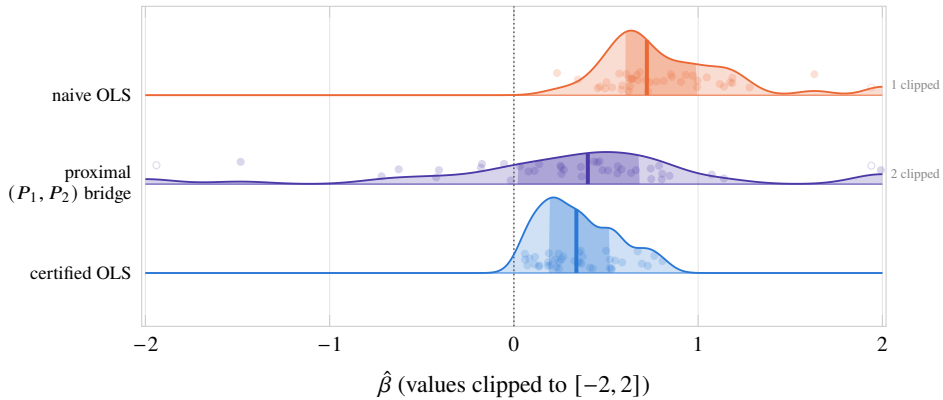}
    \caption{Estimated flows between hydrologically disconnected gauges, where the true routing coefficient is zero. The naive OLS, the proximal bridge and the certified OLS are shown, with the same conventions as in \cref{fig:app-pairs}, on the window $[-2, 2]$. If an estimator manages to remove the confounding, it should return to zero.}
\label{fig:app-placebo}
\end{figure}

An estimator that removes confounding should return zero when the true effect is zero and the confounding remains. 
We construct one such design and run the estimators of \cref{fig:app-pairs} on it (\cref{fig:app-placebo} and the location selection rules are in \cref{app:application}).
Two locations in the same basin but with no flow path between them have routing coefficient zero, but are still confounded by the shared weather; we take the $40$ closest such pairs that meet the record and proxy requirements of the main design. 
Naive OLS has median $0.72$. The certified OLS moves to $0.34$ and is closer to zero than naive OLS on $38$ of the $40$ pairs. The bridge has median $0.40$, though it also takes values down to $-2.6$ and up to $4.8$.
While the certified OLS moves towards the true value, it does not reach it. This indicates confounding that is unmonitored by the safeguards, rendering $T$ and $Y$ dependent even when $P$ is ordinary. 

\vspace{-0.5cm}
\section{Conclusion and discussion}

Extreme observations enable causal identification under unobserved confounding without a valid instrument or proxy completeness. Certification selects observations in which a source is extreme while safeguards remain ordinary, making confounding negligible relative to treatment. Graphical conditions specify which latent ancestors require monitoring, while extreme propagation or tail dominance ensures effective monitoring. The framework covers valid instruments, double-proxy structures, and some settings where invalid instruments serve as safeguards. Geometrically, certification isolates the extremal ray whose slope is the causal effect.
In simulations the certificate converges at the rate from \cref{thm:tail-index-certified-rates} and remains consistent where two-stage least squares and proximal two-stage least squares fail. On the river data the certified OLS removes a median $94\%$ of the naive bias (\cref{fig:app-pairs}), and ablations show  the safeguard drives the improvement.

Our method has limitations. 
Under light-tailed noise, cancellation among moderate deviations becomes the most plausible explanation of a certifying event, so the safeguards no longer monitor the confounder and the certificate does not converge (\cref{fig:sim-distributional}). 
The sample that survives certification is typically small, with a median of $6$ days per river location in the application, which may result in a high variance of certified estimators. Normal approximations are then unreliable, and we propose instead the rank interval of \cref{app:confidence-intervals}, which is valid for a fixed number of certified observations.
Another weakness is that safeguards must cover every latent ancestor of the source; in the placebo design the certified estimate moved towards zero but not to it, which points to unmonitored confounding.

Several questions remain open. 
Certification discards the bulk of the data; for structures in which a classical estimator is also valid, the certified and the classical estimator can be used jointly; \cref{app:combining} discusses how to combine them, but it is still unclear how to optimally use all the sample available. 
The threshold and the safeguard level were fixed in our setup, but a data-driven choice would make the method more robust in real-world applications.
Our method requires the user to assume a causal graph, but one could combine this with methods for causal discovery, to identify potential safeguards for a treatment-response pair. 
\vspace{-0.5cm}
\section*{Acknowledgements }
\addcontentsline{toc}{section}{Acknowledgements}
All authors contributed equally to the paper. We thank Niels Richard Hansen for hosting a research visit of Juraj Bodik in Copenhagen, during which the initial ideas for this work were developed. We would also like to thank Dominik Rothenhäusler for useful discussions. Nikolaj Thams is supported by the Novo Nordisk Foundation (grant NNF26OC0114284).
\vspace{-0.5cm}
\section*{Declaration of the use of generative AI}
\addcontentsline{toc}{section}{Declaration of the use of generative AI}
Almost all conceptual ideas were developed without AI and a majority of the written text was written by the authors. 
We used AI tools to help with coding, to improve the clarity of the text and some feedback. The authors reviewed and edited all content that was touched by an LLM and take full responsibility for the content of the publication.
\vspace{-0.5cm}
\section*{Supplementary material}
\addcontentsline{toc}{section}{Supplementary material overview}
The supplement includes proofs of all results, an extended version of \cref{thm:naive-certified-estimator-rate} with limiting distributions, the single-maximum estimator, a confidence interval, and simulation details, data selection rules, estimation details, per-pair results and placebo figure of the application.

\FloatBarrier
\supplementarymaterial
\pdfbookmark[0]{Supplementary material}{nav.supplement}

This supplement contains the proofs of \cref{lem:certifying-sequence} and \cref{thm:unified-certification-prob,thm:extreme-propagation-tail-race-monitoring} from \cref{Section_3}, of \cref{thm:certified-extremal-rays} from \cref{sec:causal-pot}, and of \cref{thm:naive-certified-estimator-rate,thm:tail-index-certified-rates} from \cref{sec:estimation}; an auxiliary lemma on subexponential perturbations; \cref{prop:single-certified-maximum} related to Remark~\ref{Remark_max_estimator}; a confidence interval for $\beta_0$ based on the certified sample (\cref{app:confidence-intervals}); ways of combining certified estimators with classical bulk estimators (\cref{app:combining}); the numerical details of the simulation study in \cref{sec:simulations}; and the data, selection rules and estimation details of the application in \cref{sec:application}. Numbers of the form S1 refer to the supplement; all other numbers refer to the main paper. Data and code are located at  \url{https://github.com/nikolajthams/extreme-proxy}

\section{Proofs for Section~\ref{Section_3}}

\begin{customlem}{\ref*{lem:certifying-sequence}}[Certifying sequence]
Let Assumption~\ref{ass:constant-linear-causal-effect} hold, and let $(A_v)_{v\in\mathbb{R}}$ be a family of events with $\mathbb P(A_v)>0$ for all $v$ satisfying:
\begin{enumerate}[leftmargin=2em]
    \item \textbf{$T$ is extreme.} $T\,|\,A_v \overset{d}{\to} \infty$ as $v\to\infty$. That is, for every finite $c$,
    \begin{equation}
        \mathbb P(T\leq c\mid A_v)\longrightarrow0,
        \qquad v\to\infty.
 \tag{\ref{eq:selected-treatment-diverges}}
    \end{equation}
   
    \item \textbf{$R$ is not extreme.} For every $\epsilon>0$, there is a finite $M$, not depending on $v$, such that \begin{equation}
        \limsup_{v\to\infty}
        \mathbb P\bigl(|R|>M\mid A_v\bigr)<\epsilon.
        \tag{\ref{eq:selected-remainder-controlled}}
    \end{equation}
\end{enumerate}
Then $(A_v)_{v\in\mathbb R}$ is a certifying sequence and $\beta_0$ is identified from \eqref{CONVERGENCE} with $g(y,t)=y/t$. 
\end{customlem}

\begin{proof}
For each \(v\), write $ \mathbb P_v(\,\cdot\,)
    :=
    \mathbb P(\,\cdot\mid A_v).$
Define \(Y/T\) arbitrarily on the event \(\{T=0\}\). This convention is
asymptotically irrelevant because, for every \(c>0\),
\[
    \mathbb P_v(T=0)
    \leq
    \mathbb P_v(T\leq c)
    \longrightarrow 0.
\]
Under Assumption~\ref{ass:constant-linear-causal-effect},
 $ \frac{Y}{T}-\beta_0=\frac{R}{T}.$
We show that the right-hand side converges to zero in \(\mathbb P_v\)-probability.

Fix \(\eta>0\) and \(\delta>0\). By \eqref{eq:selected-remainder-controlled}, there exists a finite
\(M\geq0\), independent of \(v\), such that
\[
    \limsup_{v\to\infty}
    \mathbb P_v(|R|>M)
    <
    \delta.
\]
Choose a finite \(c>M/\eta\). On the event
$ \{T>c,\ |R|\leq M\},$
we have
\[
    \left|\frac{R}{T}\right|
    \leq
    \frac{M}{T}
    \leq
    \frac{M}{c}
    <
    \eta.
\]
Consequently,
\[
\begin{aligned}
    \mathbb P_v\left(
        \left|\frac{Y}{T}-\beta_0\right|>\eta
    \right)
    &=
    \mathbb P_v\left(
        \left|\frac{R}{T}\right|>\eta
    \right)\\
    &\leq
    \mathbb P_v(T\leq c)
    +
    \mathbb P_v(|R|>M).
\end{aligned}
\]
Taking the limit superior as \(v\to\infty\), the first term vanishes by
\eqref{eq:selected-treatment-diverges}, while the second term has limit
superior smaller than \(\delta\). Therefore,
\[
    \limsup_{v\to\infty}
    \mathbb P_v\left(
        \left|\frac{Y}{T}-\beta_0\right|>\eta
    \right)
    \leq \delta.
\]
Since \(\delta>0\) was arbitrary, $ \mathbb P_v\left(
        \left|\frac{Y}{T}-\beta_0\right|>\eta
    \right)
    \longrightarrow0.$ As this holds for every \(\eta>0\), it follows that
\[
    \frac{Y}{T}\,\bigg|\,A_v
    \xrightarrow{\mathbb P}
    \beta_0.
\]
Thus \((A_v)_{v\in\mathbb R}\) is a certifying sequence by
Definition~\ref{def:certifying-sequence}.

Finally, the conditional laws of \(Y/T\) given \(A_v\) converge weakly to
the point mass \(\delta_{\beta_0}\). If the same conditional laws also
converged to \(\delta_b\) for some \(b\in\mathbb R\), uniqueness of weak
limits would imply
  $ \delta_b=\delta_{\beta_0},$
and hence \(b=\beta_0\). Therefore, \(\beta_0\) is uniquely identified
from the limiting relation~\eqref{CONVERGENCE}.
\end{proof}
Tail unconfoundedness also underlies extrapolation of dose–response functions to extreme treatment levels \citep{supp:bodik_mathemtaics}.

\begin{customthm}{\ref*{thm:unified-certification-prob}}[Certification by graphical coverage under monitoring]
Let Assumption~\ref{ass:constant-linear-causal-effect} hold, and let $(E,S)$ be a source--safeguard pair satisfying the graphical conditions \labelcref{ass:E-affects-T,ass:E-not-affects-SY,ass:no-unmonitored-confounding} and the monitoring condition \eqref{ass:confounders-monitored-in-probability} of Theorem~\ref{thm:unified-certification-prob}. Then $(A_v^{E,S})_v$ is a certifying sequence for $\beta_0$ with $g(y,t)=y/t$, and $\beta_0$ is identified.
\end{customthm}

\begin{myproof}
Write $A_v:=A_v^{E,S}$, $\mathbb P_v:=\mathbb P(\cdot\mid A_v)$ and $D:=\mathrm{an}_{G^-}(Y)$. The proof covers the empty safeguard $S=\emptyset$, for which we let $\{S\in\mathcal B\}$ be a probability-one event and $\mathrm{an}(S)=\emptyset$. 
In line with Definition~\ref{def:prob-monitoring}, we call a random vector $V$ \emph{$(A_v)$-tight} if its conditional laws given $A_v$ are tight, that is, if for every $\epsilon>0$ there is a compact $K$, not depending on $v$, with $\limsup_v\mathbb P_v(V\notin K)<\epsilon$; monitoring in probability of a node is $(A_v)$-tightness of that node.

Each $V_j$ is
$\sigma(\varepsilon_k:k\in\mathrm{an}(j))$-measurable, so
$A_v\in\sigma(\varepsilon_k:k\in\mathrm{an}(E)\cup\mathrm{an}(S))$ and $R$ is
$\sigma(\varepsilon_j:j\in D)$-measurable.
Moreover, $T\notin D$: a directed path into $T$ cannot use the edge $T\to Y$, so $E\in\mathrm{an}(T)$ from \labelcref{ass:E-affects-T} gives $E\in\mathrm{an}_{G^-}(T)$, and $T\in D$ would then imply $E\in D$, contradicting \labelcref{ass:E-not-affects-SY}. Since the parents of any $j\in D\setminus\{Y\}$ lie in $D$, no such parent equals $T$.

By Lemma~\ref{lem:certifying-sequence}, it suffices to verify the two conditions \eqref{eq:selected-treatment-diverges} and \eqref{eq:selected-remainder-controlled} for $(A_v)_v$. The first, together with the positivity $\mathbb P(A_v)>0$, is part of \labelcref{ass:E-affects-T}, so only the conditional tightness of $R$ remains to be shown.

\emph{Conditional tightness of $R$.} We show by induction along
a topological order of $D\setminus\{Y\}$ that every $V_j$,
$j\in D\setminus\{Y\}$, is $(A_v)$-tight. If $j\in\mathrm{an}(S)$, then
$j\in\mathrm{an}(S)\cap D=\mathcal C_S$ and $V_j$ is monitored in probability by
\eqref{ass:confounders-monitored-in-probability}, hence tight. If $j\notin\mathrm{an}(S)$, then
$j\notin\mathrm{an}(E)$: otherwise
$j\in\mathrm{an}(E)\cap D\subseteq\mathrm{an}(S)$ by \labelcref{ass:no-unmonitored-confounding}, a
contradiction. Consequently $\varepsilon_j$ is independent of $A_v$ and is tight with its
unconditional law; the parents of $j$ lie in $D$, precede $j$, and are tight by
induction; and continuity of $f_j$ transfers tightness to $V_j$. Finally,
writing, per Assumption~\ref{ass:constant-linear-causal-effect}, the assignment
of $Y$ as
$f_Y(V_{\mathrm{pa}(Y)},\varepsilon_Y)=\beta_0T+h_Y(V_{\mathrm{pa}(Y)\setminus\{T\}},\varepsilon_Y)$
with $h_Y$ continuous (since $f_Y$ is), we have
$R=h_Y(V_{\mathrm{pa}(Y)\setminus\{T\}},\varepsilon_Y)$ with
parents in $D\setminus\{Y\}$, which are tight, and with $\varepsilon_Y$ independent of $A_v$: indeed
$Y\notin\mathrm{an}(E)$, since $Y\in\mathrm{an}(E)$ together with
$E\in\mathrm{an}(T)$ from \labelcref{ass:E-affects-T} and the edge $T\to Y$ would create a
directed cycle; and $Y\notin\mathrm{an}(S)$, since $Y\in\mathrm{an}(S)$ with
$T\to Y$ would give $T\in\mathrm{an}(S)$, hence $E\in\mathrm{an}(S)$ by
\labelcref{ass:E-affects-T}, contradicting \labelcref{ass:E-not-affects-SY}. Joint conditional
tightness of finitely many components follows by a union bound over compacts,
and continuity of $h_Y$ gives: for every $\epsilon>0$ there is $M<\infty$,
independent of $v$, with $\limsup_v\mathbb P_v(|R|>M)<\epsilon$, which is \eqref{eq:selected-remainder-controlled}.

Both conditions of Lemma~\ref{lem:certifying-sequence} hold, so $(A_v)_v$ is a certifying sequence with $g(y,t)=y/t$; since $A_v$ and $Y/T$ are functions of observed variables, $\beta_0$ is identified.
\end{myproof}

\begin{customthm}{\ref*{thm:extreme-propagation-tail-race-monitoring}}[Extreme propagation or tail dominance $\implies$ monitoring]

Consider model \eqref{SCM_general}. Let $(E,S)$ be a source--safeguard pair and suppose that
$\mathbb P(A_v^{E,S})>0$ for all $v$. If one of the following holds:
\begin{enumerate}[label=(\roman*),leftmargin=2.5em]
    \item Strong extreme propagation from $U\to S$ (Definition~\ref{def:extreme-propagation}),
    \item Weak extreme propagation from $U\to S$ (Definition~\ref{def:extreme-propagation}) and the SCM is an SBJ-SCM,
    \item $E$ has tail dominance over $U$ (Definition~\ref{tail_race}),
\end{enumerate}
then $U$ is monitored in probability along $(A_v^{E,S})_v$ and \eqref{ass:confounders-monitored-in-probability} is satisfied.  
\end{customthm}

\begin{proof}
Write
\[
    A_v:=A_v^{E,S},
    \qquad
    C:=\{S\in\mathcal B\},
    \qquad 
        \tau_{\mathcal B}
    :=
    \sup_{s\in\mathcal B}\|s\|.
\]
Since \(S\in\mathcal B\) on \(A_v\), the vector \(S\) is monitored
almost surely. It therefore remains to prove that \(U\mid A_v\) is
tight.

\paragraph{Strong extreme propagation.}

\textbf{1)} First suppose that the negation form of strong extreme propagation
holds. Then there is a compact \(B_U\) such that $ C\subseteq\{U\in B_U\}$ almost surely.
Since \(A_v\subseteq C\), it follows that
\[
    \mathbb P(U\notin B_U\mid A_v)=0
\]
for every \(v\), and hence \(U\) is monitored almost surely.

\textbf{2)} Next suppose that the structural form holds:
\[
    \lim_{\substack{\|u\|\to\infty\\ u\in\operatorname{supp}(U)}}  %
    \inf_{e\in\operatorname{supp}(\varepsilon_S)}
    \|f_S(u,e)\|
    =
    \infty.
\]
Since \(\tau_{\mathcal B}<\infty\), there exists \(M<\infty\) such
that
\[
    \|u\|>M,\ u\in\operatorname{supp}(U)
    \quad\Longrightarrow\quad
    \inf_{e\in\operatorname{supp}(\varepsilon_S)}
    \|f_S(u,e)\|
    >
    \tau_{\mathcal B}.
\]
Thus \(S\in\mathcal B\) implies \(\|U\|\leq M\) almost surely (as $U\in\operatorname{supp}(U)$ almost surely), so
again \(U\) is monitored almost surely.

\textbf{3)} Finally, suppose that the bounding form holds:
\[
    \|S\|
    \geq
    \gamma(\|U\|)-\Lambda,
\]
where \(\gamma\) is nondecreasing, \(\gamma(r)\to\infty\), and
\(\Lambda\mid A_v=O_{\mathbb P}(1)\). Conditioning on \(A_v\),
\[
    \gamma(\|U\|)
    \leq
    \tau_{\mathcal B}+\Lambda.
\]
Consequently, since \(\gamma\) is nondecreasing, for every \(M>0\),
\[
\begin{aligned}
    \mathbb P(\|U\|>M\mid A_v)
    &\leq
    \mathbb P\!\left(
        \Lambda
        \geq
        \gamma(M)-\tau_{\mathcal B}
        \,\middle|\,
        A_v
    \right).
\end{aligned}
\]
Since \(\Lambda\mid A_v=O_{\mathbb P}(1)\) and
\(\gamma(M)\to\infty\), the right-hand side can be made arbitrarily
small uniformly for large \(v\) by taking \(M\) sufficiently large.
Hence \(U\mid A_v\) is tight.

\paragraph{Weak extreme propagation in an SBJ-SCM.}

Let $ \boldsymbol\varepsilon
    =
    (\varepsilon_1,\ldots,\varepsilon_m)$ denote the exogenous innovation vector, and put
$ \overline F_\star(x)
    :=
    \mathbb P(|\varepsilon_1|>x).$
By the SBJ assumptions, \(\overline F_\star\) is regularly varying
with index $\alpha$, and the upper tail of the innovations is not negligible: $\mathbb P(\varepsilon_1>x)\sim p_+\overline F_\star(x)$ for some $p_+>0$.

Write the reduced forms of \(U\) and \(E\) as
\[
    U
    =
    \Phi_U(\boldsymbol\varepsilon)
    +
    r_U(\boldsymbol\varepsilon),
    \qquad
    E
    =
    \phi_E(\boldsymbol\varepsilon)
    +
    r_E(\boldsymbol\varepsilon),
\]
where \(\Phi_U\) is the vector of the corresponding continuous
sublinear maps and \(r_U,r_E\) are bounded. By continuity and
positive homogeneity, there are finite constants \(c_U,c_E\) such
that
\[
    \|\Phi_U(x)\|
    \leq
    c_U\|x\|_\infty,
    \qquad
    |\phi_E(x)|
    \leq
    c_E\|x\|_\infty.
\]
In particular, there exists \(c_0>0\) such that
\[
    E>v
    \quad\Longrightarrow\quad
    \max_j|\varepsilon_j|>c_0v
\]
for all sufficiently large \(v\).

Let \(j_E\) denote the coordinate corresponding to the source's own
innovation. Nondegeneracy in the own innovation implies
$ a_E:=\phi_E(e_{j_E})>0$: if $a_E\leq0$, sublinearity would give $\phi_E(te_{j_E}+z)\leq ta_E+\phi_E(z)\leq\phi_E(z)$ for all $t\geq0$, so $E$ would be bounded above as $\varepsilon_{j_E}\to\infty$ with the other innovations fixed, contradicting $\lim_{e_{j_E}\to\infty}g_E(e_{-j_E},e_{j_E})=\infty$.
Moreover, \(C=\{S\in\mathcal B\}\) does not depend on
\(\varepsilon_{j_E}\), because $E\notin\mathrm{an}(S)$ in model \eqref{SCM_general}. Since \(\mathbb P(C)>0\), we can choose
\(K_0<\infty\) such that
\[
    p_0
    :=
    \mathbb P\!\left(
        C,\ 
        \max_{j\neq j_E}|\varepsilon_j|\leq K_0
    \right)
    >
    0.
\]
For \(z_{j_E}=0\), sublinearity gives
\[
    \phi_E(te_{j_E}+z)
    \geq
    t\phi_E(e_{j_E})-\phi_E(-z)
    =
    ta_E-\phi_E(-z).
\]
Uniformly over \(\|z\|_\infty\leq K_0\), the last term is bounded.
Since \(r_E\) is also bounded, there is \(c_1<\infty\) such that
\[
    \varepsilon_{j_E}>c_1v,
    \quad
    \max_{j\neq j_E}|\varepsilon_j|\leq K_0
    \quad\Longrightarrow\quad
    E>v
\]
for all sufficiently large \(v\). Therefore, by independence of $\varepsilon_{j_E}$ from the other innovations, and since $\mathbb P(\varepsilon_{j_E}>c_1v)\sim p_+c_1^{-\alpha}\overline F_\star(v)$ with $p_+>0$, there is some \(c_2>0\) with
\begin{equation}
    \mathbb P(A_v)
    \geq
    p_0\mathbb P(\varepsilon_{j_E}>c_1v)
    \geq
    c_2\overline F_\star(v)
    \label{eq:thm3-1}
\end{equation}
for all sufficiently large $v$.

For every fixed \(\delta>0\), weak extreme propagation gives
\[
\begin{aligned}
    \mathbb P(
        A_v,\ \|U\|>\delta v
    )
    &\leq
    \mathbb P(
        C,\ \|U\|>\delta v
    )
=
    \mathbb P(\|U\|>\delta v)
    \mathbb P(
        C\mid \|U\|>\delta v
    )
 =
    o\!\left(
        \mathbb P(\|U\|>\delta v)
    \right).
\end{aligned}
\]
The tail-sublinear representation implies $ \mathbb P(\|U\|>\delta v)
    =
    O\!\left(\overline F_\star(v)\right),$
and hence
\begin{equation}
    \mathbb P(
        A_v,\ \|U\|>\delta v
    )
    =
    o\!\left(\overline F_\star(v)\right).
    \label{eq:thm3-2}
\end{equation}

It remains to rule out values of \(U\) which diverge more slowly than
\(v\). Let \(\varepsilon_{(1)}\geq\varepsilon_{(2)}\) denote the largest and
second-largest values among
  $ |\varepsilon_1|,\ldots,|\varepsilon_m|.$

For \(j=1,\ldots,m\) and \(\sigma\in\{-1,1\}\), define
 $ a_{j,\sigma}
    :=
    \Phi_U(\sigma e_j).$
For \(z_j=0\), sublinearity of each coordinate of \(\Phi_U\) gives
\[
    \left\|
        \Phi_U(\sigma te_j+z)
        -
        t a_{j,\sigma}
    \right\|
    \leq
    c_3\|z\|_\infty
\]
for some finite \(c_3\).

Fix \(K<\infty\). On \(\{\varepsilon_{(2)}\leq K\}\), there is at most one
innovation whose magnitude exceeds \(K\); let $(j,\sigma)$ denote its index and its sign, so that $\varepsilon_j=\sigma t$ with $t=\varepsilon_{(1)}>K$. If
\(a_{j,\sigma}\neq0\), then
\[
    \|U\|
    \geq
    t\|a_{j,\sigma}\|-c_3K-c_4,
\]
where \(c_4\) bounds \(r_U\). Since there are only finitely many
pairs \((j,\sigma)\), there exists \(\delta>0\), not depending on $K$, such that, for all
sufficiently large \(v\),
\[
    E>v,\quad
    \varepsilon_{(2)}\leq K,\quad
    a_{j,\sigma}\neq0
    \quad\Longrightarrow\quad
    \|U\|>\delta v.
\]
If \(a_{j,\sigma}=0\), then instead
\[
    \|U\|
    \leq
    c_3K+c_4
    =:
    M_K.
\]
It follows that, for all sufficiently large \(v\),
\begin{equation}
\begin{split}
    A_v\cap\{\|U\|>M_K\}
    \subseteq\;&
    \left\{
        C,\ \|U\|>\delta v
    \right\}
   \cup
    \left\{
        \varepsilon_{(1)}>c_0v,\ \varepsilon_{(2)}>K
    \right\}.
\end{split}
\label{eq:thm3-3}
\end{equation}
By independence of the innovations,
\begin{equation}
\begin{aligned}
    \mathbb P(
        \varepsilon_{(1)}>c_0v,\ \varepsilon_{(2)}>K
    )
    &\leq
    \sum_{j\neq k}
    \mathbb P(|\varepsilon_j|>c_0v)
    \mathbb P(|\varepsilon_k|>K)
    \\
    &\leq
    c_5\overline F_\star(v)
    \overline F_\star(K).
\end{aligned}
\label{eq:thm3-4}
\end{equation}
Combining \eqref{eq:thm3-1}--\eqref{eq:thm3-4} yields
\[
    \limsup_{v\to\infty}
    \mathbb P(
        \|U\|>M_K
        \mid A_v
    )
    \leq
    c_6\overline F_\star(K).
\]
Since \(\overline F_\star(K)\to0\) as \(K\to\infty\), this proves
that \(U\mid A_v\) is tight.

\paragraph{Tail dominance: conditional form.}

Suppose that condition
\labelcref{cond:tail-race-conditional} holds. Put
\[
    G:=g_E(U,S),
    \qquad
    \overline F_E(x)
    :=
    \mathbb P(\varepsilon_E>x).
\]
The innovation \(\varepsilon_E\) is independent of \((U,S)\).
Since \(\overline F_E\) is subexponential, it is in particular
long-tailed.

Because \(\mathbb P(C)>0\) and \(G\) is finite almost surely, there
exists \(K_0<\infty\) such that
$ q_0
    :=
    \mathbb P(
        C,\ G\geq-K_0
    )
    >
    0.$
Consequently,
\begin{equation}
\begin{aligned}
    \mathbb P(A_v)
    &=
    \mathbb E\!\left[
        \mathbf 1_C
        \overline F_E(v-G)
    \right]
 \geq
    q_0\overline F_E(v+K_0)
   \sim
    q_0\overline F_E(v).
\end{aligned}
\label{eq:thm3-5}
\end{equation}

Let \(L\) be a Lipschitz constant for \(g_E\). Since
\(\mathcal B\) is compact, there exists \(c_0<\infty\) such that,
on \(C\),
\begin{equation}
    G_+
    :=
    \max(G,0)
    \leq
    c_0+L\|U\|.
    \label{eq:thm3-6}
\end{equation}
If \(L=0\), then \(G_+\) is bounded on \(C\), and the argument below
simplifies. Suppose \(L>0\). For all sufficiently large \(x\),
\begin{equation}
\begin{aligned}
    \mathbb P(G_+>x,\ C)
    &\leq
    \mathbb P\!\left(
        \|U\|>\frac{x-c_0}{L},
        \ C
    \right)
\leq
    \mathbb P\!\left(
        \|U\|>\frac{x}{2L},
        \ C
    \right)
=
    o\!\left(\overline F_E(x)\right),
\end{aligned}
\label{eq:thm3-7}
\end{equation}
where the last line follows from the conditional tail-dominance
assumption with \(a=(2L)^{-1}\).

For \(M>0\), define
\[
    D_M
    :=
    \{
        \|U\|>M,\ C
    \},
    \qquad
    q_M:=\mathbb P(D_M).
\]
If \(q_M=0\), there is nothing to prove. Otherwise, let \(H_M\) have
the conditional distribution
  $  \mathcal L(H_M)
    :=
    \mathcal L(G_+\mid D_M).$
By \eqref{eq:thm3-7},
\begin{equation}
    \mathbb P(H_M>x)
    =
    \frac{
        \mathbb P(G_+>x,\ D_M)
    }{
        q_M
    }
    \leq
    \frac{
        \mathbb P(G_+>x,\ C)
    }{
        q_M
    }
    =
    o\!\left(\overline F_E(x)\right).
\label{eq:thm3-8}
\end{equation}
Taking \(H_M\) independent of \(\varepsilon_E\),
Lemma~\ref{lem:subexponential-perturbation} gives
\begin{equation}
    \mathbb P(\varepsilon_E+H_M>v)
    \sim
    \overline F_E(v).
    \label{eq:thm3-9}
\end{equation}
Moreover, since $\overline F_E$ is nonincreasing and $G\leq G_+$,
\[
\begin{aligned}
    \mathbb P(D_M,A_v)
    &=
    \mathbb E\!\left[
        \mathbf 1_{D_M}
        \overline F_E(v-G)
    \right]
 \leq
    \mathbb E\!\left[
        \mathbf 1_{D_M}
        \overline F_E(v-G_+)
    \right]
  =
    q_M
    \mathbb P(\varepsilon_E+H_M>v).
\end{aligned}
\]
Combining this with \eqref{eq:thm3-5} and \eqref{eq:thm3-9}, and using $\{\|U\|>M\}\cap A_v=D_M\cap A_v$, gives
\begin{equation}
    \limsup_{v\to\infty}
    \mathbb P(
        \|U\|>M
        \mid A_v
    )
    \leq
    \frac{q_M}{q_0}.
    \label{eq:thm3-10}
\end{equation}
Finally,
$    q_M
    =
    \mathbb P(
        \|U\|>M,\ C
    )
    \longrightarrow0$,
as \(M\to\infty\). Equation \eqref{eq:thm3-10} therefore proves that
\(U\mid A_v\) is tight.

\paragraph{Tail dominance II: joint form.}

Suppose that condition \labelcref{cond:tail-race-primitive} holds, so that $ U=\varepsilon_U,
    S=BU+\varepsilon_S,$ where \(B\in\mathbb R^{d_S\times k}\) has full column rank. Let
$\sigma_{\min}(B)>0$
denote its smallest singular value, which is positive since the columns of $B$ are linearly independent, so that $\|Bu\|\geq\sigma_{\min}(B)\|u\|$ for every $u\in\mathbb R^k$, and recall
 $ \tau_{\mathcal B}
    =
    \sup_{s\in\mathcal B}\|s\|<\infty.$

Fix \(a>0\). On the event
$ \{\|U\|>av,\ S\in\mathcal B\},$
we have
$ \|\varepsilon_U\|=\|U\|>av,$
and
\[
\begin{aligned}
    \|\varepsilon_S\|
    &=
    \|S-BU\| \geq
    \|BU\|-\|S\| \geq
    \sigma_{\min}(B)\|U\|-\tau_{\mathcal B}.
\end{aligned}
\]
Hence, for all sufficiently large \(v\),
\[
    \{\|U\|>av,\ S\in\mathcal B\}
    \subseteq
    \left\{
        \|\varepsilon_U\|>av,\,
        \|\varepsilon_S\|>
        \frac{a\sigma_{\min}(B)}{2}v
    \right\}.
\]
Therefore, by the tail-dominance condition \labelcref{cond:tail-race-primitive},
\[
\begin{aligned}
    \mathbb P(\|U\|>av,\ S\in\mathcal B)
    &=
    o\!\left(\mathbb P(\varepsilon_E>v)\right).
\end{aligned}
\]
Since \(\mathbb P(S\in\mathcal B)>0\), it follows that
$ \mathbb P\!\left(
        \|U\|>av
        \,\middle|\,
        S\in\mathcal B
    \right)
    =
    o\!\left(\mathbb P(\varepsilon_E>v)\right).$
The remaining requirements of the conditional form, namely $E=g_E(U,S)+\varepsilon_E$ with $g_E$ globally Lipschitz and $\varepsilon_E$ independent of $(U,S)$ with a subexponential upper tail, are part of condition \labelcref{cond:tail-race-primitive}. Thus condition \labelcref{cond:tail-race-conditional} holds, and the previous case applies.
\end{proof}

\begin{lemma}[Subexponential perturbation]
\label{lem:subexponential-perturbation}
Let \(X\) have a subexponential upper tail
$ \overline F(x):=\mathbb P(X>x),$
and let \(Z\geq0\) be independent of \(X\). 
\[
\text{If } \mathbb P(Z>x)
    =
    o\!\left(\overline F(x)\right),
\quad \text{then } \quad 
    \mathbb P(X+Z>x)
    \sim
    \overline F(x),
    \qquad x\to\infty.
\]
\end{lemma}

\begin{proof}
Let \(H\) denote the distribution of \(Z\). Since
\(\overline H(x)=o(\overline F(x))\) and \(F\) is subexponential,
the standard convolution property of the subexponential class gives
\[
    \overline{F*H}(x)
    =
    \overline F(x)
    +
    \overline H(x)
    +
    o\!\left(\overline F(x)\right).
\]
Consequently,
\[
    \mathbb P(X+Z>x)
    =
    \overline{F*H}(x)
    \sim
    \overline F(x).
\]
See, for example, \citet{supp:FossKorshunovZachary2013}.
\end{proof}

\section{Proof for Section~\ref{sec:causal-pot}}
\begin{customthm}{\ref*{thm:certified-extremal-rays}}[Certification isolates the causal ray]
\label{thm_certified-extremal-rays_appendix_version}
Consider the linear SCM \(
V_j=\sum_{k\in\operatorname{pa}(j)}b_{jk}V_k+\varepsilon_j
\), for \(j=1,\ldots,d\), with i.i.d.\ regularly varying noise variables. To define the directions and weights, write
\[
\overline F(x)
:=
\mathbb P\!\left(|\varepsilon_1|>x\right)
=
x^{-\alpha}L(x),
\qquad
    p_\sigma
    :=
    \lim_{x\to\infty}
    \frac{\mathbb P(\sigma\varepsilon_1>x)}
         {\mathbb P(|\varepsilon_1|>x)},
    \qquad \sigma\in\{-1,1\},
\]
so that \(p_{-1}+p_1=1\). Let \(B=(b_{jk})_{j,k=1}^d\) and \(M=(I_d-B)^{-1}\), and write
\[
    \boldsymbol a_j
    =
    \begin{pmatrix}
        M_{Tj}\\
        M_{Yj}
    \end{pmatrix},
    \qquad
    \mathcal J
    =
    \{j:\boldsymbol a_j\neq\boldsymbol 0\}.
\]
Thus, \(\boldsymbol a_j\) contains the total effects of the structural
disturbance \(\varepsilon_j\) on \(T\) and \(Y\). Define
\[
    \boldsymbol\omega_{j,\sigma}
    =
    \frac{\sigma\boldsymbol a_j}
         {\lVert\boldsymbol a_j\rVert},
    \qquad
    \pi_{j,\sigma}
    =
    \frac{
        p_\sigma\lVert\boldsymbol a_j\rVert^\alpha
    }{
        \displaystyle
        \sum_{k\in\mathcal J}
        \lVert\boldsymbol a_k\rVert^\alpha
    },
    \qquad j\in\mathcal J,\ \sigma\in\{-1,1\}.
\]

Let \(
\boldsymbol X=(T,Y)^\top
\) be two of the nodes and
\(
\boldsymbol\Theta
=\boldsymbol X/\lVert\boldsymbol X\rVert
\), where \(\lVert\cdot\rVert\) is any norm on \(\mathbb R^2\) such that  \(\lVert(0,1)^\top\rVert=1\). Then
\begin{equation}
    \mathcal L\!\left(
        \boldsymbol\Theta
        \,\middle|\,
        \lVert\boldsymbol X\rVert>v
    \right)
    \overset{d}{\longrightarrow}
    \sum_{j\in\mathcal J}
    \sum_{\sigma\in\{-1,1\}}
        \pi_{j,\sigma}\,
        \delta_{\boldsymbol\omega_{j,\sigma}},
    \qquad v\to\infty.
    \tag{\ref*{eq:uncertified-ray-limit}}
\end{equation}

Let \((E,S)\) satisfy the conditions of
Theorem~\ref{thm:unified-certification-prob}, let
\(C=\{S\in\mathcal B\}\), and let \(\mathcal L_C\) denote laws under
\(\mathbb P_C(\,\cdot\,)=\mathbb P(\,\cdot\mid C)\). Writing
\[
    \boldsymbol\omega_{\beta_0}
    =
    \frac{(1,\beta_0)^\top}
         {\lVert(1,\beta_0)^\top\rVert},
\]
we have for \(
    \widetilde\pi_T\geq0,
    \widetilde\pi_Y\geq0,
    \widetilde\pi_T+\widetilde\pi_Y=1
\):
\begin{align}
\mathcal L_C\!\left(
    \boldsymbol\Theta
    \,\middle|\,
    \lVert\boldsymbol X\rVert>v
\right)
&\ \overset{d}{\longrightarrow}\
    \widetilde\pi_T\delta_{\boldsymbol\omega_{\beta_0}}
    +\widetilde\pi_Y\delta_{(0,1)},
\qquad v\to\infty,
\tag{\ref*{eq:conditional-ray-limit}}
\\
\mathcal L\!\left(
    \boldsymbol\Theta
    \,\middle|\,
    A_v^{E,S}
\right)
&\ \overset{d}{\longrightarrow}\
    \delta_{\boldsymbol\omega_{\beta_0}},
\qquad v\to\infty,
\tag{\ref*{eq:certified-ray-limit}}
\end{align}
where the first convergence additionally requires \(E=T\) and \(\boldsymbol X\in\mathbb R_+^2\) almost surely. 
\end{customthm}

\begin{proof}
Let \(\boldsymbol\varepsilon=(\varepsilon_1,\ldots,\varepsilon_d)^\top\). Since the graph is acyclic, the vertices may be ordered topologically, in
which case \(B\) is strictly lower triangular and
\[
    M=(I_d-B)^{-1}
      =I_d+B+\cdots+B^{d-1}.
\]
Hence
\(
    \boldsymbol V=M\boldsymbol\varepsilon,
    \boldsymbol X
    =
    \sum_{j=1}^d
        \boldsymbol a_j\varepsilon_j.
\)

\emph{Proof of \eqref{eq:uncertified-ray-limit}.}
Independence and regular variation imply that, for distinct \(j\) and \(k\),
\[
    \frac{
        \mathbb P(
            |\varepsilon_j|>x,\,
            |\varepsilon_k|>x
        )
    }{
        \overline F(x)
    }
    =
    \overline F(x)
    \longrightarrow0.
\]
Thus simultaneous extreme disturbances are negligible relative to a single
extreme disturbance. The tail measure of
\(\boldsymbol\varepsilon\) is therefore concentrated on the signed coordinate
axes and is given by
\[
    \mu_\varepsilon(\,\cdot\,)
    =
    \sum_{j=1}^d
    \sum_{\sigma\in\{-1,1\}}
        p_\sigma
        \int_0^\infty
            \mathbbm 1\{
                r\sigma\boldsymbol e_j\in\,\cdot\,
            \}
            \alpha r^{-\alpha-1}\,dr;
\]
see, for example, \citet{supp:Resnick} and \citet{supp:HultLindskog2006}.
Applying the linear map
\(
\boldsymbol z\mapsto
\sum_{j=1}^d\boldsymbol a_jz_j
\)
gives the tail measure of \(\boldsymbol X\):
\[
    \nu_{\boldsymbol X}(\,\cdot\,)
    =
    \sum_{j\in\mathcal J}
    \sum_{\sigma\in\{-1,1\}}
        p_\sigma
        \int_0^\infty
            \mathbbm 1\{
                r\sigma\boldsymbol a_j\in\,\cdot\,
            \}
            \alpha r^{-\alpha-1}\,dr.
\]

Let
\(
\mathbb S=\{\boldsymbol\theta:
\lVert\boldsymbol\theta\rVert=1\}
\),
and let \(D\subseteq\mathbb S\) be a Borel set whose boundary contains none
of the directions \(\boldsymbol\omega_{j,\sigma}\). It follows that
\begin{align*}
\lim_{v\to\infty}
\frac{
    \mathbb P\!\left(
        \lVert\boldsymbol X\rVert>v,\,
        \boldsymbol\Theta\in D
    \right)
}{
    \overline F(v)
}
&=
\sum_{j\in\mathcal J}
\sum_{\sigma\in\{-1,1\}}
    p_\sigma
    \int_{1/\lVert\boldsymbol a_j\rVert}^{\infty}
        \mathbbm 1\{
            \boldsymbol\omega_{j,\sigma}\in D
        \}
        \alpha r^{-\alpha-1}\,dr
\\
&\qquad
=
\sum_{j\in\mathcal J}
\sum_{\sigma\in\{-1,1\}}
    p_\sigma
    \lVert\boldsymbol a_j\rVert^\alpha
    \mathbbm 1\{
        \boldsymbol\omega_{j,\sigma}\in D
    \}.
\end{align*}
Taking \(D=\mathbb S\) and using \(p_{-1}+p_1=1\) gives
\[
    \mathbb P(
        \lVert\boldsymbol X\rVert>v
    )
    \sim
    \overline F(v)
    \sum_{k\in\mathcal J}
        \lVert\boldsymbol a_k\rVert^\alpha.
\]
Dividing the preceding two limits yields
\[
    \mathbb P\!\left(
        \boldsymbol\Theta\in D
        \,\middle|\,
        \lVert\boldsymbol X\rVert>v
    \right)
    \longrightarrow
    \sum_{j\in\mathcal J}
    \sum_{\sigma\in\{-1,1\}}
        \pi_{j,\sigma}
        \mathbbm 1\{
            \boldsymbol\omega_{j,\sigma}\in D
        \},
\]
which proves \eqref{eq:uncertified-ray-limit}.

\emph{Proof of 
\eqref{eq:certified-ray-limit}.}
No additional assumption that \(E=T\) or
\(\boldsymbol X\in\mathbb R_+^2\) is needed for this part. Condition
\labelcref{ass:E-affects-T} gives
\(
    \mathbb P\!\left(
        T\leq0
        \,\middle|\,
        A_v^{E,S}
    \right)
    \longrightarrow0.
\)
The conclusion of
Theorem~\ref{thm:unified-certification-prob} also gives
\[
    \frac{Y}{T}
    \,\Big|\,
    A_v^{E,S}
    \xrightarrow{\mathbb P}
    \beta_0.
\]
Since \(
    \boldsymbol\Theta
    =
    \frac{
        (1,Y/T)^\top
    }{
        \lVert(1,Y/T)^\top\rVert
    }\) (when $T>0$), and since the map
\(
    z
    \longmapsto
    \frac{(1,z)^\top}
         {\lVert(1,z)^\top\rVert}
\)
is continuous on \(\mathbb R\), the conditional continuous mapping theorem
therefore yields
\[
    \boldsymbol\Theta
    \,\Big|\,
    A_v^{E,S}
    \xrightarrow{\mathbb P}
    \frac{
        (1,\beta_0)^\top
    }{
        \lVert(1,\beta_0)^\top\rVert
    }
    =
    \boldsymbol\omega_{\beta_0}.
\]
Equivalently,
\(
    \mathcal L\!\left(
        \boldsymbol\Theta
        \,\middle|\,
        A_v^{E,S}
    \right)
    \overset{d}{\longrightarrow}
    \delta_{\boldsymbol\omega_{\beta_0}},
\)
which proves  \eqref{eq:certified-ray-limit}.

\emph{Proof of \eqref{eq:conditional-ray-limit}.}
Suppose henceforth that
\(
    E=T,
    \boldsymbol X\in\mathbb R_+^2
\) almost surely. 

\textbf{Observation 1:} because of the nonnegativity assumption, every \(\varepsilon_j\) is bounded below. Indeed, we may write
\[
    T
    =
    \varepsilon_T+T_{-T},
    \qquad \text{ where }\,\,
    T_{-T}
    :=
    \sum_{k\neq T}M_{Tk}\varepsilon_k,
\]
where \(T_{-T}\) is independent of \(\varepsilon_T\). Choose a finite
constant \(c\) such that \(
    \mathbb P(T_{-T}\leq c)>0.
\)
If the lower tail of \(\varepsilon_T\) were unbounded, then, for some
\(x>c\),
\[
\begin{aligned}
    \mathbb P(T<0)
    &\geq
    \mathbb P\!\left(
        T_{-T}\leq c,\,
        \varepsilon_T<-x
    \right)=
    \mathbb P(T_{-T}\leq c)
    \mathbb P(\varepsilon_T<-x)
    >0,
\end{aligned}
\]
contradicting \(T\geq0\) almost surely. Thus \(\varepsilon_T\), and hence
every \(\varepsilon_j\) by identical distribution, is bounded below.
Consequently, $  p_1=1,  p_{-1}=0$ and 
\(
    \overline F(x)
    =
    \mathbb P(\varepsilon_1>x)
\).

\textbf{Observation 2:} the two rows of $M$ related to $T,Y$ are non-negative. For \(q\in\{T,Y\}\), write
\[
    X_q
    =
    M_{qj}\varepsilon_j+X_{q,-j},
    \qquad
    X_{q,-j}
    :=
    \sum_{k\neq j}M_{qk}\varepsilon_k.
\]
Obviously, \(X_{q,-j}\indep \varepsilon_j\). If \(M_{qj}<0\), choose a finite \(c\) satisfying
\(
    \mathbb P(X_{q,-j}\leq c)>0
\)
and then choose \(x\) sufficiently large that
\(
    c + M_{qj}x<0.
\)
Since the upper tail of \(\varepsilon_j\) is unbounded,
\[
\begin{aligned}
    \mathbb P(X_q<0)
    &\geq
    \mathbb P\!\left(
        X_{q,-j}\leq c,\,
        \varepsilon_j>x
    \right)
=
    \mathbb P(X_{q,-j}\leq c)
    \mathbb P(\varepsilon_j>x)
    >0,
\end{aligned}
\]
again contradicting the assumed nonnegativity. Hence
\[
    M_{Tj}\geq0,
    \qquad
    M_{Yj}\geq0,
    \qquad j=1,\ldots,d.
    \label{eq:nonnegative-reduced-loadings}
\]

\textbf{More notation:} For each disturbance, write
\[
    \boldsymbol s_j
    :=
    M_{Sj},
    \qquad
    S
    =
    \sum_{j=1}^d
        \boldsymbol s_j\varepsilon_j,
        \qquad   \mathcal J_S
    :=
    \{j:\boldsymbol s_j\neq\boldsymbol0\},
    \qquad
    \mathcal J_C
    :=
    \mathcal J\setminus\mathcal J_S.
\]
Thus \(\mathcal J_C\) contains precisely those disturbances that reach \((T,Y)\) but have zero total effect on \(S\).

Condition
\labelcref{ass:E-not-affects-SY} of
Theorem~\ref{thm:unified-certification-prob} gives
\(
    T\notin\operatorname{an}(S),
    T\notin\operatorname{an}_{G^-}(Y).
\)
It follows that
\(
    \boldsymbol s_T=\boldsymbol0.
\)
Moreover, the disturbance \(\varepsilon_T\) enters \(T\) with coefficient
one and cannot enter the remainder \(R=Y-\beta_0T\). Therefore
\[
 \boldsymbol a_T
    =
    (1,\beta_0)^\top, \qquad    T\in\mathcal J_C,
    \qquad
    \mathcal J_C\neq\emptyset.
    \label{eq:treatment-shock-in-JC}
\]
For \(j\in\mathcal J\), define
\(
    \boldsymbol\omega_j
    :=
    \frac{\boldsymbol a_j}
         {\lVert\boldsymbol a_j\rVert}.
\)
We now show that, for every Borel set \(D\subseteq\mathbb S\) whose boundary
contains none of the directions
\(\boldsymbol\omega_j\), \(j\in\mathcal J_C\),
\begin{equation}
\begin{aligned}
&\lim_{v\to\infty}
\frac{
    \mathbb P\!\left(
        \lVert\boldsymbol X\rVert>v,\,
        \boldsymbol\Theta\in D,\,
        C
    \right)
}{
    \overline F(v)
}
\qquad
= \qquad
\mathbb P(C)
\sum_{j\in\mathcal J_C}
    \lVert\boldsymbol a_j\rVert^\alpha
    \mathbbm 1\{
        \boldsymbol\omega_j\in D
    \}.
\end{aligned}
\label{eq:conditional-ray-key}
\end{equation}

Put
\(
    \tau
    :=
    \sup_{\boldsymbol s\in\mathcal B}
        \lVert\boldsymbol s\rVert,
    \kappa
    :=
    \max_{1\leq j\leq d}
        \lVert\boldsymbol a_j\rVert.
\)
Let \(\varepsilon_{(1)}\geq\varepsilon_{(2)}\) denote the two largest
values among
\(
    |\varepsilon_1|,
    \ldots,
    |\varepsilon_d|.
\)

\emph{Step 1: disturbances affecting \(S\) are negligible.}
For every \(\eta>0\), independence and regular variation give
\[\mathbb P(\varepsilon_{(2)}>\eta v)
\leq\binom d2\overline F(\eta v)^2=o(\overline F(v)).\]
Fix \(j\in\mathcal J_S\) and put
\(b_j:=\eta\lVert\boldsymbol s_j\rVert/
(2\sum_k\lVert\boldsymbol s_k\rVert)>0\).
Conditioning on \(C\cap\{|\varepsilon_j|>\eta v\}\), at least one other disturbance
must exceed \(b_jv\) in absolute value for all sufficiently large \(v\);
otherwise the reverse triangle inequality would give
\(\lVert S\rVert>\eta\lVert\boldsymbol s_j\rVert v/2>\tau\).
The preceding bound, applied with \(\min\{\eta,b_j\}\), therefore yields
$$\mathbb P(C,|\varepsilon_j|>\eta v)=o(\overline F(v)).$$

\emph{Step 2: the conditional tail measure.}
Write \(\boldsymbol X=\boldsymbol X_0+\boldsymbol X_1\), where
\(\boldsymbol X_0:=\sum_{j\in\mathcal J_C}\boldsymbol a_j\varepsilon_j\)
and
\(\boldsymbol X_1:=\sum_{j\in\mathcal J_S}\boldsymbol a_j\varepsilon_j\).
Since \(C\) depends only on disturbances indexed by \(\mathcal J_S\),
\(\boldsymbol X_0\) is independent of \(C\).
Moreover, \(\lVert\boldsymbol X_1\rVert
\leq\kappa\sum_{j\in\mathcal J_S}|\varepsilon_j|\), so Step~1 and a finite
union bound give
$$\mathbb P_C(\lVert\boldsymbol X_1\rVert>\delta v)
=o(\overline F(v)) \text{ for every }\delta>0.$$

Apply the unconditional tail calculation above to \(\boldsymbol X_0\),
with the index set restricted to \(\mathcal J_C\).
Adding \(\boldsymbol X_1\) does not change this tail limit.
Indeed, outside an event of conditional probability
\(o(\overline F(v))\), the norms of \(\boldsymbol X\) and
\(\boldsymbol X_0\) differ by at most \(\delta v\).
On \(\{\lVert\boldsymbol X\rVert>v\}\), their directions then differ
by at most \(2\delta\).
Thus the conditional tail probability is bounded above and below by
the corresponding probabilities for \(\boldsymbol X_0\), using radial
thresholds \((1-\delta)v\) and \((1+\delta)v\), respectively, and angular
sets obtained by enlarging and contracting \(D\) by \(2\delta\).
For sufficiently small \(\delta\), these angular sets contain exactly
the same limiting rays as \(D\), because \(\partial D\) avoids them.
Letting \(v\to\infty\) and then \(\delta\downarrow0\) proves
\eqref{eq:conditional-ray-key}.
Taking \(D=\mathbb S\) and dividing the limits gives
\begin{equation}
    \mathcal L_C\!\left(
        \boldsymbol\Theta\mid\lVert\boldsymbol X\rVert>v
    \right)
    \overset{d}{\longrightarrow}
    H_C:=\sum_{j\in\mathcal J_C}\pi_j^C
        \delta_{\boldsymbol\omega_j},
    \qquad
    \pi_j^C:=
    \frac{\lVert\boldsymbol a_j\rVert^\alpha}
         {\sum_{k\in\mathcal J_C}\lVert\boldsymbol a_k\rVert^\alpha}.
    \label{eq:conditional-ray-weights}
\end{equation}

\emph{Step 3: identification of the surviving rays.}
Suppose that \(j\in\mathcal J_C\), \(M_{Tj}>0\), and
\(\boldsymbol\omega_j\neq\boldsymbol\omega_{\beta_0}\).
Choose a spherical neighborhood \(D\) of \(\boldsymbol\omega_j\)
whose closure excludes \(\boldsymbol\omega_{\beta_0}\), whose boundary
avoids all the finitely many rays, and on which \(\theta_T>c\) for
some \(c>0\).
Since \(E=T\), the already proved convergence
\eqref{eq:certified-ray-limit} implies
\(\mathbb P_C(\boldsymbol\Theta\in D\mid T>cv)\to0\).
The unconditional tail asymptotic and continuity of the coordinate
projection give \(\mathbb P_C(T>cv)=O(\overline F(v))\). Hence
\[
    \mathbb P_C(\lVert\boldsymbol X\rVert>v,\boldsymbol\Theta\in D)
    \leq
    \mathbb P_C(T>cv,\boldsymbol\Theta\in D)
    =o(\overline F(v)),
\]
contradicting the strictly positive limit in
\eqref{eq:conditional-ray-key}.
Consequently, every surviving ray with \(M_{Tj}>0\) must be equal to
\(\boldsymbol\omega_{\beta_0}\).

If \(M_{Tj}=0\), then \(M_{Yj}>0\), so
\(\boldsymbol\omega_j=(0,1)^\top/\lVert(0,1)^\top\rVert\).
Define
\(\widetilde\pi_T:=\sum_{j\in\mathcal J_C:\,M_{Tj}>0}\pi_j^C\)
and
\(\widetilde\pi_Y:=\sum_{j\in\mathcal J_C:\,M_{Tj}=0}\pi_j^C\).
These weights are nonnegative and sum to one.
Thus, under the normalization \(\lVert(0,1)^\top\rVert=1\),
\eqref{eq:conditional-ray-weights} is precisely
\eqref{eq:conditional-ray-limit}.
For an arbitrary norm, its vertical atom is instead at
\((0,1)^\top/\lVert(0,1)^\top\rVert\).
\end{proof}

\section{Proofs for Section~\ref{sec:estimation} and the single-maximum estimator}
\begin{customthm}{\ref*{thm:naive-certified-estimator-rate}}[Convergence rates under finite moments]
Let $v_n\to\infty$, $m_n\to\infty$, and
$A_v\subseteq\{T\geq v\}$ for all sufficiently large $v$.

\begin{enumerate}[label=(\roman*),leftmargin=2.5em]
    \item \emph{OLS through the origin.} Put
    $\mu_v(T)=\mathbb E(R\mid T,A_v)$. If, uniformly for large $v$,
    $|\mu_v(T)|\leq M$ and
    $\operatorname{Var}(R\mid T,A_v)\leq\sigma^2$ almost surely, then
    \begin{equation*}
        \widehat\beta_n^{\mathrm{OLS}}-\beta_0
        =O_{\mathbb P}(v_n^{-1})
         +O_{\mathbb P}\!\left((v_n\sqrt{m_n})^{-1}\right).
        \tag{\ref*{eq:certified-ols-rate}}
    \end{equation*}
    If $\mu_v(T)=0$ almost surely for all sufficiently large $v$, the first
    term is absent. 
     \item \emph{Ray estimator.} Let $\lVert\cdot\rVert$ be the $\ell_1$ norm, $\lVert\boldsymbol X\rVert=|T|+|Y|$. Define $b_v^{\mathrm{ray}} := v\,\mathbb E(R/\lVert\boldsymbol X\rVert\mid A_v)/\mathbb E(T/\lVert\boldsymbol X\rVert\mid A_v)$.
    Suppose, uniformly for large $v$,
    $\mathbb E(R^2\mid T,A_v)\leq M_2$ almost surely. Then
    $b_v^{\mathrm{ray}}=O(1)$ and
    \begin{equation*}
        \widehat\beta_n^{\mathrm{ray}}-\beta_0
        =
        \frac{b_{v_n}^{\mathrm{ray}}}{v_n}
        +
        O_{\mathbb P}\!\left((v_n\sqrt{m_n})^{-1}\right)
        =
        O(v_n^{-1})
        +
        O_{\mathbb P}\!\left((v_n\sqrt{m_n})^{-1}\right).
        \tag{\ref*{eq:certified-ray-rate}}
    \end{equation*}
    If $\mathbb E(R\mid T,A_v)=0$ almost surely, then
    $b_v^{\mathrm{ray}}=O(v^{-1})$, so the first term improves to
    $O(v_n^{-2})$. If $\mathbb E(R/\lVert\boldsymbol X\rVert\mid A_v)=0$, it is absent.
    
\end{enumerate}
\end{customthm}

\begin{proof}
The two statements are the corresponding items of \cref{prop:certified-rates-extended} below.
\end{proof}

\begin{proposition}[Extended convergence rates under finite moments]
\label{prop:certified-rates-extended}
In the setting of \cref{thm:naive-certified-estimator-rate}, the following hold.
Let $v_n\to\infty$, $m_n\to\infty$, and
$A_v\subseteq\{T\geq v\}$ for all sufficiently large $v$.

\begin{enumerate}[label=(\roman*),leftmargin=2.5em]
    \item \emph{OLS through the origin.} Put
    $\mu_v(T)=\mathbb E(R\mid T,A_v)$. If, uniformly for large $v$,
    $|\mu_v(T)|\leq M$ and
    $\operatorname{Var}(R\mid T,A_v)\leq\sigma^2$ almost surely, then
    \begin{equation*}
        \widehat\beta_n^{\mathrm{OLS}}-\beta_0
        =O_{\mathbb P}(v_n^{-1})
         +O_{\mathbb P}\!\left((v_n\sqrt{m_n})^{-1}\right).
        \tag{\ref*{eq:certified-ols-rate}}
    \end{equation*}
    If $\mu_v(T)=0$ almost surely for all sufficiently large $v$, the first
    term is absent. 
    Moreover, let $Z_n=T/v_n$ and suppose, under
    $\mathbb P(\,\cdot\mid A_{v_n})$,
    $(Z_n,R)\xrightarrow{d} (Z_\infty,R_\infty)$ and, for some $\delta>0$,
    \[
        \sup_n\mathbb E\!\left(
            |Z_nR|^{2+\delta}+Z_n^{4+2\delta}
            \,\middle|\,A_{v_n}
        \right)<\infty.
    \]
   Then
    \begin{equation*}
        \sqrt{m_n}\left\{
            v_n(\widehat\beta_n^{\mathrm{OLS}}-\beta_0)
            -b_n^{\mathrm{OLS}}
        \right\}
        \xrightarrow{d}
        \mathcal N\!\left(
            0,
            \frac{
                \operatorname{Var}\!\left(
                    Z_\infty R_\infty
                    -b_\infty^{\mathrm{OLS}}Z_\infty^2
                \right)
            }{
                \mathbb E(Z_\infty^2)^2
            }
        \right),
    \end{equation*} where
$ b_n^{\mathrm{OLS}}
        :=
        \frac{\mathbb E(Z_nR\mid A_{v_n})}
             {\mathbb E(Z_n^2\mid A_{v_n})},
        b_\infty^{\mathrm{OLS}}
        :=
        \frac{\mathbb E(Z_\infty R_\infty)}
             {\mathbb E(Z_\infty^2)}$.
    \item \emph{OLS with an intercept.} Suppose, uniformly for large $v$,
    $\mu_v(T)=a_v$ almost surely and
    $\operatorname{Var}(R\mid T,A_v)\leq\sigma^2$. If, for some
    $c_T>0$, $N_n^{-1}\sum_i I_{i,n}(T_i-\overline T_n)^2\geq c_Tv_n^2$
    with probability tending to one, then
    \begin{equation}
        \widehat\beta_n^{\mathrm{OLS+intercept}}-\beta_0
        =O_{\mathbb P}\!\left((v_n\sqrt{m_n})^{-1}\right).
        \label{eq:certified-ols-intercept-rate}
    \end{equation}
    More generally, if $\mu_v(T)=a_v+\delta_v(T)$ with $\sup_{t\geq v}|\delta_v(t)|\leq Mv^{-\gamma}$ for some $\gamma>0$ and all large $v$, then $\widehat\beta_n^{\mathrm{OLS+intercept}}-\beta_0=O_{\mathbb P}(v_n^{-1-\gamma})+O_{\mathbb P}\{(v_n\sqrt{m_n})^{-1}\}$. In the positive linear Pareto model of \cref{sec:certification-under-monitoring} the conditional law of $R$ given $T=t$ on $A_v$ differs from its limit by a relative error of order $1/t\leq1/v$, uniformly in $t\geq v$ (see the discussion following \cref{lem:certified-exogeneity}), so $\gamma=1$ there: the intercept reduces the bias of the certified OLS from $O(v_n^{-1})$ to $O(v_n^{-2})$, which is what \cref{app:combining} uses.

    \item \emph{Median.} Let $q_v=\operatorname{Med}(Y/T\mid A_v)$ be unique.
    Suppose that, uniformly for large $v$,
    $\mathbb P(|R|\leq M\mid A_v)\geq1/2+\eta$ and the conditional density
    $f_v$ of $Y/T$ satisfies
    $\inf_{|x-q_v|\leq\delta/v}f_v(x)\geq cv$ for some
    $M<\infty$ and $\eta,c,\delta>0$. Then
    \begin{equation}
        \widehat\beta_n^{\mathrm{med}}-\beta_0
        =O_{\mathbb P}(v_n^{-1})
         +O_{\mathbb P}\!\left((v_n\sqrt{m_n})^{-1}\right).
        \label{eq:certified-median-rate}
    \end{equation}
If $q_v=\beta_0$ for all sufficiently large $v$, the first term is absent.
    Moreover, if
    $f_{v_n}(q_{v_n}+t/(v_n\sqrt{m_n}))/v_n\to g\in(0,\infty)$ locally
    uniformly in $t$, then
    \begin{equation*}
        v_n\sqrt{m_n}\,
        (\widehat\beta_n^{\mathrm{med}}-q_{v_n})
        \xrightarrow{d}\mathcal N\!\left(0,\frac{1}{4g^2}\right).
    \end{equation*}
     \item \emph{Ray estimator.} Define $b_v^{\mathrm{ray}} := v\,\mathbb E(R/\lVert\boldsymbol X\rVert\mid A_v)/\mathbb E(T/\lVert\boldsymbol X\rVert\mid A_v)$.
    Suppose, uniformly for large $v$,
    $\mathbb E(R^2\mid T,A_v)\leq M_2$ almost surely. Then
    $b_v^{\mathrm{ray}}=O(1)$ and
    \begin{equation*}
        \widehat\beta_n^{\mathrm{ray}}-\beta_0
        =
        \frac{b_{v_n}^{\mathrm{ray}}}{v_n}
        +
        O_{\mathbb P}\!\left((v_n\sqrt{m_n})^{-1}\right)
        =
        O(v_n^{-1})
        +
        O_{\mathbb P}\!\left((v_n\sqrt{m_n})^{-1}\right).
        \tag{\ref*{eq:certified-ray-rate}}
    \end{equation*}
    If $\mathbb E(R\mid T,A_v)=0$ almost surely, then
    $b_v^{\mathrm{ray}}=O(v^{-1})$, so the first term improves to
    $O(v_n^{-2})$. If $\mathbb E(R/\lVert\boldsymbol X\rVert\mid A_v)=0$, it is absent.
    
    If, in addition, under $\mathbb P(\,\cdot\mid A_{v_n})$,
    \[
        \left(\frac{T}{v_n},R\right)
        \xrightarrow{d}
        (Z_\infty,R_\infty),
        \qquad
        Z_\infty\geq1\quad\text{almost surely},
    \]
    and $\sup_n\mathbb E(|R|^{2+\delta}\mid A_{v_n})<\infty$ for some
    $\delta>0$, then, writing
    $b_n^{\mathrm{ray}}:=b_{v_n}^{\mathrm{ray}}$,
    \begin{equation*}
        \sqrt{m_n}\left\{
            v_n(\widehat\beta_n^{\mathrm{ray}}-\beta_0)
            -b_n^{\mathrm{ray}}
        \right\}
        \xrightarrow{d}
        \mathcal N\!\left(
            0,
            \operatorname{Var}\!\left(
                \frac{R_\infty}{Z_\infty}
            \right)
        \right).
    \end{equation*}
\end{enumerate}
\end{proposition}

\begin{proof}
For OLS through the origin, decompose
\begin{equation*}
\widehat\beta_n^{\mathrm{OLS}}-\beta_0
=
\frac{\sum_i I_{i,n}T_i\mu_{v_n}(T_i)}
     {\sum_i I_{i,n}T_i^2}
+
\frac{\sum_i I_{i,n}T_i\{R_i-\mu_{v_n}(T_i)\}}
     {\sum_i I_{i,n}T_i^2}.
\end{equation*}
On the selected sample, $T_i\geq v_n$, so the first term is bounded by
$M/v_n$. Conditional on the selected treatment values, the variance of the
second term is at most
$\sigma^2/\sum_i I_{i,n}T_i^2\leq\sigma^2/(N_nv_n^2)$. Since
$N_n/m_n\xrightarrow{\mathbb P}1$, this proves
\eqref{eq:certified-ols-rate}; if $\mu_v(T)=0$, the first term vanishes.

For the limit distribution, note that $v_n(\widehat\beta_n^{\mathrm{OLS}}-\beta_0)=\overline{ZR}_n/\overline{Z^2}_n$, where $\overline{ZR}_n$ and $\overline{Z^2}_n$ denote the averages of $Z_{i,n}R_i$ and of $Z_{i,n}^2$ over the certified sample, $Z_{i,n}:=T_i/v_n$. Conditionally on the certification indicators, the certified observations are i.i.d.\ draws from $\mathbb P(\,\cdot\mid A_{v_n})$, and $\xi_{i,n}:=Z_{i,n}R_i-b_n^{\mathrm{OLS}}Z_{i,n}^2$ has conditional mean zero by the definition of $b_n^{\mathrm{OLS}}$. Since $Z_n\geq1$ on $A_{v_n}$, $|b_n^{\mathrm{OLS}}|\leq\mathbb E(|Z_nR|\mid A_{v_n})$ is bounded, so the moment condition gives $\sup_n\mathbb E(|\xi_{i,n}|^{2+\delta}\mid A_{v_n})<\infty$, and the Lyapunov central limit theorem for triangular arrays yields $\sqrt{N_n}\,\overline\xi_n\xrightarrow{d}\mathcal N(0,\sigma_\xi^2)$ with $\sigma_\xi^2=\lim_n\operatorname{Var}(\xi_{i,n}\mid A_{v_n})=\operatorname{Var}(Z_\infty R_\infty-b_\infty^{\mathrm{OLS}}Z_\infty^2)$; the convergence of the variances and $b_n^{\mathrm{OLS}}\to b_\infty^{\mathrm{OLS}}$ follow from $(Z_n,R)\xrightarrow{d}(Z_\infty,R_\infty)$ and the uniform integrability provided by the $(2+\delta)$-moments. Finally, $\overline{Z^2}_n\xrightarrow{\mathbb P}\mathbb E(Z_\infty^2)$ and $N_n/m_n\xrightarrow{\mathbb P}1$, and since $\sqrt{m_n}\{v_n(\widehat\beta_n^{\mathrm{OLS}}-\beta_0)-b_n^{\mathrm{OLS}}\}=\sqrt{m_n}\,\overline\xi_n/\overline{Z^2}_n$, Slutsky's lemma gives the stated limit.

For OLS with an intercept, put
$S_{T,n}=\sum_i I_{i,n}(T_i-\overline T_n)^2$. Since centred treatment values
sum to zero,
\begin{equation*}
    \widehat\beta_n^{\mathrm{OLS+intercept}}-\beta_0
    =
    \frac{\sum_i I_{i,n}(T_i-\overline T_n)(R_i-a_{v_n})}{S_{T,n}}.
\end{equation*}
Conditional on the selected treatment values, the numerator has mean zero
and variance at most $\sigma^2S_{T,n}$. The spread condition gives
$S_{T,n}\geq c_TN_nv_n^2$ with probability tending to one, and hence the
ratio is
$O_{\mathbb P}\{(v_n\sqrt{N_n})^{-1}\}$. Together with
$N_n/m_n\xrightarrow{\mathbb P}1$, this proves
\eqref{eq:certified-ols-intercept-rate}.
Under $\mu_v(T)=a_v+\delta_v(T)$ the numerator acquires the additional term $\sum_iI_{i,n}(T_i-\overline T_n)\delta_{v_n}(T_i)$, which by the Cauchy--Schwarz inequality is at most $S_{T,n}^{1/2}\{N_n\sup_{t\geq v_n}\delta_{v_n}(t)^2\}^{1/2}\leq S_{T,n}^{1/2}N_n^{1/2}Mv_n^{-\gamma}$; divided by $S_{T,n}\geq c_TN_nv_n^2$ this is $O_{\mathbb P}(v_n^{-1-\gamma})$.

For the median, the event $\{|R|\leq M\}$ implies
$|Y/T-\beta_0|\leq M/v$ on $A_v$. Since this event has conditional
probability strictly larger than one half, $|q_v-\beta_0|\leq M/v$. The
standard sample-quantile expansion gives
$\widehat\beta_n^{\mathrm{med}}-q_{v_n}
=O_{\mathbb P}\{(f_{v_n}(q_{v_n})\sqrt{m_n})^{-1}\}$, which is
$O_{\mathbb P}\{(v_n\sqrt{m_n})^{-1}\}$ under the density lower bound. Explicitly, for $0<t\leq\delta/v_n$ the conditional distribution function $F_{v_n}$ of $Y/T$ given $A_{v_n}$ satisfies $F_{v_n}(q_{v_n}+t)-\tfrac12\geq cv_nt$, so Hoeffding's inequality applied to the empirical distribution function of the certified sample gives $\mathbb P(\widehat\beta_n^{\mathrm{med}}>q_{v_n}+t\mid N_n)\leq\exp(-2N_nc^2v_n^2t^2)$, and symmetrically below $q_{v_n}$; taking $t=K/(cv_n\sqrt{N_n})$, which is below $\delta/v_n$ for large $n$, and using $N_n/m_n\xrightarrow{\mathbb P}1$ gives the rate. The
local uniform density condition yields the stated triangular-array median
limit: with $t_n:=t/(v_n\sqrt{m_n})$ it gives $\sqrt{m_n}\{F_{v_n}(q_{v_n}+t_n)-\tfrac12\}\to gt$, so $\mathbb P(\widehat\beta_n^{\mathrm{med}}\leq q_{v_n}+t_n\mid N_n)=\mathbb P\{\mathrm{Bin}(N_n,F_{v_n}(q_{v_n}+t_n))\geq N_n/2\}\to\Phi(2gt)$ by the normal approximation to the binomial and $N_n/m_n\xrightarrow{\mathbb P}1$, where $\Phi$ is the standard normal distribution function; this is the distribution function of $\mathcal N(0,1/(4g^2))$ at $t$.

For the ray estimator, let $G=R/S$, $H=T/S$, with $S:=\lVert\boldsymbol X\rVert=|T|+|Y|$,
$r_v=\mathbb E(G\mid A_v)$, and $d_v=\mathbb E(H\mid A_v)$. Since
\[
    H
    =
    \frac{1}{1+|\beta_0+R/T|}
\]
and $\mathbb E\{(R/T)^2\mid A_v\}\leq M_2/v^2$, we have
$d_v\to(1+|\beta_0|)^{-1}>0$. Moreover,
$\mathbb E(G^2\mid A_v)\leq M_2/v^2$. Conditional on $N_n$, the selected
sample averages therefore satisfy
\[
    \overline G_n-r_{v_n}
    =
    O_{\mathbb P}\!\left((v_n\sqrt{N_n})^{-1}\right),
    \qquad
    \overline H_n-d_{v_n}
    =
    O_{\mathbb P}(N_n^{-1/2}).
\]
Since
$\widehat\beta_n^{\mathrm{ray}}-\beta_0=\overline G_n/\overline H_n$,
a ratio expansion and
$N_n/m_n\xrightarrow{\mathbb P}1$ give
\[
    \widehat\beta_n^{\mathrm{ray}}-\beta_0
    =
    \frac{r_{v_n}}{d_{v_n}}
    +
    O_{\mathbb P}\!\left((v_n\sqrt{m_n})^{-1}\right)
    =
    \frac{b_{v_n}^{\mathrm{ray}}}{v_n}
    +
    O_{\mathbb P}\!\left((v_n\sqrt{m_n})^{-1}\right).
\]
Also, $|r_v|\leq M_2^{1/2}/v$, so $b_v^{\mathrm{ray}}=O(1)$.

If $\mathbb E(R\mid T,A_v)=0$, write
$g(x)=(1+|\beta_0+x|)^{-1}$. Since $g$ is Lipschitz,
\[
    r_v
    =
    \mathbb E\!\left[
        \frac{R}{T}
        \left\{
            g(R/T)-g(0)
        \right\}
        \,\middle|\,
        A_v
    \right]
    =
    O(v^{-2}),
\]
and hence $b_v^{\mathrm{ray}}=O(v^{-1})$. Exact weighted centring,
$\mathbb E(R/S\mid A_v)=0$, gives $b_v^{\mathrm{ray}}=0$.

For the limit distribution of the ray estimator, write $Z_n:=T/v_n$, so that on $A_{v_n}$ we have $H=g(R/T)$ and $v_nG=g(R/T)\,R/Z_n$ with $g$ as above. Then $v_n(\widehat\beta_n^{\mathrm{ray}}-\beta_0)-b_n^{\mathrm{ray}}=\overline\xi_n/\overline H_n$, where $\xi_{i,n}:=v_nG_i-b_n^{\mathrm{ray}}H_i=H_i(R_i/Z_{i,n}-b_n^{\mathrm{ray}})$ has conditional mean zero by the definition of $b_n^{\mathrm{ray}}$. Since $Z_n\geq1$ and $H\leq1$ on $A_{v_n}$, $|\xi_{i,n}|\leq|R_i|+|b_n^{\mathrm{ray}}|$ with $|b_n^{\mathrm{ray}}|\leq\mathbb E(|R|\mid A_{v_n})/d_{v_n}$ bounded, so the $(2+\delta)$-moment condition on $R$ makes the Lyapunov central limit theorem applicable to the certified sample, conditionally on the certification indicators. Under $\mathbb P(\,\cdot\mid A_{v_n})$, $R/T=R/(v_nZ_n)\xrightarrow{\mathbb P}0$, hence $H\xrightarrow{\mathbb P}(1+|\beta_0|)^{-1}$ and $R/Z_n\xrightarrow{d}R_\infty/Z_\infty$ (using $Z_\infty\geq1$), and by uniform integrability $b_n^{\mathrm{ray}}\to\mathbb E(R_\infty/Z_\infty)$ and $\operatorname{Var}(\xi_{i,n}\mid A_{v_n})\to\operatorname{Var}(R_\infty/Z_\infty)/(1+|\beta_0|)^2$. Together with $\overline H_n\xrightarrow{\mathbb P}(1+|\beta_0|)^{-1}$ and $N_n/m_n\xrightarrow{\mathbb P}1$, Slutsky's lemma gives $\sqrt{m_n}\{v_n(\widehat\beta_n^{\mathrm{ray}}-\beta_0)-b_n^{\mathrm{ray}}\}\xrightarrow{d}\mathcal N(0,\operatorname{Var}(R_\infty/Z_\infty))$.
\end{proof}

\begin{customcor}{\ref*{thm:tail-index-certified-rates}}
[Tail-index-dependent rates]
Assume that the certifying events $A_v$ are nested, satisfying $A_v\subseteq\{T\geq v\}$ for all $v$, and being regularly varying:
\[
    p(v):=\mathbb P(A_v)\sim c_Av^{-\alpha},
    \qquad c_A>0,\quad \alpha>0.
\]
Let $k_n\to\infty$, $k_n/n\to0$, and \(
    v_n:=\left(\frac{c_An}{k_n}\right)^{1/\alpha}.
\)

\begin{enumerate}[label=(\roman*),leftmargin=2.5em]
    \item \emph{OLS and median estimators.}
    Each estimator in
    \cref{prop:certified-rates-extended}, under its corresponding
    conditions, satisfies
    \begin{equation*}
        \widehat\beta_n-\beta_0
        =
        O_{\mathbb P}\!\left(
            \left(\frac{k_n}{n}\right)^{1/\alpha}
        \right).
    \end{equation*}
    \item \emph{Ray estimator.}
 Suppose that $A_v=A_v^{T,S}=\{T>v,\ S\in\mathcal B\}$ and that the conditions of
    Theorem~\ref{thm:naive-certified-estimator-rate} for the ray estimator hold. Then the radial limit is standard Pareto,
    \(
        \mathbb P(Z_\alpha>z)=z^{-\alpha},\ z\geq1,
    \) and, writing $b_n^{\mathrm{ray}}:=b_{v_n}^{\mathrm{ray}}$, under the joint convergence $(T/v_n,R)\mid A_{v_n}\overset{d}{\to}(Z_\alpha,R_\infty)$ assumed in \cref{prop:certified-rates-extended}, $b_n^{\mathrm{ray}}\to
    b^{\mathrm{ray}}
    :=\mathbb E(R_\infty/Z_\alpha)$, and \begin{equation*}
        \widehat\beta_n^{\mathrm{ray}}-\beta_0
        =
        b_n^{\mathrm{ray}}
        \left(\frac{k_n}{c_An}\right)^{1/\alpha}
        +
        O_{\mathbb P}\!\left(
            n^{-1/\alpha}k_n^{1/\alpha-1/2}
        \right).
        \tag{\ref*{eq:ray-tail-index-expansion}}
    \end{equation*}
    Thus its generic rate is also $O_{\mathbb P}((k_n/n)^{1/\alpha})$. If $\sqrt{k_n}\,b_n^{\mathrm{ray}}\to0$, its rate improves to $O_{\mathbb P}(n^{-1/\alpha}k_n^{1/\alpha-1/2})$.
\end{enumerate}
\end{customcor}

\begin{proof}
By the choice of $v_n$,
\[
    m_n
    =
    n p(v_n)
    \sim
    nc_Av_n^{-\alpha}
    =
    k_n.
\]
Part~(i) therefore follows directly from
Theorem~\ref{thm:naive-certified-estimator-rate}.

For part~(ii), for every $z\geq1$,
\[
    \mathbb P\!\left(
        \frac{T}{v_n}>z
        \,\middle|\,
        A_{v_n}
    \right)
    =
    \frac{p(zv_n)}{p(v_n)}
    \longrightarrow z^{-\alpha}.
\]
Hence the radial limit in
Theorem~\ref{thm:naive-certified-estimator-rate} is $Z_\alpha$, that is, $Z_\infty=Z_\alpha$ in law under the joint convergence assumed there, and $b_n^{\mathrm{ray}}\to b^{\mathrm{ray}}=\mathbb E(R_\infty/Z_\alpha)$ by the proof of that theorem.
Substituting $v_n=(c_An/k_n)^{1/\alpha}$ and $m_n\sim k_n$ into that theorem shows that
\[
\sqrt{k_n}\Bigl\{
\left(\frac{c_An}{k_n}\right)^{1/\alpha}
(\widehat\beta_n^{\mathrm{ray}}-\beta_0)
-b_n^{\mathrm{ray}}
\Bigr\}
\xrightarrow{d}
\mathcal N\left(
0,
\operatorname{Var}\left(
\frac{R_\infty}{Z_\alpha}
\right)
\right),
\]
and yields \eqref{eq:ray-tail-index-expansion}.
\end{proof}

\begin{proposition}[A single certified maximum is sufficient]
\label{prop:single-certified-maximum}
Let $C=\{S\in\mathcal B\}$. On the event that at least one observation
is certified, let $i_n^\star\in\argmax\{T_i:1\leq i\leq n,\,S_i\in\mathcal B\}$ and
\begin{equation*}
    \widehat\beta_n^{\max}
    =\frac{Y_{i_n^\star}}{T_{i_n^\star}}.
    \tag{\ref*{eq:single-maximum-estimator}}
\end{equation*}
Suppose $\mathbb P(T>t,C)=t^{-\alpha}L(t)$ for some $\alpha>0$, and, for
some $q>0$ and $K<\infty$,
$\mathbb E(|R|^q\mid T,C)\leq K$ almost surely on
$\{T\geq t_0,C\}$. Define the estimator arbitrarily when no observation is
certified. Then, with
$U_C(s)=\inf\{t>0:\mathbb P(T>t,C)\leq s^{-1}\}$,
\begin{equation}
    \widehat\beta_n^{\max}-\beta_0
    =O_{\mathbb P}\!\left(U_C(n)^{-1}\right).
    \label{eq:single-maximum-general-rate}
\end{equation}
If $\mathbb P(T>t,C)\sim c_A t^{-\alpha}$, this becomes
$\widehat\beta_n^{\max}-\beta_0=O_{\mathbb P}(n^{-1/\alpha})$; in particular,
it is $o_{\mathbb P}(n^{-1/2})$ when $\alpha<2$.
\end{proposition}

\begin{proof}
Let $T_n^\star=T_{i_n^\star}$. Regular variation of the certified tail gives
\begin{equation*}
    \frac{T_n^\star}{U_C(n)}\xrightarrow{d}Z_\alpha^{\max},
    \qquad
    \mathbb P(Z_\alpha^{\max}\leq z)=\exp(-z^{-\alpha}),\quad z>0,
\end{equation*}
a Fr\'echet limit (not to be confused with the Pareto variable $Z_\alpha$ of Corollary~\ref{thm:tail-index-certified-rates}); indeed, $\mathbb P(T_n^\star\leq U_C(n)z)=\{1-\mathbb P(T>U_C(n)z,\,C)\}^n$ and $n\,\mathbb P(T>U_C(n)z,\,C)\to z^{-\alpha}$.
The probability that $T_n^\star<t_0$ tends to zero. Conditional on all
certification indicators and treatment values, the selected index
$i_n^\star$ is fixed; the uniform conditional moment bound and Markov's
inequality therefore imply $R_{i_n^\star}=O_{\mathbb P}(1)$. Consequently,
\begin{equation*}
    U_C(n)(\widehat\beta_n^{\max}-\beta_0)
    =
    \frac{R_{i_n^\star}}{T_n^\star/U_C(n)}
    =O_{\mathbb P}(1),
\end{equation*}
which proves \eqref{eq:single-maximum-general-rate}. The remaining claims
follow from $U_C(n)\sim(c_A n)^{1/\alpha}$.
\end{proof}

\begin{proposition}[A fixed number of certified observations]
\label{prop:fixed-k-rate}
Let $k\geq1$ be fixed and let the certified sample consist of the $k$ observations with the largest source values among those with $S_i\in\mathcal B$, with $v_n$ the $(k+1)$th largest such value, as in \cref{sec:simulations}. Suppose that $\mathbb P(E>t,\ S\in\mathcal B)\sim c_At^{-\alpha}$ for some $\alpha,c_A>0$, that the conditional distribution of $E$ given $S\in\mathcal B$ is continuous, that $A_v\subseteq\{T\geq c_0v\}$ for some $c_0>0$ and all large $v$, and that $R\mid A_v$ is tight in the sense of \eqref{eq:selected-remainder-controlled}. Then
\[
    \widehat\beta_n^{\mathrm{OLS}}-\beta_0=O_{\mathbb P}(n^{-1/\alpha}),
\]
and the same holds for $\widehat\beta_n^{\mathrm{med}}$ and $\widehat\beta_n^{\mathrm{ray}}$, and for $\widehat\beta_n^{\mathrm{OLS+intercept}}$ if in addition $\sum_iI_{i,n}(T_i-\overline T_n)^2\geq c_Tv_n^2$ with probability tending to one. No moment condition on $R$ is required.
\end{proposition}

\begin{proof}
Put $X_i=E_i$ if $S_i\in\mathcal B$ and $X_i=-\infty$ otherwise. The $X_i$ are independent and identically distributed with $\mathbb P(X_1>t)\sim c_At^{-\alpha}$, and $v_n$ is their $(k+1)$th largest value, so $v_n/(c_An)^{1/\alpha}$ converges in distribution to the $(k+1)$th largest point of a Poisson process with intensity $\alpha x^{-\alpha-1}$ on $(0,\infty)$ \citep{supp:Resnick}, which is positive almost surely; hence $1/v_n=O_{\mathbb P}(n^{-1/\alpha})$ and $v_n\to\infty$ in probability. Conditionally on $v_n$ and on the certified index set, the certified observations are independent with law $\mathbb P(\,\cdot\mid A_{v_n})$, by the conditional representation of the upper order statistics used in the proof of \cref{prop:rank-interval}. The tightness of $R\mid A_v$ therefore gives $\max_i|R_i|=O_{\mathbb P}(1)$ over the certified sample, while $\min_iT_i\geq c_0v_n$. Hence
\[
    \bigl|\widehat\beta_n^{\mathrm{OLS}}-\beta_0\bigr|
    =\frac{\bigl|\sum_iI_{i,n}T_iR_i\bigr|}{\sum_iI_{i,n}T_i^2}
    \leq\frac{\max_i|R_i|}{\min_iT_i}
    \leq\frac{\max_i|R_i|}{c_0v_n}
    =O_{\mathbb P}(n^{-1/\alpha}),
\]
and the same bound holds for the median of the ratios $Y_i/T_i=\beta_0+R_i/T_i$ and for the ray estimator, which is a weighted average of these ratios with non-negative weights. For the version with an intercept, the numerator $\sum_iI_{i,n}(T_i-\overline T_n)(R_i-\overline R_n)$ is at most $2\max_i|R_i|\sum_iI_{i,n}|T_i-\overline T_n|\leq2k^{1/2}\max_i|R_i|\{\sum_iI_{i,n}(T_i-\overline T_n)^2\}^{1/2}$, and dividing by $\sum_iI_{i,n}(T_i-\overline T_n)^2\geq c_Tv_n^2$ gives $O_{\mathbb P}(v_n^{-1})$ again.
\end{proof}

\section{Confidence intervals for certified estimates}
\label{app:confidence-intervals}

We construct a confidence interval for $\beta_0$ from the certified sample. The limit laws of \cref{prop:certified-rates-extended} are not a convenient basis for this. The estimators through the origin carry the bias $b_n^{\mathrm{OLS}}/v_n$ or $b_n^{\mathrm{ray}}/v_n$, which is of the same order as their standard deviation $(v_n\sqrt{m_n})^{-1}$ when $m_n$ stays bounded and dominates it when $m_n\to\infty$, so a normal interval centred at $\widehat\beta_n^{\mathrm{OLS}}$ or $\widehat\beta_n^{\mathrm{ray}}$ with a plug-in standard error does not attain its nominal coverage in either regime. The normal limit of $\widehat\beta_n^{\mathrm{OLS}}$ moreover requires $\mathbb E(Z_\infty^{4+2\delta})<\infty$, that is $\alpha>4$ under the regular variation of \cref{thm:tail-index-certified-rates}, whereas the application in \cref{app:application} has $\alpha\leq2$; and all limits require $m_n\to\infty$, whereas the certified samples of \cref{sec:simulations,sec:application} contain between three and twelve observations. 
We therefore invert a rank test on the certified sample instead. The resulting interval is exact when the certified treatment values are independent of the certified remainders, which certification delivers in the limit. The test requires no moments and no variance estimate, and for a fixed number of certified observations its coverage converges to the exact level of the rank test, which is at least the nominal level. We verify the coverage in a simulation.

Throughout, $A_v=A_v^{E,S}=\{E>v,\ S\in\mathcal B\}$ are the certifying events of a source--safeguard pair with $\mathbb P(A_v)>0$ for all $v$, $C=\{S\in\mathcal B\}$, the observations $(E_i,S_i,T_i,Y_i)$ are independent copies of $(E,S,T,Y)$, and $R_i=Y_i-\beta_0T_i$ is the remainder of observation $i$.

\begin{assumption}[Certified exogeneity]
\label{ass:certified-exogeneity}
There are independent random variables $Z_\infty$ and $R_\infty$ with continuous distribution functions such that
\begin{equation}
    (T/v,R)\mid A_v\overset{d}{\longrightarrow}(Z_\infty,R_\infty),
    \qquad v\to\infty.
    \label{eq:certified-exogeneity}
\end{equation}
\end{assumption}

The joint convergence in \eqref{eq:certified-exogeneity} is the same one as assumed in \cref{prop:certified-rates-extended}; under the regular variation of \cref{thm:tail-index-certified-rates} the radial limit $Z_\infty$ is the Pareto variable $Z_\alpha$ there. The additional content of Assumption~\ref{ass:certified-exogeneity} is the independence $Z_\infty \indep R_\infty$ of the two limits. It holds whenever the exceedance of the source is produced by the source's own noise while the confounder is bounded on the safeguard event.

\begin{lemma}
\label{lem:certified-exogeneity}
Consider model \eqref{SCM_general} with $E=T$, so that $A_v=\{T>v,\ S\in\mathcal B\}$ and $R=h_Y(U,S,\varepsilon_Y)$, and suppose that $T=g(U,S)+\varepsilon_T$, where $\varepsilon_T$ is independent of $(U,S,\varepsilon_Y)$ and has a regularly varying upper tail with index $\alpha>0$, and that $|g(U,S)|\leq M$ almost surely on $\{S\in\mathcal B\}$ for some finite $M$. Then
\[
    (T/v,R)\mid A_v\overset{d}{\longrightarrow}(Z_\alpha,R_\infty),
    \qquad Z_\alpha\indep R_\infty,
\]
where $Z_\alpha$ is standard Pareto, $\mathbb P(Z_\alpha>z)=z^{-\alpha}$ for $z\geq1$, and $R_\infty$ is distributed as $R$ given $S\in\mathcal B$. In particular, Assumption~\ref{ass:certified-exogeneity} holds if the conditional distribution of $R$ given $S\in\mathcal B$ is continuous.
\end{lemma}

\begin{proof}
Write $\overline F(x)=\mathbb P(\varepsilon_T>x)$ and $G=g(U,S)$. The variables $G$, $R$ and $\mathbbm 1_C$ are functions of $(U,S,\varepsilon_Y)$ and hence independent of $\varepsilon_T$. For $z\geq1$ and a Borel set $B$, conditioning on $(U,S,\varepsilon_Y)$ gives
\[
    \mathbb P(T/v>z,\ R\in B\mid A_v)
    =
    \frac{\mathbb E\{\overline F(zv-G)\,\mathbbm 1\{R\in B\}\,\mathbbm 1_C\}}
         {\mathbb E\{\overline F(v-G)\,\mathbbm 1_C\}}.
\]
On $C$ we have $|G|\leq M$, so $\overline F(zv+M)\leq\overline F(zv-G)\leq\overline F(zv-M)$, and $\overline F(zv\pm M)/\overline F(v)=\overline F\{v(z\pm M/v)\}/\overline F(v)\to z^{-\alpha}$ by the uniform convergence theorem for regularly varying functions \citep[Theorem~1.5.2]{supp:BinghamGoldieTeugels1987}. Hence $\overline F(zv-G)/\overline F(v)\to z^{-\alpha}$ on $C$, and these ratios are bounded by $\overline F(v-M)/\overline F(v)$, which converges to one. Dividing numerator and denominator by $\overline F(v)$ and applying dominated convergence,
\[
    \mathbb P(T/v>z,\ R\in B\mid A_v)\longrightarrow z^{-\alpha}\,\mathbb P(R\in B\mid C).
\]
Since $T>v$ on $A_v$, the case $z=1$ gives $\mathbb P(R\in B\mid A_v)\to\mathbb P(R\in B\mid C)$, and subtracting the display with $B=(-\infty,r]$ yields $\mathbb P(T/v\leq z,\ R\leq r\mid A_v)\to(1-z^{-\alpha})\,\mathbb P(R\leq r\mid C)$ for all $z\geq1$ and $r\in\mathbb R$, which is the distribution function of $(Z_\alpha,R_\infty)$ with $Z_\alpha\indep R_\infty$.
\end{proof}

\paragraph{Constructing intervals.}
Fix $k\geq2$ and let the certified sample consist of the $k$ observations with the largest source values among those with $S_i\in\mathcal B$; relabel them $(T_1,Y_1),\ldots,(T_k,Y_k)$.  For $i<j$ let
\[
    s_{ij}=\frac{Y_j-Y_i}{T_j-T_i}
\]
be the slope of the line through the $i$th and $j$th certified points, let $m=\binom{k}{2}$ be the number of pairs, and let $s_{(1)}\leq\cdots\leq s_{(m)}$ be the ordered pairwise slopes. For a candidate value $\beta$, Kendall's statistic between the treatments $T_i$ and the residuals $Y_i-\beta T_i$ is
\begin{equation}
    K(\beta)
    =\sum_{i<j}\operatorname{sgn}(T_j-T_i)\operatorname{sgn}\{(Y_j-\beta T_j)-(Y_i-\beta T_i)\}
    =\sum_{i<j}\operatorname{sgn}(s_{ij}-\beta),
    \label{eq:kendall-slopes}
\end{equation}
since $(Y_j-\beta T_j)-(Y_i-\beta T_i)=(T_j-T_i)(s_{ij}-\beta)$: it is the number of pairwise slopes above $\beta$ minus the number below. At $\beta=\beta_0$ the residuals are the remainders $R_i$, and $K(\beta_0)=P_c-P_d$, where $P_c$ is the number of concordant pairs, $i<j$ with $(T_j-T_i)(R_j-R_i)>0$, and $P_d$ the number of discordant pairs, with $(T_j-T_i)(R_j-R_i)<0$; a concordant pair has $s_{ij}>\beta_0$ and a discordant pair $s_{ij}<\beta_0$. Suppose that the pairs $(T_i,R_i)$ are independent with $T_i\indep R_i$ and continuous marginals. By continuity, there are no ties, so $P_c+P_d=m$ and $K(\beta_0)=m-2P_d$. The distribution of $P_d$ does not depend on the marginals: order the observations by increasing $T$ and let $\pi(i)$ be the rank of the remainder of the $i$th observation in this order. By the independence, $\pi$ is a uniformly distributed random permutation of $\{1,\ldots,k\}$, and a pair is discordant exactly when $i<j$ and $\pi(i)>\pi(j)$, an inversion of $\pi$. Hence $P_d$ is distributed as the number $D_k$ of inversions of a uniform random permutation \citep{supp:Kendall1938}. The distribution of $D_k$ is obtained by inserting the elements one at a time: the $j$th element creates between $0$ and $j-1$ new inversions, each with probability $1/j$ and independently of the earlier ones, so that $D_k$ has probability generating function $\prod_{j=1}^k(1+q+\cdots+q^{j-1})/k!$, is symmetric about $m/2$ and takes every value in $\{0,\ldots,m\}$ with positive probability; for $k=3$, for instance, $D_3$ takes the values $0,1,2,3$ with probabilities $1/6,1/3,1/3,1/6$. For an integer $r\geq0$ with $2r<m$ let
\[
    \gamma_{k,r}=\mathbb P(|m-2D_k|\geq m-2r)=2\,\mathbb P(D_k\leq r)
\]
be the level of the test that rejects $\beta$ when $|K(\beta)|\geq m-2r$, that is, when at most $r$ of the pairwise slopes lie on one side of $\beta$. The attainable levels are discrete, $\gamma_{k,0}=2/k!$ and $\gamma_{k,1}=2k/k!$ being the two smallest. Given $\gamma\in[2/k!,1)$, let $r_{k,\gamma}=\max\{r:\gamma_{k,r}\leq\gamma\}$ and define
\begin{equation}
    \mathcal C_{k,\gamma}=\bigl[s_{(r+1)},\,s_{(m-r)}\bigr],
    \qquad r=r_{k,\gamma}.
    \label{eq:rank-interval}
\end{equation}
This is the interval of \citet{supp:Sen1968}, and it is the inversion of the test: the set of $\beta$ that the test of level $\gamma_{k,r}$ does not reject. Indeed, $K$ is a nonincreasing step function of $\beta$, equal to $m$ below $s_{(1)}$ and to $-m$ above $s_{(m)}$, and when the pairwise slopes are distinct it decreases by two across each of them, so that $|K(\beta)|<m-2r$ exactly when $s_{(r+1)}\leq\beta\leq s_{(m-r)}$. Since the interval is read off order statistics of the pairwise slopes, it requires no moments of $T$, $Y$ or $R$ and no estimate of a variance. At $\gamma=0.05$ it exists for $k\geq5$, with $r_{k,\gamma}=0,1,3,5,8,11,14,18$ for $k=5,\ldots,12$. For $k=10$, for instance, there are $m=45$ pairwise slopes, $\gamma_{10,11}=0.047$ and $\gamma_{10,12}=0.073$, so that $r_{10,0.05}=11$ and $\mathcal C_{10,0.05}=[s_{(12)},s_{(34)}]$ runs from the twelfth smallest to the twelfth largest slope; at $\gamma=0.1$, the level of the intervals in \cref{sec:application}, $r_{10,0.1}=12$ and $\mathcal C_{10,0.1}=[s_{(13)},s_{(33)}]$, with exact level $0.073$. 

\begin{proposition}[Asymptotic coverage of the rank interval]
\label{prop:rank-interval}
Suppose that the conditional distributions of $E$ and of $T$ given $S\in\mathcal B$ are continuous and that Assumption~\ref{ass:certified-exogeneity} holds. Fix $k$ and $\gamma\in[2/k!,1)$, and let $\mathcal C_{k,\gamma}$ be the interval \eqref{eq:rank-interval} computed from the $k$ observations with the largest source values among those with $S_i\in\mathcal B$. Then, as $n\to\infty$,
\[
    \mathbb P(\beta_0\in\mathcal C_{k,\gamma})
    \longrightarrow
    1-\gamma_{k,r_{k,\gamma}}
    \geq1-\gamma.
\]
\end{proposition}

\begin{proof}
Let $n_C$ be the number of observations with $S_i\in\mathcal B$, which tends to infinity almost surely since $\mathbb P(C)>0$, and on $\{n_C\geq k+1\}$ let $v_n$ be the $(k+1)$th largest source value among them; $v_n\to\infty$ almost surely, since $\mathbb P(A_v)>0$ for all $v$. Conditionally on $n_C$, the safeguarded observations are independent with law $\mathbb P(\,\cdot\mid C)$, and conditionally on $v_n$ in addition, by the conditional representation of the upper order statistics of an independent sample, the $k$ observations with source values above $v_n$ are independent with law $\mathbb P(\,\cdot\mid E>v_n,\ C)=\mathbb P(\,\cdot\mid A_{v_n})$. Given $v_n$, the certified sample is thus an independent sample from the certifying event at the random threshold $v_n$; here it matters that the selection depends on $(E_i,S_i)$ only.

Let $\mu_v$ be the law of $(T/v,R)$ given $A_v$ and $\mu_\infty$ the law of $(Z_\infty,R_\infty)$. By Assumption~\ref{ass:certified-exogeneity}, $\mu_v\to\mu_\infty$ weakly, and hence $\mu_v^{\otimes k}\to\mu_\infty^{\otimes k}$. For a bounded continuous $\phi$ on $\mathbb R^{2k}$ put $\psi(v)=\int\phi\,\mathrm d\mu_v^{\otimes k}$, so that $\psi(v)\to\int\phi\,\mathrm d\mu_\infty^{\otimes k}$ as $v\to\infty$. By the first paragraph, $\mathbb E\{\phi(T_1/v_n,R_1,\ldots,T_k/v_n,R_k)\mid v_n\}=\psi(v_n)$ on $\{n_C\geq k+1\}$, and since $v_n\to\infty$ almost surely, bounded convergence gives $\mathbb E\{\phi(T_1/v_n,R_1,\ldots,T_k/v_n,R_k)\}\to\int\phi\,\mathrm d\mu_\infty^{\otimes k}$. Thus the normalised certified sample $(T_i/v_n,R_i)_{i\leq k}$ converges in distribution to $k$ independent copies of $(Z_\infty,R_\infty)$.

Let $P_c$ and $P_d$ be the numbers of concordant and discordant pairs of the certified sample, so that $P_c$ is the number of pairwise slopes above $\beta_0$ and $P_d$ the number below, and
\[
    \{\beta_0\in\mathcal C_{k,\gamma}\}
    =\{s_{(r+1)}\leq\beta_0\}\cap\{\beta_0\leq s_{(m-r)}\}
    =\{P_c\leq m-r-1\}\cap\{P_d\leq m-r-1\}.
\]
The pair $(P_c,P_d)$ is a function of the normalised certified sample, unchanged by the scaling of the $T_i$ by $v_n$, and continuous except where two treatment values or two remainders coincide, an event of probability zero under $\mu_\infty^{\otimes k}$ because $Z_\infty$ and $R_\infty$ have continuous distributions. By the continuous mapping theorem, $(P_c,P_d)$ converges in distribution to $(m-D_k,D_k)$, where $D_k$ is the number of discordant pairs among $k$ independent copies of $(Z_\infty,R_\infty)$, which is distributed as the number of inversions of a uniform random permutation because $Z_\infty\indep R_\infty$, as shown above. As $(P_c,P_d)$ is integer-valued,
\[
    \mathbb P(\beta_0\in\mathcal C_{k,\gamma})
    \longrightarrow
    \mathbb P(r+1\leq D_k\leq m-r-1)
    =1-2\,\mathbb P(D_k\leq r)
    =1-\gamma_{k,r},
\]
where the middle equality uses that $D_k$ is symmetric about $m/2$. Finally, $\gamma_{k,r}\leq\gamma$ by the choice of $r=r_{k,\gamma}$.
\end{proof}

\paragraph{Simulation.}
We check the coverage in designs (b) and (c) of \cref{sec:sim-iv-proxy}, using the same settings. 
 We take $\gamma=0.1$. \Cref{tab:rank-interval-sim} reports the empirical coverage of $\beta_0$, its limit $1-\gamma_{k,r_{k,\gamma}}$ of \cref{prop:rank-interval}, and the median length of the interval. In every Pareto setting the coverage differs from its limit by at most $2.3$ percentage points at every $n$. The last two rows repeat design (b) with Gaussian noise, where \eqref{ass:confounders-monitored-in-probability} fails and the certified OLS is inconsistent (\cref{sec:sim-distributional}). On the certified sample the confounder then grows with the threshold and the remainders are associated with the treatment values; the coverage is $78\%$ for $k=5$ and $54\%$ for $k=10$ at $n=10^6$, and it decreases in $n$. 

\begin{table}[htp]
\centering
\caption{Coverage of $\beta_0$, in per cent, by the $90\%$ rank interval $\mathcal C_{k,0.1}$ on the certified sample, with the median length of the interval in parentheses, over $1{,}000$ repetitions; limit is the asymptotic coverage $1-\gamma_{k,r_{k,0.1}}$ of \cref{prop:rank-interval}. Designs (b) and (c) are those of \cref{fig:sim-iv-proxy}; the last two rows use Gaussian noise.}
\label{tab:rank-interval-sim}
\small
\begin{tabular}{llccccc}
\hline
Design & $\alpha$ & $k$ & limit & $n = 10^{4}$ & $n = 10^{5}$ & $n = 10^{6}$ \\
\hline
(b) linear double proxy & 1.5 & 5 & 91.7 & 90.1 (0.34) & 91.5 (0.079) & 92.1 (0.016) \\
 & 1.5 & 10 & 92.7 & 92.3 (0.14) & 92.7 (0.034) & 91.9 (0.0069) \\
 & 2.5 & 5 & 91.7 & 92.2 (1.0) & 92.4 (0.47) & 91.0 (0.20) \\
 & 2.5 & 10 & 92.7 & 92.9 (0.41) & 93.3 (0.18) & 93.3 (0.079) \\
\hline
(c) invalid instrument & 1.5 & 5 & 91.7 & 92.6 (0.19) & 91.4 (0.042) & 90.7 (0.0095) \\
 & 1.5 & 10 & 92.7 & 92.1 (0.085) & 92.4 (0.019) & 93.0 (0.0038) \\
 & 2.5 & 5 & 91.7 & 91.3 (0.64) & 91.9 (0.25) & 92.0 (0.11) \\
 & 2.5 & 10 & 92.7 & 92.3 (0.28) & 93.2 (0.11) & 95.0 (0.045) \\
\hline
(b) with Gaussian noise & -- & 5 & 91.7 & 81.9 (1.8) & 79.5 (1.6) & 77.7 (1.7) \\
 & -- & 10 & 92.7 & 64.8 (0.65) & 59.6 (0.56) & 54.0 (0.53) \\
\hline
\end{tabular}

\end{table}

\section{Combining certified and bulk estimators}
\label{app:combining}

Certification keeps a subset of observations and discards everything else. When the causal structure also admits a classical estimator, such as proximal two-stage least squares in the double-proxy graph of \cref{sec:sim-iv-proxy}, the discarded observations are informative about $\beta_0$ too. We call such an estimator a \emph{bulk} estimator: it is fitted to the body of the distribution and rests on a structural model of the whole system, such as exclusion or a linear bridge, and on moment conditions, though not specifically on the tails. The certified estimators of \cref{sec:estimation} on the other hand rest on the tails, on monitoring, and on linearity of the single edge $T\to Y$. The two use different observations and different assumptions, and we here discuss how to combine them in a weighted average. 

We keep the classical notation of \cref{app:confidence-intervals}: $A_v=\{E>v,\ S\in\mathcal B\}$, the observations $O_i=(E_i,S_i,T_i,Y_i)$ are independent copies of $O=(E,S,T,Y)$, and $v_n$ is the smallest threshold at which exactly $k$ observations are certified, that is, the $(k+1)$th largest source value among the observations with $S_i\in\mathcal B$, attained at the index $i_n$. The threshold splits the remaining observations into three parts: the \emph{certified tail} $\mathcal I_n=\{i:E_i>v_n,\ S_i\in\mathcal B\}$, which has $k$ elements; the \emph{uncertified tail} $\mathcal K_n=\{i:E_i>v_n,\ S_i\notin\mathcal B\}$, the exceedances of the source that may have a confounded explanation; and the \emph{body} $\mathcal J_n=\{i:E_i<v_n\}$. The certified estimator $\widehat\beta_n^{\mathrm{cert}}$ is one of the estimators of \cref{sec:estimation} computed from the certified tail, and the bulk estimator $\widehat\beta_n^{\mathrm{bulk}}$ is a classical estimator computed from the body. Neither $\widehat\beta_n^{\mathrm{cert}}$ nor $\widehat\beta_n^{\mathrm{bulk}}$ uses the uncertified tail. 

\begin{lemma}[Conditional independence of the body and the certified tail]
\label{lem:body-tail-independence}
Suppose that the conditional distribution of $E$ given $S\in\mathcal B$ is continuous and that $\mathbb P(A_v)>0$ for all $v$. Conditionally on $v_n$ and on the index sets $\mathcal I_n$, $\mathcal J_n$ and $\mathcal K_n$, the observations $O_i$, $i\neq i_n$, are independent, with $O_i\sim\mathbb P(\,\cdot\mid A_{v_n})$ for $i\in\mathcal I_n$, $O_i\sim\mathbb P(\,\cdot\mid E<v_n)$ for $i\in\mathcal J_n$ and $O_i\sim\mathbb P(\,\cdot\mid E>v_n,\ S\notin\mathcal B)$ for $i\in\mathcal K_n$. In particular, $\widehat\beta_n^{\mathrm{cert}}$ and $\widehat\beta_n^{\mathrm{bulk}}$ are conditionally independent given $v_n$.
\end{lemma}

\begin{proof}
Put $X_i=E_i$ if $S_i\in\mathcal B$ and $X_i=-\infty$ otherwise. The $X_i$ are independent and identically distributed, $v_n$ is their $(k+1)$th largest value, and $\mathcal I_n=\{i:X_i>v_n\}$. Given that the $(k+1)$th largest value is attained at $i_n$ and equals $x$, the other $n-1$ observations are independent copies of $O$ conditioned on the event that exactly $k$ of them satisfy $X_i>x$; given in addition which $k$ indices these are, those $k$ observations are independent with law $\mathbb P(\,\cdot\mid X>x)=\mathbb P(\,\cdot\mid A_x)$, the remaining $n-k-1$ are independent with law $\mathbb P(\,\cdot\mid X\le x)$, and the two groups are independent of each other; this is the conditional representation of order statistics used in the proof of \cref{prop:rank-interval}. Up to null sets, $\{X\le x\}\cap\{E<x\}=\{E<x\}$ and $\{X\le x\}\cap\{E>x\}=\{E>x,\ S\notin\mathcal B\}$, so splitting the second group according to whether $E_i<x$ or $E_i>x$, which is a split of an independent sample by a measurable condition, preserves independence and gives the laws $\mathbb P(\,\cdot\mid E<x)$ and $\mathbb P(\,\cdot\mid E>x,\ S\notin\mathcal B)$.
\end{proof}

\paragraph{Precision weighting.}
Let $\widehat\beta_w=w\widehat\beta_n^{\mathrm{bulk}}+(1-w)\widehat\beta_n^{\mathrm{cert}}$ for $w\in[0,1]$, and write $\sigma_{\mathrm{bulk}}^2$ and $\sigma_{\mathrm{cert}}^2$ for the variances of the two estimators given $v_n$. If both are centred at $\beta_0$, independence (\cref{lem:body-tail-independence}) gives $\operatorname{Var}(\widehat\beta_w\mid v_n)=w^2\sigma_{\mathrm{bulk}}^2+(1-w)^2\sigma_{\mathrm{cert}}^2$, which is minimised by the precision weight $w^\ast=\sigma_{\mathrm{cert}}^2/(\sigma_{\mathrm{cert}}^2+\sigma_{\mathrm{bulk}}^2)$ with minimum $(\sigma_{\mathrm{cert}}^{-2}+\sigma_{\mathrm{bulk}}^{-2})^{-1}$ \citep{supp:GraybillDeal1959,supp:BatesGranger1969}. Stacking the estimating equations of the two estimators in a single generalised method of moments problem with the optimal weight matrix \citep{supp:hansen1982large} yields the same combination to first order, because the two sets of equations are conditionally independent. Three points deserve attention. First, averaging does not remove bias: the certified OLS through the origin can have a bias of the order of its standard deviation (\cref{prop:certified-rates-extended}), which a precision weight treats as noise, so the combination should use a certified estimator that is centred to first order, such as the certified OLS with an intercept, whose limit under Assumption~\ref{ass:certified-exogeneity} has conditional mean zero given the certified treatment values. Second, the weights require variance estimates, which are noisy on a certified sample of a few observations; the loss from a wrong weight is only quadratic, $\operatorname{Var}(\widehat\beta_w\mid v_n)-\operatorname{Var}(\widehat\beta_{w^\ast}\mid v_n)=(w-w^\ast)^2(\sigma_{\mathrm{cert}}^2+\sigma_{\mathrm{bulk}}^2)$, and $\widehat\beta_w$ improves on both components for every $w\in[2w^\ast-1,\,2w^\ast]$. Third, under the regular variation of \cref{thm:tail-index-certified-rates} with a fixed $k$, $\sigma_{\mathrm{cert}}=O(n^{-1/\alpha})$ whereas $\sigma_{\mathrm{bulk}}=O(n^{-1/2})$ for a root-$n$ consistent bulk estimator, so $w^\ast\to1$ when $\alpha>2$ and $w^\ast\to0$ when $\alpha<2$: the combination attains the faster of the two rates without knowledge of $\alpha$, and its gain over the better component is a finite-sample one. 
Finally, precision weighting presumes a consistent bulk estimator. When the bulk estimator may be biased, as under the nonlinear bridge of design (d) in \cref{sec:sim-iv-proxy}, the weight can instead be chosen from the disagreement $\widehat\beta_n^{\mathrm{bulk}}-\widehat\beta_n^{\mathrm{cert}}$, whose variance is $\sigma_{\mathrm{cert}}^2+\sigma_{\mathrm{bulk}}^2$ by \cref{lem:body-tail-independence}, trusting the bulk estimate only to the extent that the disagreement is explained by the noise of the certified estimate, as in the Stein-like combinations of \citet{supp:GreenStrawderman1991} and \citet{supp:Hansen2017Stein}; such a weight tends to zero when the bulk estimator is inconsistent, so that the combination remains consistent whenever the certified estimator is.

\paragraph{Simulation.}
We illustrate the precision-weighted combination in design (b) of \cref{sec:sim-iv-proxy}. We fit a certified OLS with an intercept and a linear proximal bridge on the body $\{Z_i<v_n\}$, as well as a precision weighted average. The weight is $\widehat w^\ast=\widehat\sigma_{\mathrm{cert}}^2/(\widehat\sigma_{\mathrm{cert}}^2+\widehat\sigma_{\mathrm{bulk}}^2)$, where $\widehat\sigma_{\mathrm{cert}}^2=\{k/(k-2)\}\sum_{i\in\mathcal I_n}(T_i-\overline T_n)^2\widehat u_i^2\big/\{\sum_{i\in\mathcal I_n}(T_i-\overline T_n)^2\}^2$ is the heteroscedasticity-consistent sandwich variance of the certified slope, with residuals $\widehat u_i$ and the small-sample factor $k/(k-2)$, and $\widehat\sigma_{\mathrm{bulk}}^2$ is the sandwich variance of the coefficient of $T$ in a two-stage least squares fit. \Cref{tab:combining-sim} reports the median absolute error of the three estimators. In every cell the combination has a smaller median absolute error than either component. Above we discussed that $w^\ast\to 0$ for $\alpha<2$, and indeed the median weight decreases with $n$ at $\alpha=1.5$ and stays near $0.9$ at $\alpha=2.5$. 

\begin{table}[htp]
\centering
\caption{Design (b) of \cref{fig:sim-iv-proxy} with $k=10$ certified observations: median absolute error $|\widehat\beta-\beta_0|$ over $1{,}000$ repetitions, with the $0.9$-quantile of the absolute error in parentheses, of the certified OLS with an intercept, the proximal bridge fitted to the body, and their precision-weighted combination; the last row of each block is the median weight $w^\ast$ on the bridge.}
\label{tab:combining-sim}
\small
\begin{tabular}{llccc}
\hline
$\alpha$ & Estimator & $n = 10^{4}$ & $n = 10^{5}$ & $n = 10^{6}$ \\
\hline
1.5 & certified OLS with intercept & 0.018 (0.14) & 0.0037 (0.024) & 0.00089 (0.0065) \\
 & bridge on body & 0.0051 (0.025) & 0.0015 (0.0080) & 0.00040 (0.0022) \\
 & precision-weighted combination & 0.0037 (0.020) & 0.00099 (0.0047) & 0.00025 (0.0013) \\
 & median weight on the bridge & 0.88 & 0.77 & 0.72 \\
\hline
2.5 & certified OLS with intercept & 0.051 (0.29) & 0.026 (0.14) & 0.0096 (0.052) \\
 & bridge on body & 0.014 (0.055) & 0.0052 (0.031) & 0.0025 (0.013) \\
 & precision-weighted combination & 0.012 (0.046) & 0.0050 (0.020) & 0.0021 (0.0088) \\
 & median weight on the bridge & 0.89 & 0.91 & 0.88 \\
\hline
\end{tabular}

\end{table}

\section{Details of the simulation study in \texorpdfstring{\cref{sec:simulations}}{Section~\getrefnumber{sec:simulations}}}
\label{app:simulations}

This section collects the numbers behind \cref{sec:simulations}; the designs, the estimators and the certified sample ($k = 3$ observations with the largest source value inside the safeguard ball) are as described there, with $1{,}000$ repetitions per setting. Panels are lettered as in \cref{fig:sim-iv-proxy}: (a) valid instrument, (b) linear double proxy, (c) invalid instrument, (d) nonlinear double proxy.

\paragraph{Structural equations.}
In every design the noise variables $\varepsilon_j$ are independent with $\mathbb P(\varepsilon_j>x)=(1+x)^{-\alpha}$ for $x\geq0$, and every coefficient $a_{jk}$ other than $\beta_0=1$ is drawn from the uniform distribution on $[0.2,1]$, once per repetition. The designs are:
(a) valid instrument, the graph of \cref{fig:iv-structure}, $Z=\varepsilon_Z$, $U=\varepsilon_U$, $T=a_{TZ}Z+a_{TU}U+\varepsilon_T$, $Y=\beta_0T+a_{YU}U+\varepsilon_Y$;
(b) linear double proxy, the graph of \cref{fig:double-proxy-structure}, $U=\varepsilon_U$, $Z=a_{ZU}U+\varepsilon_Z$, $W=a_{WU}U+\varepsilon_W$, $T=a_{TZ}Z+a_{TU}U+\varepsilon_T$, $Y=\beta_0T+a_{YU}U+a_{YW}W+\varepsilon_Y$;
(c) invalid instrument, the graph of \cref{fig:invalid-iv-structure}, as (a) but with $Z=a_{ZU}U+\varepsilon_Z$;
(d) nonlinear double proxy, the graph of (b) in which every parent enters through its square, except $T$ in the assignment of $Y$: $U=\varepsilon_U$, $Z=a_{ZU}U^2+\varepsilon_Z$, $W=a_{WU}U^2+\varepsilon_W$, $T=a_{TZ}Z^2+a_{TU}U^2+\varepsilon_T$, $Y=\beta_0T+a_{YU}U^2+a_{YW}W^2+\varepsilon_Y$.
The edge $W\to Y$ belongs to the double-proxy graph of \cref{fig:double-proxy-structure}, so $a_{YW}$ is drawn like the other coefficients. \Cref{sec:sim-distributional} uses the equations of (b) with standard Gaussian or Student-$t$ noise in place of the Pareto noise, and \cref{app:confidence-intervals,app:combining} use designs (b) and (c) unchanged.

\paragraph{Designs in which the classical assumption holds.}
In (a), two-stage least squares is more accurate than the certified OLS by a factor of three to five at $\alpha = 2.5$, and less accurate by a factor of up to $1.5$ at $\alpha = 1.5$. In (b), proximal two-stage least squares is more accurate by a factor of $2.5$ to $4$ at $\alpha = 2.5$ and of $1.5$ to $2.4$ at $\alpha = 1.5$. For the classical estimators the median of $\hat\beta - \beta_0$ is at most $3\cdot 10^{-3}$ in magnitude in every setting, so their median absolute error measures dispersion around $\beta_0$. The certified OLS overshoots in every repetition, since $R \geq 0$ and $T \geq 0$ imply $\widehat\beta_n^{\mathrm{OLS}} - \beta_0 = \sum_i I_{i,n} T_i R_i / \sum_i I_{i,n} T_i^2 \geq 0$, so its median absolute error equals its median bias.

\paragraph{Designs in which the classical assumption fails.}
In (c), two-stage least squares has median absolute error between $0.26$ and $0.28$ and misses $\beta_0$ by more than $0.1$ in $80$--$95\%$ of the repetitions; in (d), proximal two-stage least squares has median absolute error $1.01$ and misses by more than $0.1$ in $95\%$ of the repetitions, in both cases without any trend in $n$. The disqualified pair $(Z, \emptyset)$ in (c) misses $\beta_0$ by more than $0.1$ in $47$--$62\%$ of the repetitions at every $n$; its error distribution is bimodal, because the certified $Z$ values are driven by $\varepsilon_Z$ in some repetitions and by $U$ in others, so its median absolute error is erratic. The correctly specified quadratic bridge in (d) is consistent, with fitted slopes $-1.20$ and $-0.74$, and is more accurate than the certified OLS by a factor of $1.7$ to $4.2$ across $n$ and both tail indices.

\paragraph{Rates.}
\Cref{tab:sim-slopes} reports the least-squares slope of $\log$ median absolute error on $\log n$ for the certified OLS, against the rate $n^{-1/\alpha}$ of \cref{thm:tail-index-certified-rates}. In (d), $T \geq a_{TZ} Z^2$, so the certified treatment values grow like $n^{2/\alpha}$ and the slope is compared with $-2/\alpha$.

\begin{table}[t]
\centering
\caption{Fitted slopes of the certified OLS curves in \cref{fig:sim-iv-proxy}, with the theoretical slope in parentheses.}
\label{tab:sim-slopes}
\begin{tabular}{lcc}
\hline
Design & $\alpha = 1.5$ & $\alpha = 2.5$ \\
\hline
(a) valid instrument & $-0.70$ \ ($-0.67$) & $-0.39$ \ ($-0.40$) \\
(b) linear double proxy & $-0.68$ \ ($-0.67$) & $-0.40$ \ ($-0.40$) \\
(c) invalid instrument & $-0.66$ \ ($-0.67$) & $-0.41$ \ ($-0.40$) \\
(d) nonlinear double proxy & $-1.32$ \ ($-1.33$) & $-0.82$ \ ($-0.80$) \\
\hline
\end{tabular}
\end{table}

\paragraph{Failure of the distributional assumption.}
Under Gaussian noise the median absolute error of the certified OLS decreases only from $0.25$ to $0.24$ between $n = 10^4$ and $n = 10^6$, and its median bias stays between $0.23$ and $0.24$, while the error of proximal two-stage least squares decreases from $3.5\cdot 10^{-2}$ to $3.3\cdot 10^{-3}$ on the same data. Under Student-$t$ noise, from $n = 10^4$ to $n = 10^6$ the median absolute error of the certified OLS falls by a factor $21$ at $\nu = 1.5$ but only by a factor $1.1$ at $\nu = 30$; at $n = 10^6$ it increases monotonically in $\nu$, from $8\cdot 10^{-4}$ at $\nu = 1.5$ to $0.20$ at $\nu = 30$. At $n = 10^6$ the certified OLS is approximately median-unbiased for $\nu \leq 3$, whereas its median bias is $0.01$, $0.08$ and $0.19$ at $\nu = 5$, $10$ and $30$. The bridge estimator has error about $3\cdot 10^{-3}$ at $n = 10^6$ for every $\nu \geq 2.5$.

\section{Application: data, selection rules and estimation}
\label{app:application}

This section gives the details of the river-discharge application in \cref{sec:application}. 
The code and the cached data are in the \texttt{hydro/} directory of the repository \url{https://github.com/nikolajthams/extreme-proxy}, and every constant quoted below is set in \texttt{hydro/config.py}.

\paragraph{Data.} We use daily mean discharge (USGS NWIS parameter 00060) from $1930$ to $2024$ for every gauge in three basins, which we identify by their outlet gauge: the Susquehanna at Harrisburg, the Wabash at Mt.\ Carmel and the Iowa at Wapello. The network topology, that is, which gauges lie upstream of which, comes from the USGS NLDI navigation service. We keep a gauge if it is a stream gauge with a published drainage area of at least $200$ km$^2$ and a record of at least $25$ years with at least $7300$ daily values, and we remove gauges whose station name indicates an impoundment or diversion, as well as gauges on a curated list of sites immediately below flood-control dams. This leaves $71$ gauges in the Susquehanna, $87$ in the Wabash and $26$ in the Iowa/Cedar basin. We restrict each basin to one flood-generating season, May--November for the Susquehanna, March--August for the Wabash and April--September for the Iowa/Cedar, so that snowmelt floods and convective floods are not mixed. We drop days whose NWIS qualifier marks them as provisional, estimated, ice-affected, equipment-affected, seasonal, under maintenance, discontinued or unknown. We call a season of a gauge's record complete if at most $5\%$ of its in-season days are missing; completeness and the tail index below are evaluated on the record before the qualifier filter.

\paragraph{Pair selection.} Within a basin, every pair $(T, Y)$ of screened gauges with $T$ upstream of $Y$ is a candidate; there are $955$ candidates. We admit a pair if it satisfies six rules. The drainage-area ratio $A_Y/A_T$ lies in $[1.2, 2.5]$, so that the catchment between the gauges contributes to $Y$ but $T$ is not negligible; without this, the confounding $U$ in the causal graph \cref{fig:app-map} would mostly be non-existent. The two records share at least $40$ complete seasons. The river distance between the gauges, which we approximate by the great-circle distance times a sinuosity factor of $1.35$, is at most $200$ km. The safeguard rule below yields three gauges. The upstream gauge is heavy-tailed: the Hill estimate of the tail index of $T$'s in-season daily discharge, computed from the upper $10\%$ of the order statistics of the whole record, is at most $2$, so that the variance is infinite; this rule removes, among others, the Wabash main stem and the larger gauges on the Susquehanna main stem. Finally, the travel-time lag passes the celerity check described below. The rules are applied in this order, and each pair is counted at the first rule it fails: of the $955$ candidates, $838$ fail the area-ratio rule, $31$ the record rule, $5$ the distance rule, $9$ have fewer than three proxy gauges, $31$ fail the tail rule and one fails the celerity check, so that $40$ pairs remain. No pair with the basin outlet as $Y$ qualifies, since every gauge in the basin is upstream of the outlet and none is disconnected from it.

\paragraph{Safeguard.} The safeguard $P$ of a pair consists of the three gauges nearest to $Y$ by river distance among those that are hydrologically disconnected from both $T$ and $Y$, drain at least $2\%$ of $T$'s area and have a complete record in at least $80\%$ of the pair's common complete seasons. When two such gauges lie on the same stream, we use only the downstream one. We write $P_1$ for the nearest gauge and $P_2$ for the second nearest.

\paragraph{Alignment.} The panel of a pair consists of the in-season days on which $T$, $Y$ and the three proxy gauges are all observed. We estimate the travel time from $T$ to $Y$ as the lag in $\{0, \dots, 6\}$ days that maximises the cross-correlation of $\log T$ and $\log Y$, computed over windows of $\pm 10$ days around the days on which $T$ exceeds its $0.95$-quantile. We read $Y$ at that lag and the proxies on the same day as $T$. The lag is $0$ days for $31$ pairs, $1$ day for $7$ and $2$ days for $2$. For the celerity check, we treat a lag of $d$ days as a travel time within half a day of $d$ and require the range of flood-wave speeds this implies over the river distance to overlap $0.3$--$5$ m/s; the one pair that fails, with a lag of $6$ days over $20$ km, is excluded. The daily record of a pair is this panel: a median of $10{,}436$ days per pair, between $6{,}593$ and $18{,}205$.

\paragraph{Estimation.} 
The certifying event is $A_v = \{T > v,\ P \in \mathcal B\}$ of \cref{sec:app-design}, where $\mathcal B$ bounds each proxy gauge by its own $q_P$-quantile over the pair's days. The threshold $v$ is the $q_T$-quantile of $T$ over the days on which $P \in \mathcal B$, and we choose $q_T$ as the largest value in $\{0.9, 0.95, 0.98, 0.99, 0.995, 0.999\}$ that leaves at least one certified day; the grid reaches deep into the tail because a daily $0.995$-quantile corresponds roughly to a flood peak. At the tabulated level $q_P = 0.75$ this selects $q_T = 0.999$ on every pair, with a median threshold of $89$ m$^3$/s and a median of six certified days (range four to twelve). If the three proxies and $T$ were independent, a fraction of about $q_P^3 (1 - q_T)$ of the days would be certified; neighbouring gauges are positively dependent, so the fraction is larger. The certified OLS is least squares through the origin of $Y$ on $T$ over the certified days; with a single certified day it is the ratio $Y/T$ at that day, the single-maximum estimator \eqref{eq:single-maximum-estimator}. Naive OLS is least squares through the origin of $Y$ on $T$ over all days. The proximal bridge is the linear bridge of \cref{sec:sim-iv-proxy}, with $P_1$ as treatment-inducing and $P_2$ as outcome-inducing proxy, fitted with an intercept as there, over all days.

The intervals in \cref{fig:app-headline}b are $90\%$ percentile intervals from a year-level bootstrap. We resample whole seasons with replacement $400$ times, which preserves the within-season dependence of daily data, hold the threshold $v$ fixed and recompute the safeguard quantiles on each resample. We drop resamples on which no day is certified. With a handful of certified days the intervals are indicative only. The intervals of V1--V3 are computed in the same way; for V2 every resample also draws a fresh permutation of the proxy block, so that its interval reflects the permutation as well as the sampling variability.

\paragraph{Highlighted pair and ablations.} The pair of \cref{fig:app-map,fig:app-headline}, the Juniata from Huntingdon to Mapleton Depot, is the admitted pair with the largest drainage-area ratio, that is, the most room for confounding inflow between the gauges. The ablations V1--V3 of \cref{fig:app-headline}b are computed at each safeguard level at the threshold $v$ of the certified OLS. V1 fits least squares through the origin on $\{T > v\}$ without the safeguard, so that V1 and the certified OLS differ in the safeguard alone. V2 permutes the proxy vector $P$ across the days, jointly for the three gauges, $50$ times; on each permutation we recompute $v$ at the same $q_T$, and we report the median of the certified OLS over the permutations. V3 replaces $P$ by three gauges in the Iowa/Cedar basin, whose storms are unrelated to those of the Juniata: the Iowa River at Marshalltown and near Rowan, and Salt Creek near Elberon. The panel and the threshold $v$ of V3 are those defined by this safeguard. The pair has $15{,}840$ aligned in-season days over $83$ years and a daily tail index of $2.0$. Naive OLS gives $2.16$ ($90\%$ interval $2.05$--$2.26$), the bridge $1.80$ ($1.56$--$2.10$) and the certified OLS $1.27$ ($1.16$--$1.54$) on $11$ certified days above $v = 77$ m$^3$/s. The ablations give $2.09$ ($1.94$--$2.23$) for V1 on the $612$ days above $v$, $1.81$ ($1.41$--$2.13$) for V2 and $1.72$ ($1.48$--$2.02$) for V3 on $10$ certified days. Along the safeguard grid of \cref{fig:app-headline}a the certified OLS falls from $1.58$ at $q_P = 0.9$ on $14$ certified days to $1.27$ at $q_P = 0.75$ and $1.24$ at $q_P = 0.5$ on six days; at $q_P \leq 0.25$ only one or two days are certified and the estimate is no longer stable. On this pair a quarter of the confounding remains at the sample sizes the safeguard leaves; across pairs, less than a tenth remains in the median.

\paragraph{Placebo design.} For the placebo of \cref{sec:app-placebo}, a disconnected pair is two screened gauges of the same basin with no flow path between them; there are $5596$ such pairs. We admit a pair if the drainage-area ratio lies in $[0.5, 2]$, the records share at least $40$ complete seasons and the gauges are at most $100$ km apart by the distance measure above, which leaves $274$ pairs. Ordering these by distance, we take the $40$ nearest pairs that have three proxy gauges by the rule above. The two gauges are read on the same day, and the estimators and the rules for $q_T$ and $v$ are those of the main design.

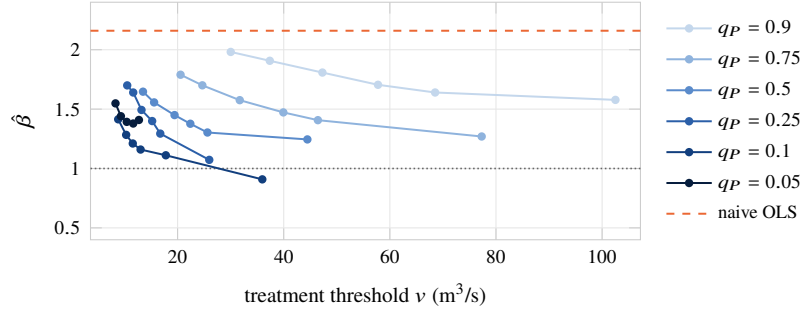
\begin{figure}[]
\centering
\begin{tikzpicture}
\definecolor{appblue}{HTML}{2A78D6}
\definecolor{apporange}{HTML}{EB6834}
\definecolor{appgreen}{HTML}{1BAF7A}
\definecolor{appviolet}{HTML}{4A3AA7}
\definecolor{appgrey}{HTML}{8A8A8A}
\definecolor{applight}{HTML}{CCCCCC}
\definecolor{appq0}{HTML}{C4D7EC}
\definecolor{appq1}{HTML}{8FB3DD}
\definecolor{appq2}{HTML}{5A8BCB}
\definecolor{appq3}{HTML}{2B5FA8}
\definecolor{appq4}{HTML}{123F7E}
\definecolor{appq5}{HTML}{061D3D}

\begin{axis}[
  width=0.60\textwidth,
  height=0.32\textwidth,
  ymin=0.4, ymax=2.4,
  ytick={0.5,1,1.5,2},
  xtick={20,40,...,100},
  enlarge x limits=0.05,
  xlabel={treatment threshold $v$ (m$^3$/s)},
  ylabel={$\hat\beta$},
  xmajorgrids,
  ymajorgrids,
  grid style={applight!50},
  axis line style={applight},
  tick style={applight},
  tick label style={font=\scriptsize},
  label style={font=\footnotesize},
  title style={font=\footnotesize, yshift=-3pt},
  every axis plot/.append style={
    line width=0.75pt,
    mark size=1.2pt
  },
  legend columns=1,
  legend style={
    at={(1.03,0.5)},
    anchor=west,
    draw=none,
    fill=none,
    font=\scriptsize,
    legend cell align=left,
    row sep=1pt
  },
  clip=false,
]

\draw[densely dotted, black!60, line width=0.6pt]
  ({rel axis cs:0,0}|-{axis cs:0,1.0}) --
  ({rel axis cs:1,0}|-{axis cs:0,1.0});

\draw[dashed, apporange, line width=0.75pt]
  ({rel axis cs:0,0}|-{axis cs:0,2.1600}) --
  ({rel axis cs:1,0}|-{axis cs:0,2.1600});

\addplot[appq0, mark=*] coordinates {
  (30.016,1.9815) (37.378,1.907) (47.289,1.8079)
  (57.766,1.7048) (68.527,1.64) (102.51,1.5778)
};
\addlegendentry{$q_P = 0.9$}

\addplot[appq1, mark=*] coordinates {
  (20.558,1.7896) (24.664,1.7004) (31.715,1.5754)
  (39.927,1.4723) (46.44,1.407) (77.305,1.2695)
};
\addlegendentry{$q_P = 0.75$}

\addplot[appq2, mark=*] coordinates {
  (13.479,1.6468) (15.574,1.5566) (19.425,1.4497)
  (22.399,1.3768) (25.627,1.3031) (44.457,1.2449)
};
\addlegendentry{$q_P = 0.5$}

\addplot[appq3, mark=*] coordinates {
  (10.506,1.6998) (11.638,1.6399) (13.196,1.4931)
  (15.206,1.3995) (16.735,1.2938) (25.967,1.0727)
};
\addlegendentry{$q_P = 0.25$}

\addplot[appq4, mark=*] coordinates {
  (8.8065,1.4142) (10.307,1.2832) (11.553,1.2106)
  (13.026,1.1596) (17.783,1.1113) (35.962,0.90854)
};
\addlegendentry{$q_P = 0.1$}

\addplot[appq5, mark=*] coordinates {
  (8.2968,1.5483) (9.3162,1.4401) (10.421,1.3927)
  (11.638,1.3788) (12.714,1.4087)
};
\addlegendentry{$q_P = 0.05$}

\addlegendimage{apporange, dashed}
\addlegendentry{naive OLS}

\end{axis}
\end{tikzpicture}
\caption{Huntingdon to Mapleton Depot. The certified OLS against the treatment threshold $v$, one line per safeguard level, darker as the safeguard tightens. Horizontal lines are the naive OLS (dashed) and the ground-truth target (dotted).}
\label{fig:app-threshold}
\end{figure}

\paragraph{Per-pair results.} \Cref{fig:app-threshold} shows that on the highlighted pair the certified OLS moves towards the target as the threshold $v$ rises within each safeguard level, and the safeguard levels order as in \cref{fig:app-headline}a. \Cref{tab:app-pairs} lists every pair with its number of days, the drainage-area ratio, naive OLS, the certified OLS with its number of certified days and the selected $q_T$, and the two-proxy bridge.

\begin{table}[htp]
\centering
\scriptsize
\setlength{\tabcolsep}{3.5pt}
\caption{The $40$ pairs of the application, ordered by naive OLS. For each pair, days is the number of days in the panel; cert.\ OLS is the certified OLS at $q_P = 0.75$, with the number of certified days and the selected $q_T$ in parentheses; bridge is the linear bridge with $(P_1, P_2)$. All estimators are fitted on the daily record. Gauges in the same town name are shown with their USGS site numbers.}
\label{tab:app-pairs}
\resizebox{\textwidth}{!}{\begin{tabular}{llrrrrr}
\toprule
Pair & Basin & days & $A_Y/A_T$ & naive OLS & cert.\ OLS ($n$, $q_T$) & bridge $(P_1,P_2)$ \\
\midrule
Tioga $\to$ Tioga Junction & Susq. & 8914 & $1.58$ & $1.14$ & $1.03$ (6, 0.999) & $1.06$ \\
Bower $\to$ Hyde & Susq. & 9356 & $1.50$ & $1.18$ & $0.67$ (6, 0.999) & $0.18$ \\
Towanda $\to$ Wilkes-Barre & Susq. & 17562 & $1.28$ & $1.24$ & $0.75$ (11, 0.999) & $1.16$ \\
Meshoppen $\to$ Danville & Susq. & 9695 & $1.29$ & $1.24$ & $1.06$ (6, 0.999) & $1.11$ \\
Curwensville $\to$ Hyde & Susq. & 9087 & $1.29$ & $1.25$ & $1.09$ (6, 0.999) & $0.97$ \\
Nora $\to$ Indianapolis & Wabash & 14400 & $1.34$ & $1.28$ & $0.87$ (8, 0.999) & $0.99$ \\
Cross Fork $\to$ Westport & Susq. & 11709 & $1.71$ & $1.31$ & $1.11$ (8, 0.999) & $-2.13$ \\
Noblesville $\to$ Nora (03349000) & Wabash & 11173 & $1.42$ & $1.36$ & $1.08$ (7, 0.999) & $1.25$ \\
Erwins $\to$ Corning & Susq. & 8936 & $1.46$ & $1.37$ & $1.42$ (6, 0.999) & $1.16$ \\
Towanda $\to$ Danville & Susq. & 17563 & $1.44$ & $1.40$ & $0.87$ (11, 0.999) & $1.20$ \\
Kokomo $\to$ Owasco & Wabash & 9471 & $1.62$ & $1.40$ & $0.40$ (6, 0.999) & $0.40$ \\
Zionsville $\to$ Indianapolis & Wabash & 10535 & $1.64$ & $1.40$ & $0.55$ (6, 0.999) & $-0.01$ \\
Columbus $\to$ Seymour & Wabash & 13255 & $1.37$ & $1.41$ & $0.93$ (8, 0.999) & $1.37$ \\
Seymour $\to$ Bedford & Wabash & 10665 & $1.65$ & $1.41$ & $0.61$ (6, 0.999) & $1.16$ \\
Noblesville $\to$ Nora (03348500) & Wabash & 8065 & $1.47$ & $1.41$ & $1.05$ (5, 0.999) & $1.08$ \\
Jerome $\to$ Kokomo & Wabash & 11145 & $1.67$ & $1.45$ & $0.81$ (7, 0.999) & $0.85$ \\
Indianapolis $\to$ Centerton & Wabash & 10346 & $1.49$ & $1.46$ & $1.01$ (6, 0.999) & $0.47$ \\
Charles City $\to$ Janesville & Iowa/Cedar & 9391 & $1.58$ & $1.50$ & $0.92$ (6, 0.999) & $0.96$ \\
Muncie $\to$ Anderson & Wabash & 12912 & $1.68$ & $1.51$ & $1.48$ (8, 0.999) & $1.05$ \\
St. Paul $\to$ Columbus & Wabash & 9865 & $1.76$ & $1.54$ & $0.82$ (7, 0.999) & $10.10$ \\
Edinburgh $\to$ Columbus & Wabash & 10406 & $1.61$ & $1.54$ & $1.14$ (7, 0.999) & $1.49$ \\
Lindley $\to$ Erwins & Susq. & 15331 & $1.79$ & $1.56$ & $1.24$ (9, 0.999) & $0.94$ \\
Mapleton Depot $\to$ Newport & Susq. & 16525 & $1.65$ & $1.59$ & $1.26$ (11, 0.999) & $1.13$ \\
Fortville $\to$ Millersville & Wabash & 11268 & $1.76$ & $1.61$ & $0.69$ (7, 0.999) & $-0.55$ \\
Indianapolis $\to$ Spencer & Wabash & 7472 & $1.83$ & $1.72$ & $1.00$ (5, 0.999) & $-0.05$ \\
Noblesville $\to$ Indianapolis (03349000) & Wabash & 10491 & $1.91$ & $1.73$ & $1.28$ (6, 0.999) & $1.49$ \\
Anderson $\to$ Noblesville (03349000) & Wabash & 9178 & $2.11$ & $1.80$ & $1.67$ (6, 0.999) & $1.12$ \\
Noblesville $\to$ Indianapolis (03348500) & Wabash & 8065 & $1.97$ & $1.82$ & $1.05$ (5, 0.999) & $1.31$ \\
Owasco $\to$ Lafayette & Wabash & 9583 & $2.01$ & $1.83$ & $1.22$ (6, 0.999) & $0.85$ \\
Anderson $\to$ Noblesville (03348500) & Wabash & 6593 & $2.04$ & $1.83$ & $1.05$ (4, 0.999) & $1.41$ \\
Nora $\to$ Centerton & Wabash & 10862 & $2.00$ & $1.84$ & $1.03$ (6, 0.999) & $0.50$ \\
Edinburgh (03362500) $\to$ Edinburgh (03363000) & Wabash & 10467 & $2.24$ & $1.98$ & $1.42$ (7, 0.999) & $2.06$ \\
Columbus $\to$ Bedford & Wabash & 10683 & $2.26$ & $1.99$ & $1.04$ (6, 0.999) & $1.57$ \\
Meshoppen $\to$ Sunbury & Susq. & 10082 & $2.10$ & $2.04$ & $0.92$ (7, 0.999) & $1.53$ \\
Edinburgh $\to$ Seymour & Wabash & 10483 & $2.21$ & $2.08$ & $1.75$ (7, 0.999) & $1.92$ \\
Carthage $\to$ Shelbyville & Wabash & 9545 & $2.29$ & $2.14$ & $0.59$ (6, 0.999) & $0.52$ \\
Huntingdon $\to$ Mapleton Depot & Susq. & 15840 & $2.49$ & $2.16$ & $1.27$ (11, 0.999) & $1.80$ \\
Nora $\to$ Spencer & Wabash & 7472 & $2.45$ & $2.19$ & $0.84$ (5, 0.999) & $-0.03$ \\
Towanda $\to$ Sunbury & Susq. & 18205 & $2.35$ & $2.29$ & $1.35$ (12, 0.999) & $1.30$ \\
Butlerville $\to$ Vernon & Wabash & 9269 & $2.31$ & $2.30$ & $1.07$ (6, 0.999) & $1.78$ \\
\bottomrule
\end{tabular}
}
\end{table}

\end{document}